\documentclass{article}[11pt]

\usepackage{custom}
\usepackage{pdfpages}
\usepackage[export]{adjustbox}
\usepackage{marginnote}
\usepackage{comment}
\usepackage{todonotes}   % todos visible
\setuptodonotes{fancyline, color=blue!10, size=\small}

\begin{document}

\newlength{\originalVOffset}
 \newlength{\originalHOffset}
 \setlength{\originalVOffset}{\voffset}   
 \setlength{\originalHOffset}{\hoffset}

 \setlength{\voffset}{0cm}
 \setlength{\hoffset}{0cm}
%\includepdf[pages={1}]
%{LongBaselineJE.png}%{Long_Baseline_Workshop.pdf}
 \setlength{\voffset}{\originalVOffset}
 \setlength{\hoffset}{\originalHOffset}

%\setlength{\voffset}{0cm}
%\setlength{\hoffset}{0cm}
%\includepdf[pages={1}]{Community-Summary-Coverpage.pdf}
%\end{figure}
%\setlength{\voffset}{-2.54cm}
%\setlength{\hoffset}{-2.54cm}

\title{Terrestrial Very-Long-Baseline Atom Interferometry: Workshop Summary
}

%\begin{figure}
%\centering 
%\includegraphics[width=1\textwidth]{LongBaselineJE.png}
%\end{figure}

\abstract{
%This document provides a summary of the 2023 Terrestrial Very-Long-Baseline Atom Interferometry Workshop hosted by CERN, which aimed to establish a pathway for forming an international TVLBAI proto-collaboration. The collaboration will bring together researchers from various institutions to plan and secure funding for terrestrial large-scale AI projects. The aim is to develop a roadmap that outlines the design and technology choices for one or several km-scale detectors, which will be ready for operation in the mid-2030s. The report includes an introduction, physics case, technical challenges, and a summary of the discussions and conclusions from the workshop. The workshop brought together experts from around the world to discuss the exciting developments in large-scale atom interferometer prototypes and their potential for detecting ultralight dark matter and gravitational waves. \todo{John: maybe reword?}
This document presents a summary of the 2023 Terrestrial Very-Long-Baseline Atom Interferometry Workshop hosted by CERN. 
The workshop brought together experts from around the world to discuss the exciting developments in large-scale atom interferometer (AI) prototypes and their potential for detecting ultralight dark matter and gravitational waves.
The primary objective of the workshop was to lay the groundwork for an international TVLBAI proto-collaboration. This collaboration aims to unite researchers from different institutions to strategize and secure funding for terrestrial large-scale AI projects.
The ultimate goal is to create a roadmap detailing the design and technology choices for one or more km-scale detectors, which will be operational in the mid-2030s. The key sections of this report %encompasses key sections, including an introduction, 
present the physics case and technical challenges, together with a comprehensive overview of the discussions at the workshop together with the main conclusions.}
% \todo{John: tweaked}

%\newpage
%\listoftodos

%\section*{How to make comments to the text:}
%You can add any comment in the text by typing %\todo{YourName: @ToPerson Comment.}  \begin{verbatim}
 %   \todo{YourName: @ToPerson Comment.}
%\end{verbatim}

\newpage
\affiliation[@]{Contact Person}
\affiliation[*]{Section Editor/Contributor, and/or Workshop Organiser}
\emailAdd{Oliver.Buchmueller@cern.ch} \emailAdd{John.Ellis@cern.ch} 
\author[1,*]{Sven~Abend,}
\author[2]{Baptiste~Allard,}
\author[3]{Iván~Alonso,}
\author[4]{John~Antoniadis,}
\author[5]{Henrique~Ara\'ujo,}
\author[6,*]{Gianluigi~Arduini,}
\author[7]{Aidan~S.~Arnold,}
\author[8]{Tobias~Aßmann,}
\author[9]{Nadja~Augst,}
\author[10,11,*]{Leonardo~Badurina,}
% ...
\affiliation[1]{Leibniz Universität Hannover, Welfengarten 1, 30167 Hannover, Germany}
\affiliation[2]{Laboratoire Collisions Agrégats Réactivité, CNRS, Université Toulouse III - Paul Sabatier, Toulouse, France}
\affiliation[3]{Department of Industrial Engineering, Higher Polytechnic School, University of the Balearic Islands, Palma de Mallorca, Spain}
\affiliation[4]{IFORTH Institute of Astrophysics, N. Plastira 100, 70013, Heraklion, Greece}
\affiliation[5]{Physics Department, Imperial College, Prince Consort Road, London, SW7 2AZ, UK}
\affiliation[6]{CERN, CH-1211 Geneva 23, Switzerland}
\affiliation[7]{SUPA Department of Physics, University of Strathclyde, Glasgow, G4 0NG, UK}
\affiliation[8]{Institut f{\"u}r Quantenphysik and Center for Integrated Quantum Science and Technology (IQST), Universit{\"a}t Ulm, Albert-Einstein-Allee 11, 89081 Ulm, Germany}
\affiliation[9]{Deutsches Zentrum für Luft- und Raumfahrt (DLR), Institut für Quantentechnologien, Wilhelm-Runge-Straße 10, 89081 Ulm, Germany}
\affiliation[10]{Physics Department, King's College London, London, WC2R 2LS, UK}
\affiliation[11]{Walter Burke Institute for Theoretical Physics, California Institute of Technology, Pasadena, CA 91125, USA}
% ...

\maketitle
\newpage
\section{Preface}
\label{sec:preface}

Atom Interferometry (AI) is a well-established quantum sensor concept based on the superposition and interference of atomic wave packets, which affords exceptionally high sensitivity, for example, to inertial/gravitational effects. AI experimental designs take advantage of features used by state-of-the-art atomic clocks in combination with established techniques for building inertial sensors. 

The experimental landscape of AI projects has expanded significantly in recent years, ranging from ultra-sensitive experimental setups to portable devices and even commercially available gravimeters. Several large-scale terrestrial AI projects based on cold atom technologies are currently under construction, in planning stages, or being proposed.

Five large-scale fully funded prototype projects are currently under construction, namely a 10~m fountain at Stanford~\cite{Dickerson2013} \& MAGIS-100~\cite{Abe2021} at FNAL in the US, MIGA~\cite{Canuel2018} in France, VLBAI~\cite{schlippert2020matter} at Hannover in Germany, AION-10~\cite{Badurina2020} at Oxford with possible 100~m sites at Boulby in the UK and at CERN under investigation, and a 10~m fountain~\cite{Zhou2011} \& ZAIGA~\cite{Zhan2019} in China. These projects will demonstrate the feasibility of AI at large scales, paving the way for terrestrial km-scale experiments as the next steps.

There have already been discussions of projects to build one or more km-scale detectors, including ELGAR in Europe~\cite{Canuel2020}, MAGIS-km at the Sanford Underground Research facility (SURF) in the US~\cite{Abe2021}, AION-km at the STFC Boulby facility in the UK~\cite{Badurina2020},  and advanced ZAIGA in China~\cite{Zhan2019}.  

The goal is that by about 2035 at least one km-scale detector will have entered operation. These km-scale experiments would not only be able to explore systematically the mid-frequency band of gravitational waves and probe possible ultralight dark matter, but would also demonstrate the readiness of key technologies ahead of a space-based AI mission such as AEDGE~\cite{ElNeaj2020,AEDGEWORKSHOP}.

%\subsection{Scope of the Workshop}
The main goals of the workshop were to establish the pathway for a proto-collaboration that could develop a Roadmap for the design and technology choices for one or several km-scale detectors to be ready for operation in the mid-2030s. This initiative is supported by the cold atom community and the potential user communities interested in its science goals. This Roadmap will outline technological milestones as well as refine interim and long-term scientific goals. 

The Workshop brought together members of the cold atom, astrophysics, cosmology, and fundamental physics communities and built upon the Community Workshop on Cold Atoms in Space held in September 2021~\cite{COMMUNITYWORKSHOP}, which reviewed the cold atom experiment landscape for space and established a corresponding Roadmap for cold atoms in space~\cite{Alonso2022}.

%\todo{John: Have commented out previous Executive Summary and moved it to Abstract}\todo{Oliver: OK - can retire this comment, I think}
%\section{Executive Summary}
%\label{sec:ES}
%{\it \textcolor {red}{Shift to Abstract?}
%This document provides a summary of the Terrestrial Very-Long-Baseline Atom Interferometry Workshop, which aimed to establish a pathway for forming an international TVLBAI proto-collaboration. The collaboration will bring together researchers from various institutions to plan and secure funding for terrestrial large-scale AI projects. The aim is to develop a roadmap that outlines the design and technology choices for one or several km-scale detectors, which will be ready for operation in the mid-2030s. The report includes an introduction, physics case, technical challenges, and a summary of the discussions and conclusions from the workshop. The workshop brought together experts from around the world to discuss the exciting developments in large-scale atom interferometer prototypes and their potential for detecting ultralight dark matter and gravitational waves.}

%\section{Steps towards a TVLBAI Proto-Collaboration}
%\label{sec:proto}
One of the main goals of the workshop was to establish a pathway for forming an international TVLBAI proto-collaboration. This collaboration will bring together researchers from various institutions to plan and secure funding for terrestrial large-scale AI projects. The aim is to develop a roadmap that outlines the design and technology choices for one or several km-scale detectors, which will be ready for operation in the mid-2030s.

The Workshop Summary, presented here, represents an important step towards the formation of the TVLBAI Proto-Collaboration. This will be followed by a town hall meeting in fall 2023, where interested communities, technology experts, and potential host site stakeholders will gather to discuss the formation of the collaboration. Templates from previous collaborations will be presented to guide the structured discussion and reach a consensus.

%Once the TVLBAI Proto-Collaboration is formed, work on the actual roadmap will begin, with follow-up workshops planned for summer 2024 to further refine the roadmap and its implementation strategies.   

%\todo{John: some repetition in the previous and subsequent Sections?} \todo{Oliver: removed some - better now? }

\section{Introduction}
%: Editors Mark Kasevich, Oliver Buchmueller, Jonathan R. Ellis}
\label{sec:intro}

Atom interferometers have potential applications
to some of the central topics in fundamental physics, namely the
nature of dark matter and measurements of gravitational waves.

One of the prominent schools of thought about
dark matter~\cite{Zwicky2009,Zwicky1937,Rubin1970,Bertone2016}
is that it may consist of waves of ultralight bosonic
fields forming coherent waves that move non-relativistically through
the Universe. Many models of such ultralight dark matter
(ULDM) candidates postulate that they would have very weak interactions with the particles of the
Standard Model (SM) that make up the visible matter in the Universe. 
As we discuss below, large-scale atom interferometers are extremely sensitive to such interactions of
ULDM fields with SM particles~\cite{Graham2016}, thanks to the very precise measurements of atomic properties that they enable. 

Atom interferometers are also sensitive to the small distortions of space-time induced by the passage of gravitational waves (GWs)~\cite{Graham2013}. The
LIGO and Virgo laser interferometers discovered GWs emitted during the mergers of black holes (BHs)
and neutron stars (NSs)~\cite{Abbott2016} (and have now been joined by the KAGRA detector~\cite{Akutsu2020}), pulsar timing arrays have recently published evidence of nHz GWs that may be emitted by binary systems of supermassive BHs
(SMBHs)~\cite{Agazie2023,Antoniadis2023,Reardon2023,Xu2023}, and mergers of SMBHs are among the prime target of the planned LISA~\cite{AmaroSeoane2017}, TianQin~\cite{Luo2016} and Taiji~\cite{Ruan2020} space-borne
laser interferometers. As we discuss in more detail below,
large-scale atom interferometers are sensitive to GWs in a frequency range intermediate between LISA and
LIGO/Virgo, offering complementary measurements of GWs that might cast light on the mechanisms
for forming supermassive BHs including the detection of mergers of intermediate mass BHs~\cite{ArcaSedda2020} - the missing link between stellar-mass and supermassive BHs.

Atom interferometers can in addition test the limits of quantum mechanics~\cite{Bilardello2016},
probe quantum effects in gravitational fields~\cite{Overstreet2022}, test QED and the Standard Model~\cite{Morel2020}  and provide
stringent tests of general relativity by probing
Einstein's Equivalence Principle~\cite{Asenbaum2020}.
Beyond these significant contributions to fundamental physics, the unprecedented sensitivity and accuracy
of atom interferometry has important applications in the measurement of many physical quantities, such as acceleration, local gravity, and rotation, which can be used to create highly sensitive sensors for navigation, gravimetry and geodesy~\cite{Alonso2022}.

Motivated by these many potential applications of atom interferometry to fundamental physics and beyond,
it is a rapidly-developing field of research that is moving out of the laboratory and up to larger scales. Many of the technological challenges involved in the design and operation of large-scale atom interferometers are being addressed, and research into its possibilities is advancing rapidly.

The study of matter waves and their interference  dates back to the early days of quantum mechanics, 
following the proposal of wave-particle duality by Louis de Broglie~\cite{deBroglie1925},
which was confirmed for electrons in experiments by Davisson and Germer~\cite{Davisson1927}  and by  Thomson~\cite{Thomson1927}, and for atoms by Esterman and Stern~\cite{Estermann1930}. These experiments suggested that the principle of interferometry could be extended to matter waves,
but the development of laser cooling and trapping techniques in the 1980s and 1990s greatly expanded the possibilities for this area of research. 

The advent of lasers enabled precise control over the internal states and momenta of atoms, creating a new field of research using cold atoms and the creation of Bose-Einstein condensates (BECs), which opened different possibilities for the study of matter waves.
The first complete atom interferometers~\cite{Carnal1991,Keith1991} were prepared using microfabricated diffraction gratings, while the first atom interferometers based on laser pulses were constructed by the group of Bord\'e~\cite{Riehle1991} and by Kasevich and Chu~\cite{Kasevich1991}.
See~\cite{Cronin2009,Buchmueller2023} for introductions and reviews of atom interferometry and some applications. 

\begin{figure}[h]
\centering 
\includegraphics[width=7cm]{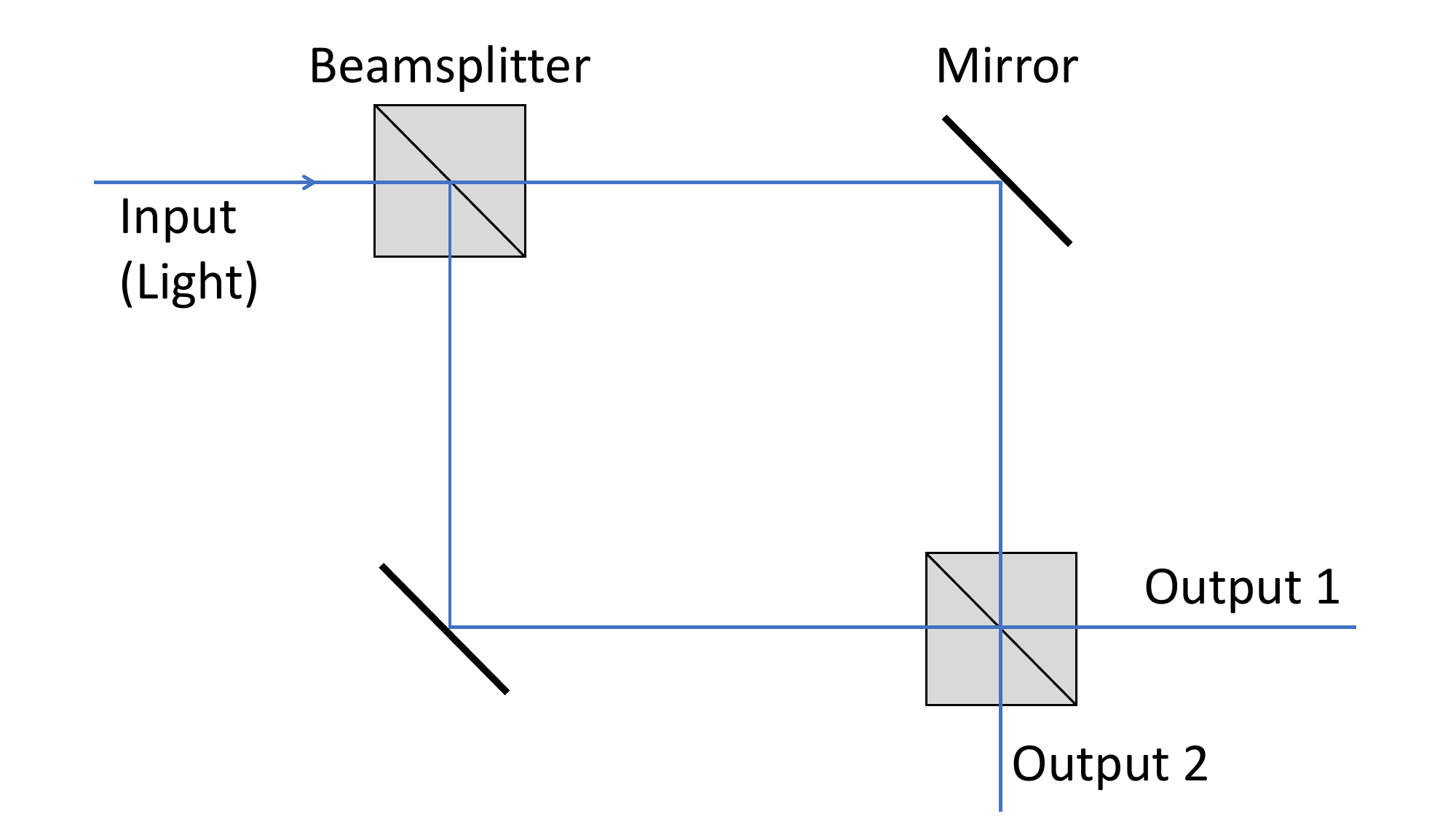}
\includegraphics[width=7cm]{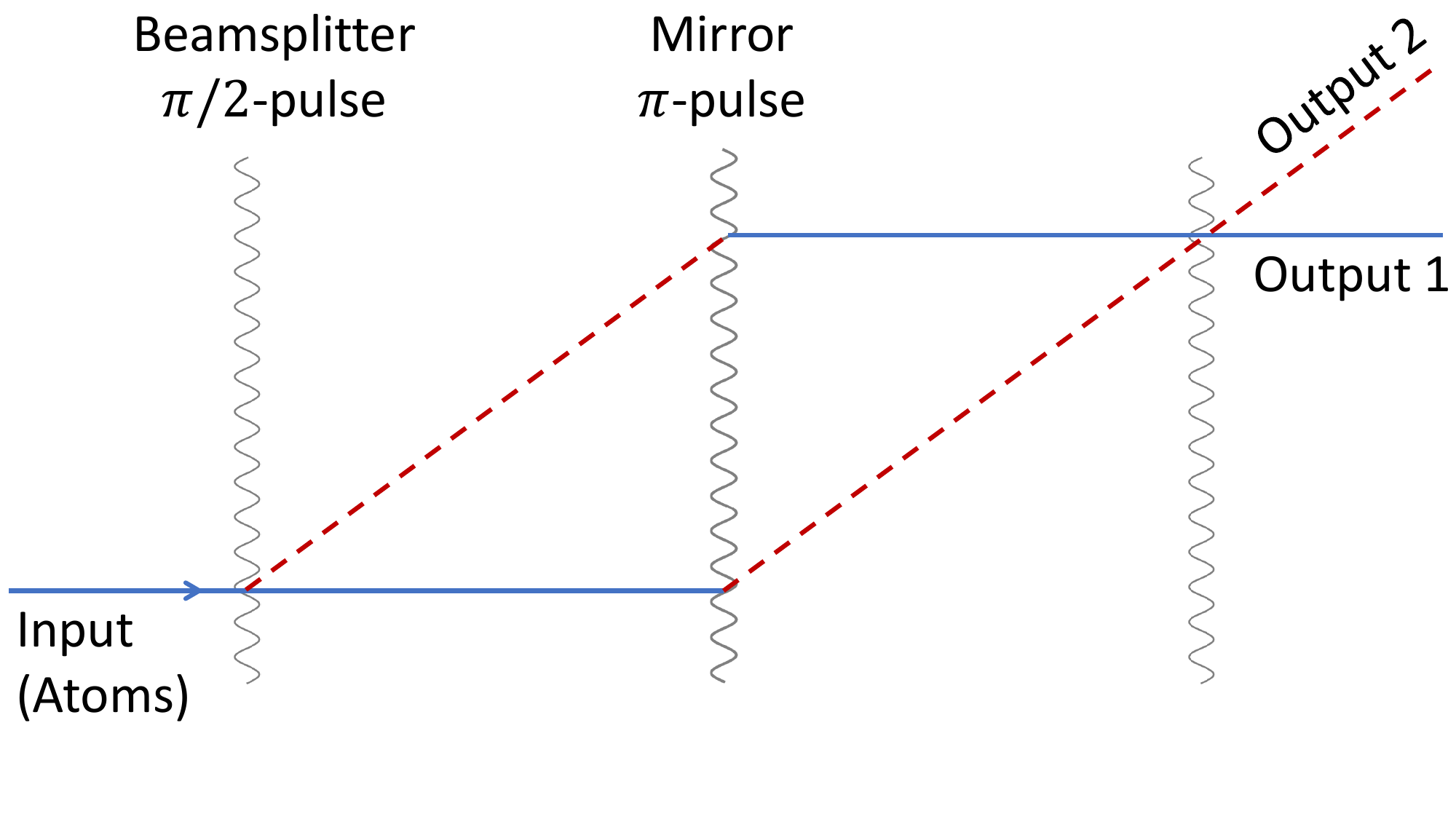}
%\equalSubFigs{MZ Optics.pdf}{MZ Atoms.pdf}[\textwidth]
%\includegraphics[width=10cm]{MZ Optics.pdf}
%\vspace{-3mm}
\caption{\it Left: Outline of the principle of a Mach-Zehnder laser interferometer~\cite{Zehnder1891,Mach1892}.
Right: Outline of an analogous atom interferometer. Atoms in the ground state, $\ket{g}$, are represented by solid blue lines, the dashed red lines represent atoms in the excited state, $\ket{e}$, and laser pulses are represented by wavy lines.}
\label{fig:Interferometers}
\end{figure}

As illustrated in Fig.~\ref{fig:Interferometers}, the principle of atom interferometers is similar to that of optical interferometers, but with physical beamsplitters and mirrors replaced by laser-atom interactions. Clouds of cold atoms are addressed by a
laser beam. In the simplest case, an initial $\pi/2$ laser pulse splits the cloud into equal populations of ground- and excited-state atoms, the latter with a momentum kick due to the absorption of a photon, separating it from the ground-state cloud. A subsequent, longer $\pi$ laser pulse inverts the populations of ground and excited states, which are brought back
together, and a final $\pi/2$ pulse then acts as another beamsplitter and the numbers of atoms in the ground and excited states are read out by, e.g., fluorescence imaging. These are sensitive to phase shifts induced by interactions with ULDM and the passage of GWs.

Achievements of atom interferometers already include a sensitive terrestrial test of
the Einstein Equivalence Principle (EEP) by observing the free fall of clouds of $^{85}$Rb and $^{87}$Rb,
which verified the EEP at the $10^{-12}$ level~\cite{Asenbaum2020}. They also include a precise determination of the fine structure constant at the $10^{-11}$ level through precise determination of the recoil velocity induced by coherent scattering of a photon from a $^{87}$Rb atom~\cite{Morel2020}. Another experiment used pairs of $^{87}$Rb clouds launched simultaneously to different heights in a vertical interferometer
to verify a gravitational analogue of the Aharonov-Bohm effect~\cite{Overstreet2022}, 
namely a phase shift
induced by the gravitational field of a tungsten source mass placed close to one of the clouds.

The future sensitivities of atom interferometers to ULDM and GWs will depend, in particular, on the intensities of the atom sources and the separations between their trajectories generated by the laser pulses. The former control the level of atom shot noise and the latter are increased by repeating the laser pulse sequence many times to obtain large momentum transfers (LMT). These are key R\&D targets on the path towards realizing the potential of very large atom interferometers.

In the following Sections of this report we first summarize the physics case for such very large atom interferometers, and then discuss the synergies between them and laser interferometers. The following Sections review the cold atom technology developments that will be required for realizing this science programme, describe the principal detector options that are under consideration, and review possible site options. The final Sections discuss supplementary topics and summarize the contents of the report.

\section{Physics Case }
\label{sec:physics}
%{\bf 8 pages guideline }

\subsection{Introduction}
%(Christopher McCabe, Naceur Gaaloul, Surjeet Rajendran) ($\sim 1$p)
\label{sec:physics-intro}
The Standard Model of particle physics is one of the most successful theories constructed by human beings. It is able to describe the physics of matter from sub-nuclear distances as small as $\sim 10^{-18}$\,m to the scale of the cosmos $\sim 10^{28}$\,m. It has also withstood every direct experimental test that it has been subjected to over the past thirty years. Despite this unprecedented success, we know that the Standard Model is not a complete theory of nature. This failure is manifest both on observational and theoretical fronts. On the observational side, the Standard Model cannot account for the matter/anti-matter asymmetry. Nor can it explain the nature of dark matter. On the theoretical side, we know that the Standard Model cannot describe the physics of strong gravity such as gravitational singularities encountered in black holes and the Big Bang singularity of the early Universe. It is also beset with theoretical puzzles such as the hierarchy, cosmological constant and strong CP problems where estimates of values of physical parameters based on well understood calculational principles are in massive contradiction with observational data. How can we make experimental progress on these issues given the obstinate agreement between experiment and theory when we directly probe the Standard Model?

Over the past century, human mastery over electromagnetism has enabled us to probe physics at higher and higher energies through a variety of particle colliders. These colliders are the right technology to probe physics at high energies that have reasonably large interactions with  particles such as electrons and protons. However, they are statistically unable to probe new physics of {\it any} mass that interacts weakly with these particles. This raises the interesting possibility that there might be weakly coupled new physics at low energies, opening a hitherto under-explored frontier in the hunt for new physics beyond the Standard Model. This possibility is bolstered by a compelling theoretical case for probing low-mass, weakly-coupled particles. 

For example, the universal nature of gravitation implies that every object in the Universe, past or present, is able to send signals to us through gravitational waves. Direct detection of these gravitational waves enables us to probe the dynamics of black hole space-times as well as the physics of the infant universe prior to era of recombination. This unique opportunity permits us to unveil the secrets of gravity in regimes that have never been observationally probed. Given the historic discoveries of gravitational waves by the LIGO/Virgo collaboration, there is a strong case for further exploration of the gravitational wave spectrum. Experiments that are able to explore other parts of this spectrum are guaranteed to make discoveries. In addition to this, there is also a strong theoretical case to look for new, weakly-coupled particles. The existence of dark matter strongly suggests that the new physics likely interacts weakly with the Standard Model. Popular solutions to outstanding theoretical puzzles such as the strong CP, hierarchy and cosmological constant problems also feature the existence of such particles. Given these exciting possibilities, what technology gives access to this space?

The key technological requirement is a sensing platform that combines high precision with superior noise cancellation capabilities, in order to be sensitive to the weak signals expected from this kind of physics. Atom interferometry offers an interesting solution to this technological challenge. It combines the pristine quantum-mechanical nature of atomic properties with noise cancellation abilities inherent to interferometry to make possible sensors that have exquisite sensitivity to accelerations, energy shifts and spin precession - the dominant ways in which new physics can affect Standard Model sensors. The demonstrated performance of these sensors is sufficient to probe new regions of parameter space and foreseeable advances made possible by focused efforts will lead to the creation of advanced sensors that can expand significantly the searches for gravitational waves and new physics.

In the following, we discuss some of the physics opportunities that can be exploited by a variety of atom interferometry configurations. We begin in Section~\ref{sec:physics-gw} by focusing on gravitational waves in the frequency band 0.01 - 10 Hz (the ``dHz range") where terrestrial atom interferometers have unique capabilities for gravitational wave detection. Following this, in Section~\ref{sec:physics-dm} we discuss the ability of atom interferometers to probe a variety of ultralight dark matter candidates. We then discuss new tests of quantum mechanics in Section~\ref{sec:physics-qm}, followed by discussions of probes of new fundamental interactions in Section~\ref{sec:physics-newint} and probes of the charge neutrality of atoms in Section~\ref{sec:physics-neut}. 

\subsection{Gravitational wave signals}
%(Marek Lewicki) ($\sim 1.5$p)
\label{sec:physics-gw}

\begin{figure}
\centering 
\includegraphics[width=0.7\textwidth]{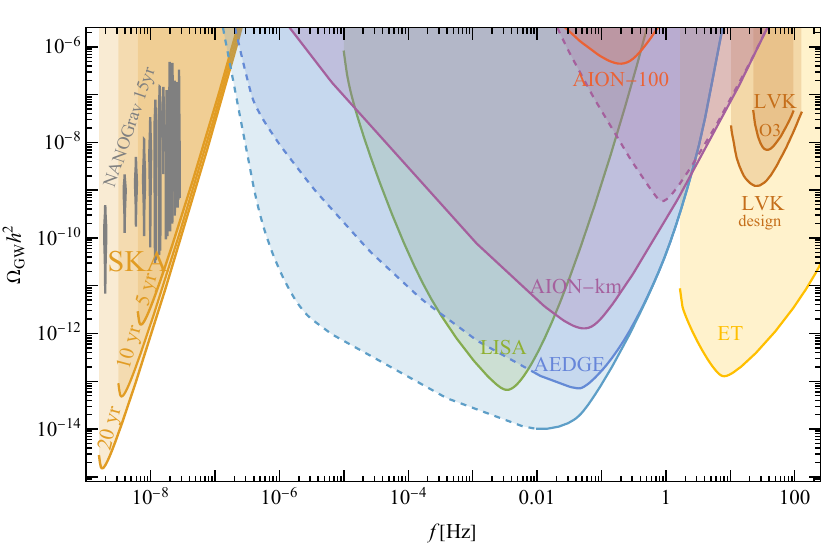}
\vspace{-3mm}
\caption{\it Sensitivities to the energy density of GWs, $\Omega_{\rm GW} h^2$, using power-law integration of the proposed terrestrial atom interferometers AION-100, AION-km, as well as the space-borne incarnations of the technology AEDGE and AEDGE+, together with other existing and planned experiments LIGO/Virgo/KAGRA (LVK), ET, PTAs and SKA. Also shown in gray are likelihood distributions in each frequency bin for the GW signal reported by the NANOGrav Collaboration in their 15-year data~\cite{Agazie2023}.
}
\label{fig:GW_exp_abundance}
\end{figure}

Fig.~\ref{fig:GW_exp_abundance} illustrates the potential sensitivities of terrestrial atom interferometers to gravitational waves based on studies of the Atom Interferometer Observatory and Network (AION) in its planned 100~m and 1~km versions~\cite{Badurina2020,Badurina2021}. The dashed line labelled GGN indicates the potential impact of gravitational gradient noise assuming the Peterson new low-noise model (NLNM) model~\cite{Peterson1993} without implementing any of the mitigation strategies discussed in Section~\ref{sec:Battling}. We also include other planned and currently running experiments starting from high frequencies with LIGO/Virgo/KAGRA and their latest O3 data~\cite{Abbott2021} and design sensitivity~\cite{Aasi2015}, as well as the planned ET detector~\cite{Maggiore2020}. Below the optimal frequencies of atom interferometers we see the sensitivity of LISA~\cite{AmaroSeoane2017} and at much lower frequencies those of pulsar timing arrays (PTAs) and SKA~\cite{Combes2021}. We also include grey violins indicating the fit to a stochastic gravitational wave background reported by the NANOGrav PTA in their 15-year data~\cite{Agazie2023}, which is corroborated by data from other PTAs~\cite{Antoniadis2023,Reardon2023,Xu2023,Agazie2023a,Afzal2023,Agazie2023b,Antoniadis2023c,Antoniadis2023a,Antoniadis2023b,Reardon2023a,Zic2023}, and also indicate the prospective sensitivities of the space-borne atom interferometer experiments AEDGE and AEDGE+~\cite{Bertoldi2020,Badurina2021}.  

The peak sensitivity of atom interferometers such as AION, positioned between the terrestrial interferometers LIGO/Virgo and the LISA space mission, makes them ideal tools for probing many astrophysical phenomena otherwise beyond our reach. It would be ideal for measuring the high-frequency tails of mergers and
ringdown stages of intermediate-mass black hole binaries. It can also observe the early infall stages of mergers that subsequently merge within the LIGO/Virgo and ET frequency band. These prospects were discussed in detail in~\cite{Badurina2020,Badurina2021}, for more details and updated predictions see also Section~\ref{sec:SMBHs}. Proposals for space-borne laser interferometer projects beyond LISA to target similar frequency ranges include BBO~\cite{Yagi2011} and DECIGO~\cite{Kawamura2021}.

\begin{figure}
\centering 
\includegraphics[width=0.32\textwidth]{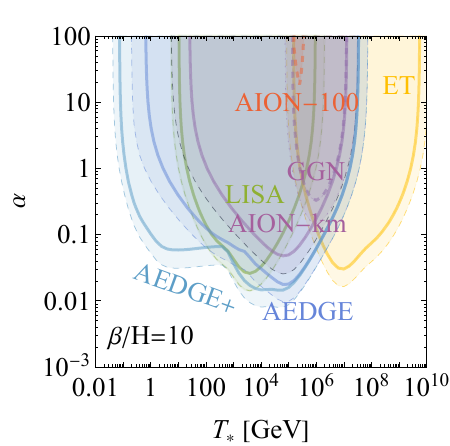}
\includegraphics[width=0.32\textwidth]{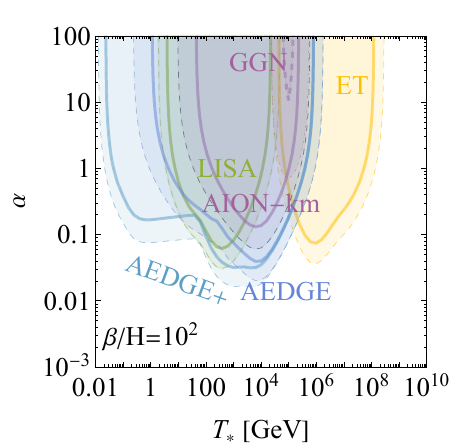}
\includegraphics[width=0.32\textwidth]{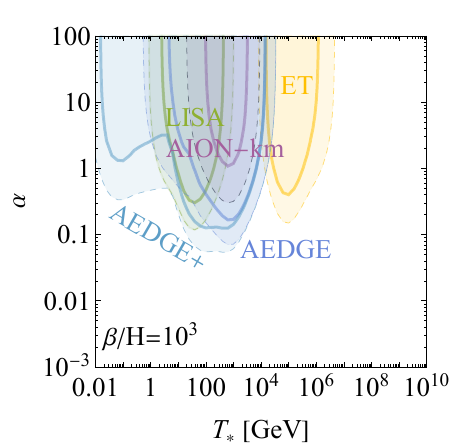}
\caption{\it Sensitivities in the $(T_*, \alpha)$ plane of AION-100 and -km, as well as other planned experiments, to the SGWB spectrum from sound waves in the plasma that could be formed in the aftermath of bubble collisions. Dashed lines show $SNR=1$ while solid lines $SNR=10$  except for AION-km GGN for which $SNR=10$ is depicted by a thick dashed line while the dotted line corresponds to $SNR=1$. Figure taken from ref~\cite{Badurina2021}.}
\label{fig:PT}
\end{figure}

Another interesting target unique to GW observatories is the direct observation of early Universe phenomena through primordial stochastic backgrounds of GWs~\cite{Auclair2022}. One of the possible sources at such early times is a first-order phase transition, as appears in a plethora of scenarios for physics beyond the SM (BSM). The dynamics of such transitions are largely described by their energy compared to the background, parameterised by $\alpha$, and the ratio of their inverse time duration to the Hubble time, $\beta/H$, as well as the transition temperature $T_*$. Fig.~\ref{fig:PT} shows the sensitivity of AION and other planned experiments to phase transitions in terms of these parameters. We show fixed values of $\beta/H=10$, $10^2$ and $10^3$ using, for simplicity, the sound-wave spectra calculated in~\cite{Hindmarsh2013,Hindmarsh2015,Hindmarsh2017}. Note that the latter are most appropriate for relatively weak phase transitions, i.e., $\alpha\lesssim 0.1$. We see that AION-km has good sensitivity for transition temperatures between $T_*=10^3$ and $10^5$ GeV, provided $\alpha\geq 0.1$  and $\beta/H\leq10^2$. It is important to point out that probing the spectrum in a large range of frequencies would be necessary to measure well the shape of the spectrum and identify a phase transition as the source behind the signal. 

The electroweak symmetry breaking (EWSB) transition, which occurs at an energy scale $T_* \sim 100 GeV$, offers a particularly interesting test-bed in this context, since it certainly took place in the early Universe, and represents the frontier between ``known” high-energy physics - the Standard Model of particle physics that has been tested on Earth in particle experiments - and its untested higher-energy extensions. Searching for a SGWB signal from such a phase transition therefore provides a probe of possible physics beyond the Standard Model (BSM). 
BSM scenarios for which the electroweak phase transition is of first order typically predict {\it weakly} first-order (and consequently brief) phase transitions, characterised by $10^2<\beta/H_*<10^4$~\cite{Caprini2016,Caprini2020}. The corresponding SGWBs are outside the sensitivity range of LISA, but could be detected by an interferometer operating at higher frequency. Consequently, the detection frequency range of an atom interferometer such as AION-km has the advantage of probing favoured regions in the EWSB parameter space. Furthermore, as can be appreciated from Fig.~\ref{fig:PT}, AION-km would be sensitive to first-order phase transitions occurring also at higher temperatures than EWSB, providing the opportunity to test the fundamental high-energy theory beyond the reach of any present or near-future particle collider.

\begin{figure}
\centering 
\includegraphics[height=4.4cm]{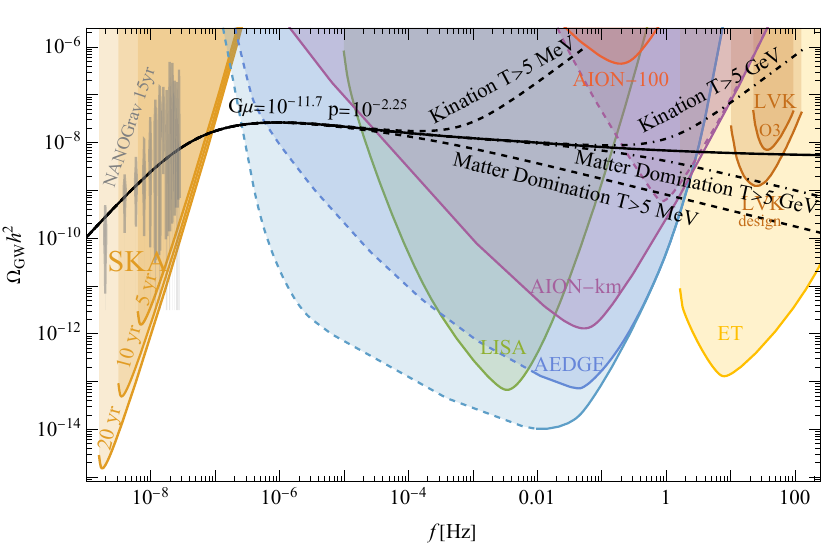}
\includegraphics[height=4.6cm]{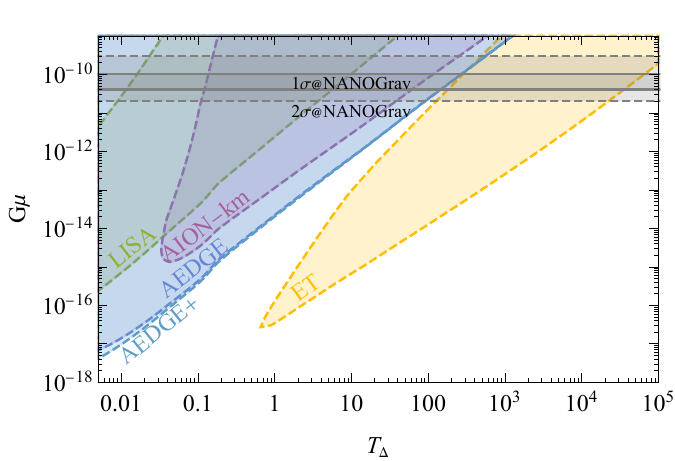}
\caption{\it Left panel: Cosmic super string spectrum with $G\mu=10^{-11.75}$ and intercommutation probability $p=10^{-2.25}$ in standard cosmology together with its possible modifications by a period of kination or
matter domination (MD) ending at temperatures $T > 5$~MeV and $5$~GeV. The grey violins indicate the spectra capable of explaining the NANOGrav 15yr data.
Right panel: Sensitivity of various experiments to a modification of the expansion rate at a temperature $T_\Delta$ for a given value of the string tension $G\mu$ with $p=1$. The gray bands indicate values favoured by the NANOGrav 12.5yr data~\cite{Arzoumanian2020,Ellis2021}. The right panel was taken from ref~\cite{Badurina2021}.}
\label{fig:CSexpansion}
\end{figure}

Another potential source of gravitational waves from the Early Universe is a cosmic string network. If produced in a phase transition at very high energies it would continue emitting GWs until today, producing a spectrum featuring a relatively flat plateau over a large range of frequencies~\cite{BlancoPillado2017,BlancoPillado2018,Auclair2020}. The recent GW signal in the 15-year data from NANOGrav~\cite{Agazie2023} could in fact be fitted very well with such a signal~\cite{Ellis2023c,Afzal2023} (for earlier analysis of this potential source see~\cite{Ellis2021,Blasi2021,BlancoPillado2021}) provided the mass per unit length of the network lies within $G\mu \sim \times 10^{-11} - 10^{-12}$ with intercommutation probability $p \sim 10^{-3} - 10^{-1}$~\cite{Ellis2023}. This interpretation would indicate that the signal could also be measured in AION-km as well as LISA, ET and AEDGE, although not necessarily in upcoming runs of the LIGO, Virgo and KAGRA experiments~\cite{Ellis2023}. Should this interpretation prevail, measurement of the spectrum over a wide range of frequencies would also enable the mapping of the expansion rate of the Universe, as any modification would leave its imprint on the spectrum. We show examples of modifications of the spectrum coming from an early period of matter domination and kination in the left panel of Fig.~\ref{fig:CSexpansion}. However, it is important to point out that even much smaller modifications such as a change in the number of relativistic degrees of freedom could be measured. The right panel of the same figure shows the reach of experiments in terms of the temperature at which the expansion rate is modified. 

\subsection{Dark matter signals} 
%(Christopher McCabe) ($\sim 1.5$p)
\label{sec:physics-dm}

\begin{figure}
\centering 
\includegraphics[width=0.45\textwidth]{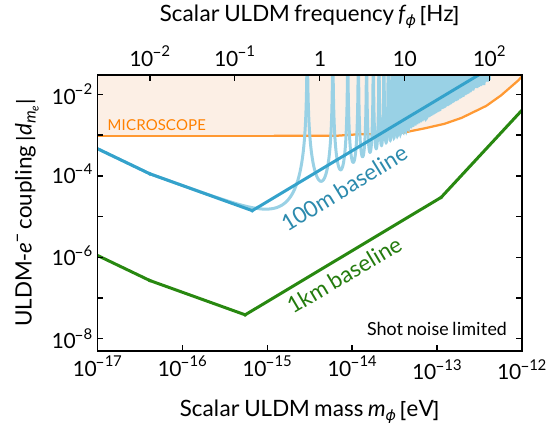}
\includegraphics[width=0.45\textwidth]{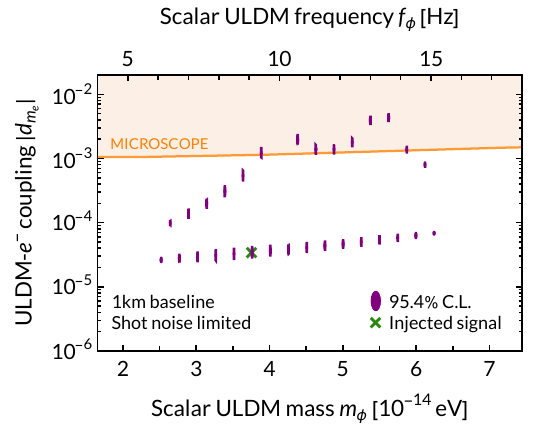}
\caption{\it Left panel: Projections for sensitivities to scalar ULDM linearly coupled to electrons (shot noise limited and assuming $\mathrm{SNR}=1$). The lighter-blue 100\,m baseline curve shows the oscillatory nature of the sensitivity projections, while the darker-blue and green curves show the envelope of the oscillations. 
Right panel: Parameter reconstruction, adapted from ref.~\cite{Badurina2023}, of an injected signal with $f_{\phi}=9.1\,\mathrm{Hz}$ and $d_{m_e}=3.7\times 10^{-5}$ (green cross) for a 1\,km baseline assuming a constant sampling frequency of $0.3$\,Hz.
The purple contours show the islands of parameter space compatible with the signal at 95.4\% CL.
In both panels, the shaded orange region shows constraints from MICROSCOPE~\cite{Touboul2022,Hees2018}.}
\label{fig:scalarULDM}
\end{figure}

For masses below approximately $1\,\mathrm{eV}$, a bosonic ultra-light dark matter (ULDM) field within our galaxy could be effectively described as a superposition of classical waves~\cite{Hui2021}. 
The coherent oscillations of these ULDM waves would give rise to a diverse range of time-dependent signals that could be explored using atom interferometers.
These signals encompass various phenomena, including the time-dependent oscillations of fundamental `constants' in the context of scalar ULDM candidates~\cite{Geraci2016,Arvanitaki2018,Badurina2021,Badurina2023},
the time-dependent differences in accelerations between atoms in theories involving vector candidates~\cite{Graham2016}, 
and the time-dependent precession of nuclear spins in the case of pseudoscalar candidates~\cite{Graham2018}.
In general, these signals have a frequency determined by the ULDM mass, an amplitude proportional to the local ULDM density and mass, and a coherence time that depends on both the ULDM mass and the ULDM virial velocity within our galaxy~\cite{Derevianko2018}.

The left panel of Fig.~\ref{fig:scalarULDM} shows sensitivity projections for 100\,m and 1\,km baseline vertical gradiometers to the linear coupling of a scalar ULDM field to electrons ($d_{m_e}$) as a function of the ULDM mass ($m_{\phi}$). This linear coupling induces an effective time-dependent correction to the electron mass, which in turn, induces small time-dependences in the energy levels of atoms~\cite{Arvanitaki2015}.~\footnote{There are similar sensitivities for scalar ULDM couplings to the photon and to quarks.} The projections in the left panel of Fig.~\ref{fig:scalarULDM} have been calculated for the ‘clock’ transition in $^{87}\mathrm{Sr}$ and follow the procedure outlined in ref.~\cite{Badurina2020}. For the 100\,m baseline, we assume $1000\,\hbar k$ atom optics and $10^{-4}\,\mathrm{rad}/\sqrt{\mathrm{Hz}}$ phase resolution; while for the 1\,km baseline we assume $2500\,\hbar k$ atom optics and $10^{-5}\,\mathrm{rad}/\sqrt{\mathrm{Hz}}$ phase resolution. 
The atom interferometer sensitivity oscillates as a function of the ULDM mass, as illustrated by the light-blue 100\,m curve in the left panel of Fig.~\ref{fig:scalarULDM}. However, the peaks and troughs can be shifted by running with slightly different interrogation times~\cite{Badurina2022}, so it is usually only the envelope of the oscillations that is plotted (darker blue and green lines).
It is important to note that these projections are atom-shot-noise limited and therefore do not take into account gravity gradient noise, which is expected to impact the 1\,km baseline projections below approximately 1\,Hz in the absence of mitigation strategies: see the discussion in Section~\ref{sec:Battling}.

The right panel of Fig.~\ref{fig:scalarULDM} shows an example of ULDM parameter reconstruction in the event of a (simulated) $5\sigma$ discovery assuming a 1\,km baseline, $2500\,\hbar k$ atom optics, $10^{-5}\,\mathrm{rad}/\sqrt{\mathrm{Hz}}$ phase resolution, $10^8$\,s integration time, and a constant sampling frequency of $0.3$\,Hz~\cite{Badurina2023}.
The green cross shows the parameters of the injected signal, while the purple contours show the islands of parameter space compatible with the signal at 95.4\% CL. The injected signal falls within the 95.4\% CL\ contour of one island.
The multiple islands arise from aliasing, since the injected signal is above the Nyquist frequency~\cite{Badurina2023}. As the signal strength increases, or if the sampling frequency is increased, the number of islands decreases.
Each island has a width $\delta f_{\phi}\sim 10^{-6}$\,Hz, which is determined by the integration time and the degree to which the time-series data is stacked~\cite{Badurina2023}. 

\begin{figure}
\centering 
\includegraphics[width=0.45\textwidth]{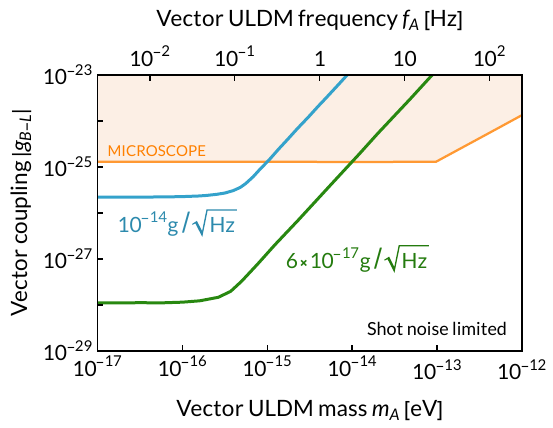}
\includegraphics[width=0.45\textwidth]{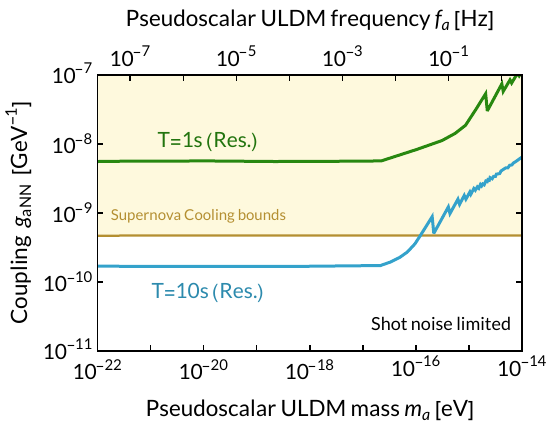}
\caption{\it Left panel: Shot noise limited projection, adapted from ref.~\cite{Abe2021}, to $B-L$ coupled vector ULDM for a dual-species interferometer (\,$^{87}\mathrm{Sr}$ and $^{88}\mathrm{Sr}$).
The projections are given in terms of the acceleration sensitivities achievable with VLBAI (see text). The shaded orange region shows constraints from MICROSCOPE~\cite{Touboul2022}.
Right panel: Shot noise limited projection, adapted from ref.~\cite{Graham2018}, to the spin coupling of pseudoscalar ULDM to atoms. The projections are given in terms of the interrogation time. The shaded yellow region shows bounds from supernova cooling.}
\label{fig:vecpsULDM}
\end{figure}

Next, we consider the shot-noise-limited sensitivity to a $B-L$ coupled vector ULDM candidate. 
The projected sensitivity, shown in the left panel of Fig.~\ref{fig:vecpsULDM} in terms of the vector mass ($m_A$) and coupling ($g_{B-L}$)~\cite{Fayet2018,Fayet2019}, arises from comparing the accelerometer signals from two simultaneous atom interferometers run with  dual species (here, $^{87}\mathrm{Sr}$ and $^{88}\mathrm{Sr}$). The sensitivity is quoted in terms of the acceleration sensitivity, where the blue line could be achieved with a 100\,m baseline, $100\,\hbar k$ atom optics and $10^{-3}\,\mathrm{rad}/\sqrt{\mathrm{Hz}}$ phase resolution, while the green line could 
be achieved with a 100\,m baseline, $1000\,\hbar k$ atom optics and $10^{-4}\,\mathrm{rad}/\sqrt{\mathrm{Hz}}$ phase resolution.

Finally, the right panel of Fig.~\ref{fig:vecpsULDM} shows the shot-noise-limited sensitivity to a pseudoscalar ULDM candidate. 
The spin coupling of the pseudoscalar field to atoms creates a spurious phase shift due to the precession of the spin induced by the pseudoscalar field, which can be measured by interfering two atoms in different nuclear spin states. The projections shown in Fig.~\ref{fig:vecpsULDM} in terms of the pseudoscalar mass ($m_a$) and coupling ($g_{aNN}$) assume the use of a resonant sequence that amplifies the phase shift at the pseudoscalar frequency~\cite{Graham2016b}, and a phase resolution of $10^{-4}\,\mathrm{rad}/\sqrt{\mathrm{Hz}}$. The sensitivity is shown for two values of the interrogation time. While 1\,s can be achieved with a $\mathcal{O}(10)\,\mathrm{m}$ baseline, 10\,s is expected to be achievable with a 1\,km baseline.

\subsection{Tests of Quantum Mechanics}
%(Surjeet Rajendran) ($\sim 1.5$p)
\label{sec:physics-qm}

Quantum Mechanics is the bedrock of physics. However, its axioms were derived phenomenologically, leading to the enticing possibility that these axioms are an approximation to a more complete theory. There is a strong need to develop formalisms for consistent deviations from these axioms and probe them experimentally. \\
~~\\
\noindent
{\bf Linearity:} One of the central axioms of quantum mechanics is the linearity of time evolution.  In every other system, linearity is an approximation. Why should quantum mechanics be perfectly linear? Following the failure of prior attempts \cite{Weinberg1989,Polchinski1991}, it has been widely believed that linearity is necessary for causality.
However, it was shown in \cite{Kaplan2022} that this belief is false. Furthermore, this work showed that causal nonlinear evolution that preserves other symmetries of nature (such as gauge invariance and conservation laws) was possible in quantum mechanics. The key point of \cite{Kaplan2022}  is as follows. In quantum mechanics, when we consider a particle like an electron moving from one point to another, the particle takes all possible paths in physical space in going between these points. In linear quantum mechanics, when we consider the motion of two electrons, the paths taken by one electron affects the paths taken by the other electron via the electromagnetic force. This occurs causally. However, in linear quantum mechanics, the paths of a single electron do not influence each other via the electromagnetic force. Such an interaction would be nonlinear. Given that the paths of distinct electrons are able to affect each other causally, why should causality preclude  nonlinearity, i.e., the paths of the single electron influencing each other? In linear quantum mechanics, the natural way to describe interactions in a causal way is quantum field theory. Thus, in  \cite{Kaplan2022}, nonlinearities were directly introduced into quantum field theory and it was shown that a wide class of causal and consistent nonlinear time evolution was possible in quantum mechanics. This work also naturally realized the general structure that  \cite{Polchinski1991} showed must be present in any causal nonlinear quantum mechanical theory.

Excitingly, this work also showed that prior experimental limits on causal nonlinear quantum mechanical systems from probes of pristine quantum mechanical systems such as atomic and nuclear physics were weak.  But, this physics can be readily probed in dedicated experiments.  A preliminary suite of experiments was performed  to test  the electromagnetic aspects of this framework \cite{Polkovnikov2023,Broz2023}. Present and future atom interferometers will be able to test the gravitational aspects of this framework with significant sensitivity, well beyond present limits on such non-linearities in the gravitational sector.

In linear quantum mechanics, the time evolution of a quantum state $|\chi\rangle$ is described by a convolution of the state $|\chi\rangle$ with a path integral that describes the evolution of the basis elements of the Hilbert space of the theory. This path integral is independent of the state $|\chi\rangle$. In \cite{Kaplan2022}, nonlinear evolution is introduced by  making this path integral depend on the state $|\chi\rangle$ by changing the Lagrangian in a state-dependent way.  For example, in linear quantum mechanics, the interaction of the gravitational field $g_{\mu \nu}$ with a scalar field $\phi$ is described by the term $g_{\mu \nu} \partial^{\mu}\phi \partial^{\nu}\phi$. In nonlinear quantum mechanics, the interaction is instead described by the term $\left(g_{\mu \nu} + \epsilon \langle \chi|\hat{g}_{\mu \nu}|\chi\rangle \right) \partial^{\mu}\phi \partial^{\nu}\phi$, where $\langle \chi|\hat{g}_{\mu \nu}|\chi\rangle$ is the expectation value of the metric operator $\hat{g}_{\mu \nu}$ in the quantum state $|\chi\rangle$. How can these terms be probed experimentally? Nonlinearities permit different arms of a superposition to interact with each other. In linear quantum mechanics, the effects of a quantum superposition can only be measured if the system is quantum coherent. Bizarrely, in nonlinear quantum mechanics, the effects of a quantum superposition can be detected even in the presence of decoherence \cite{Kaplan2022, Polchinski1991}. This leads to the following experimental concept.

First, one creates a macroscopic quantum superposition. This superposition need not be quantum coherent and is thus easily created by performing a measurement on a quantum system such as a spin $\frac{1}{2}$ system.  When such a measurement is made, the Schrodinger equation predicts that this creates a macroscopic superposition of two “worlds”: one in which the spin was measured to be up and in another where the spin was measured to be down. At a fixed location in the laboratory, we place a test mass when  spin up is obtained. At the same location in the laboratory, if we get spin down, we place an accelerometer such as an atom interferometer.  In this quantum state, even in linear quantum mechanics, the expectation value $\langle \chi|\hat{g}_{\mu \nu}|\chi\rangle$ at this particular point in the laboratory is non-zero. But, since the evolution is independent of $\langle \chi|\hat{g}_{\mu \nu}|\chi\rangle$, this cannot be observed - the accelerometer that is entangled with spin down will not respond to the test mass that is entangled with spin up. 

This is not the case in nonlinear quantum mechanics where $\langle \chi|\hat{g}_{\mu \nu}|\chi\rangle$ directly affects the evolution of the physical states. Thus, the accelerometer that is entangled with  spin down will register an acceleration arising from the test mass that is entangled with spin up. Due to this nonlinearity, there is interaction between the two arms (i.e “worlds”) of the superposition.  This phenomenon was dubbed the “Everett phone” by \cite{Polchinski1991} and it is a requirement of any causal nonlinear modification of quantum mechanics.  

An experiment of this kind can be naturally incorporated into the science program envisaged for a long baseline interferometry setup. The atom interferometers are natural accelerometers and, when arranged in a gradiometer configuration, they are naturally insensitive to a variety of systematics such as vibrational noise that affect accelerometry measurements. Test masses are also envisaged in these setups in order to search for new interactions (see Section~\ref{sec:physics-newint}). All that is missing is a quantum spin whose outcome determines where the mass and the accelerometer are placed. This can be readily obtained by either accessing a variety of publicly available quantum randomizers such as the IBM quantum processor or through in-house quantum randomizers constructed either with atomic systems or simple, low-activity radioactive sources. Conservatively, an accelerometer with sensitivity $\sim 10^{-13} \frac{\text{g}}{\sqrt{\text{Hz}}}$ and a 100 kg test mass, operating for about $\sim 10^{6}$\,s can probe the nonlinear parameter $\epsilon$ down to $\epsilon \sim 10^{-6}$, a million-fold improvement over the current $\epsilon \sim \mathcal{O}\left(1\right)$ on this parameter from \cite{Page1981}. A null result in this experiment would also be hugely important, since it would prove the quantum nature of gravity \cite{Page1981}.  

Atom interferometers also have a unique ability to probe another key aspect of nonlinear quantum mechanics.  Unlike linear quantum mechanics, the observational phenomenology of nonlinear quantum mechanics is fundamentally tied to the full quantum state \cite{Kaplan2022}, including its cosmological history. In conventional inflationary cosmology, where our  observed universe is a tiny part of the entire wave-function of the universe, laboratory experiments of the kind described above will give null results even if the fundamental nonlinear term $\epsilon$ is large simply because of the small overlap of our universe with the full quantum state of the cosmos. But, in such a scenario, nonlinear terms can still be tested since these terms naturally violate the equivalence principle. Tests of the equivalence principle are one of the key goals of long baseline atom interferometers (see Section~\ref{sec:physics-newint}) and these tests are directly applicable in constraining these aspects of nonlinear quantum mechanics. 

It is exciting that there is a consistent parameterized deviation from quantum mechanics~\cite{Kaplan2022}. If nonlinear quantum mechanics is discovered in the experiments proposed above, it could lead to transformative effects on science and society. For example, if nonlinear quantum mechanics is discovered using the ``Everett-phone'' setup described above~\cite{Polchinski1991}, it will lead to a technological revolution. This discovery would enable us to parallelize  readily-available classical resources to solve a wide variety of problems, including computing problems. Moreover, any discovery of nonlinear quantum mechanics is likely to open new avenues for solving the black hole information problem in an experimentally testable manner.  There is thus a very strong scientific and technological case for testing these theories.\\
~~\\
\noindent
{\bf Superposition:} Among the tests of the foundations of Quantum Mechanics, it is also of fundamental importance to the limits of validity of the quantum superposition principle for larger systems. The reason why  quantum properties of microscopic systems (in particular, the possibility of being in the superposition of two states at once) do not carry over to macroscopic objects has been subject of intense debate during recent decades \cite{Hawking1982,Ellis1984,Ghirardi1986,Diosi1987,Adler2004,Leggett2005,Weinberg2017}. Its possible resolution  could be a progressive breakdown of the superposition principle when moving from the microscopic to the macroscopic regime. The most important consequence would be to change fundamentally our understanding of Quantum Mechanics --- now commonly considered as the fundamental theory of Nature --- as an effective theory appearing only as the limiting case of a more general one \cite{Adler2009}. Several models have been proposed to account for such a breakdown of the quantum superposition principle. They go under the common name of (wavefunction) collapse models~\cite{Bassi2003,Adler2009,Bassi2013}, and modify the standard Schr\"odinger dynamics by adding collapse terms whose action leads to the localization of the wavefunction in a chosen basis. 

Another suggested motivation for collapse models, beyond having a universal theory whose validity stretches from the microscopic world to the macroscopic world,
comes from a cosmological perspective. Collapse models have been proposed to justify the emergence of cosmic structures in the Universe, whose signatures are imprinted in the Cosmic Microwave Background (CMB) in the form of temperature anisotropies \cite{Perez2006,Landau2012,Das2013}. Moreover, collapse models were also proposed as possible candidates to implement an effective cosmological constant, thus explaining the acceleration of the expansion of the Universe~\cite{Josset2017}. The application of collapse models to cosmology is however not straightforward, as it requires a relativistic generalization of the non-relativistic models discussed below. How to build these relativistic generalisations of collapse models is still not clear: several proposals have been suggested \cite{Martin2020,Gundhi2021,Bengochea2020,Jones2021,Jones2021a}, but each has limitations and the debate in the theoretical community is still open.

The most studied collapse model is the Continuous Spontaneous Localisation (CSL) model \cite{Pearl1989,Ghirardi1990}, a phenomenological model that treats the system under scrutiny as fundamentally quantum but subject to the weak and continuous action of some measurement-like dynamics that occurs universally. 
The CSL model is characterized by two free parameters: the collapse rate $\lambda$, which characterizes the strength of the collapse, and the correlation length of the collapse noise $r_\text{\tiny C}$, which is the length-scale  defining the spatial resolution of the collapse and thus characterizing the transition between the micro and macro domains. Although extensive research over the past 20 years has set ever stronger upper bounds on these parameters  \cite{Arndt2014,Carlesso2022}, there is still a wide unexplored region in the parameter space. 
Since the structure of the CSL dynamics resembles that of a weak continuous Gaussian measurement (at zero efficiency, since the outcome of the measurement is not recorded) --- which is a quite general framework --- one typically regards it as a figure of merit for a wide class of collapse models.   

Also worthy of mention is the Di\'osi-Penrose (DP) model \cite{Diosi1987,Penrose1996}, which is also considered among the most important collapse models.
 The DP model predicts the breakdown of the superposition principle when gravitational effects are strong enough. Penrose provided several arguments why there is a fundamental tension between the principle of general covariance in General Relativity and the superposition principle of Quantum Mechanics~\cite{Penrose1996,Penrose2014}, suggesting that systems in spatial superposition should collapse spontaneously to localised states and that this effect should get stronger the larger the mass of the system.
 A model that describes this effect was introduced by Di\'osi in \cite{Diosi1989} and is known as the DP model. It is fully characterised by the Newtonian kernel $\tfrac{G}{\hbar}\tfrac{1}{|{\bf x}-{\bf y}|}$, where $G$ is the gravitational constant, so that the model is free from any fitting parameter. However, due to the standard divergences of the Newtonian potential   at small distances, the collapse rate for a point-like particle diverges, irrespective of its mass. This implies an instantaneous collapse even for microscopic particles, in contrast to the requirements of the model.  
To avoid this divergence, one takes a Gaussian smearing of the Newtonian kernel of width $R_0$, which becomes the free parameter of the DP model. Several experiments set lower bounds on $R_0$ \cite{Carlesso2022,Vinante2021},  and the strongest bound is given currently by a search for spontaneous radiation emission from germanium \cite{Donadi2020}.

The direct way to test collapse models is to quantify the loss of quantum coherence in interferometric experiments with particles as massive as possible, so as to magnify the collapse effects on the superposition \cite{Arndt2014}. Currently, the most massive particle that has been placed in a superposition has had a mass around $2.5\times 10^4$\,amu \cite{Fein2019}. The corresponding bound is, however, around 9 orders of magnitude away from ruling out the CSL model. With the aim of testing such values, one would need to prepare superpositions with masses around $10^9$\,amu on a time-scale of 10\,s \cite{Gasbarri2021}, which is far beyond the current capabilities of the state-of-the-art and near-future technology. 

In parallel to the interferometric approach, alternative strategies have been developed, which provide stronger bounds, without necessarily requiring the creation of a superposition state. They are based on indirect effects of the modifications collapse models introduce into quantum dynamics \cite{Carlesso2022}, such as extra heating and diffusion or spontaneous radiation emission. Among them, the measurement of the variance in position $\sigma_t^2$ of a non-interacting BEC in free fall can be considered. It may be expressed as
\begin{equation}\label{eq.csl.position}
\sigma_t^2=\sigma_{\text{\tiny QM},t}^2+\frac{\hbar^2}{6m_0^2 r_\text{\tiny C}^2}\lambda t^3.
\end{equation}
The variance is enhanced by the action of collapse models on the BEC with respect to that predicted by quantum mechanics $\sigma_{\text{\tiny QM},t}^2\propto t^2$, exhibiting a different scaling that is proportional to the cube of the free evolution time.
This test can be implemented directly  without requiring additional instrumentation beyond what is already envisioned for the interferometric experiment.    

A study of BEC expansion has already set a competitive bound on CSL \cite{Bilardello2016}, which provides bounds four orders of magnitude stronger than interferometric experiments. This experiment was performed on the ground \cite{Kovachy2015}, where the major limitation was provided by gravity, which constrains the total duration of the experiments to a few seconds. In such an experiment, a BEC is created in a vertically-oriented quadrupole trap, allowed to evolve freely and cooled down through the use of a delta-kick technique to make $\sigma_{\text{\tiny QM},t}$ as small as possible. Finally, it is again allowed to evolve freely, and eventually its position variance is measured.

\subsection{Fundamental interactions}
%(Enno Giese) ($\sim 1$p)
\label{sec:physics-newint}

As discussed in Section~\ref{sec:physics-dm}, dark matter candidates may induce a signal in atom interferometers~\cite{Geraci2016}, where two spatially-separated devices are operated by common laser beams~\cite{Arvanitaki2018}.
The differential measurement suppresses common mode noise and compares the signal induced at two different points in space-time.
Dark matter may induce signatures in both the motion and the internal (electronic) energy states of the atom~\cite{DiPumpo2022}.
The working principle of atom-based GW detectors is similar, but the signature of a GW is encoded in the phase of the atomic cloud and read out by the atom interferometer~\cite{Dimopoulos2008,Hogan2010,Graham2013}.
These examples highlight the point that light-pulse atom interferometers are based on the propagation of atoms \emph{and} light, as well as on their interaction at different points in space time.
Different approaches to model atom interferometers have mainly focused on the propagation of atoms~\cite{Storey1994,Marzlin1996,Antoine2003,Dimopoulos2008,Kleinert2015,Bertoldi2019,Ufrecht2020}, but also detailed studies of diffraction induced by atom-light interactions have been performed~\cite{Wicht2005,Riou2017,Hartmann2020,Siemss2020,Siemss2022}.
These efforts have to be re-evaluated when designing detectors for BSM physics.

To include such fundamental interactions, one starts by introducing additional fields that give rise to novel physics.
The hypothetical interactions with all known constituents of the Standard Model need to be described, that is, the interactions with elementary particles such as electrons, photons, gluons, quarks, and others.
Typically, a complementary approach to high-energy or particle physics is chosen:
classical light and weak BSM fields are assumed instead of second-quantized particles.
The next step is to solve perturbatively the equation of motion of the field and its interaction with other particles.
When the coupling between the new field and known particles is introduced, light-pulse atom interferometers require the treatment of light and atoms.
The latter are composite particles, as are their nuclei, and require an effective description.
One should consider both the centre-of-mass motion of the composite particle and its internal states, as both are manipulated by light and are central to atom interferometry.
One finds an effective coupling of the novel field to the centre-of-mass motion as well as to the internal states, which in general differs from that to the individual elementary particles.
Simultaneously, new fields may directly couple the photons that make up the light pulses through and modify Maxwell's equations.

In principle, the described procedure is generic and can be applied to various types of fields with different symmetries and properties, such as scalar fields like dilatons~\cite{Damour2010}, but also to pseudoscalar fields~\cite{Ni1977,Ni2010} like axions~\cite{Sikivie2014,DiLuzio2020}, or dark photons~\cite{Co2019,Filippi2020}, which could all in principle account for dark matter.
There is a variety of possible extensions of the Standard Model, and a comprehensive overview is beyond the scope of this article.
Instead, we focus on the example of a dilaton field.

In general, the dilaton couples non-trivially to all constituents of the Standard Model, and its couplings can be linearized since it is a weakly-coupled field.
Via this procedure, all coupling constants are modified linearly by the dilaton, including the fine structure constant or the electron and nucleon masses~\cite{Damour1994,Damour2010}.
In particular, both the atom’s mass and its internal energy structure are modified and depend, to lowest order, linearly on the dilaton.
Maxwell's equations are modified similarly, whereas no modification of Einstein's equations is found to lowest order, since the field is ultralight and couples weakly.
The specific form of the dilaton field depends in general on the local environment and may be modified, e.g., by source masses such as the Earth.
It obeys an inhomogeneous wave equation, where the homogeneous part gives rise to plane waves with a specific wave vector that corresponds to its momentum~\cite{DiPumpo2022}.
This contribution serves as a model for dark matter, where the velocity distribution inferred from galactic observations is matched to the plane waves.
The inhomogeneous solution arises from the metric and depends on the local mass-energy density. 
The modified Maxwell's equations lead only to an effect on the amplitude of a classical propagating light ray, not on its phase to leading order, so that no signature arises from light propagation in dark-matter backgrounds~\cite{DiPumpo2022}.
In contrast, the atomic motion and the resulting interferometric phase are sensitive to the dilaton, and the internal energies \emph{and} the mass of the atom depend on it~\cite{Geraci2016,Arvanitaki2018}.
The inhomogeneous solution sourced by a local mass reflects itself in an apparent violation of the equivalence principle~\cite{Roura2020,DiPumpo2021,DiPumpo2023,Asenbaum2020,Schlippert2014}.
It induces a mass- and internal-state-dependent acceleration.
In contrast, the plane-wave solution that models dark matter gives rise to oscillations of internal energies, and also the total mass of the atom.
Consequently, there are two read-out strategies.
To measure an effect on the centre of mass, the atom remains in the same internal state~\cite{Geraci2016,Canuel2018,Canuel2020} using Bragg diffraction~\cite{Giltner1995,Hartmann2020,Siemss2020,Siemss2022}. 
Such an experiment needs two counterpropagating beams so that laser phase noise couples for large spatial separations.
In contrast, single photon transitions~\cite{Graham2013,Rudolph2020} change the internal state, whose energy difference depend on the dilaton field.
Similar to atomic clocks~\cite{Ushijima2015,Bloom2014,Marti2018}, this contribution is dominant~\cite{Kennedy2020}, as it originates from the rest energy of the atom~\cite{Arvanitaki2018,Abe2021,DiPumpo2022}. 
The treatment sketched above has to be reviewed for other possible dark matter candidates, that might differ in their fundamental interactions with atom interferometers. 

\subsection{Tests of Atom Neutrality} 
%(Alexander Gauguet) ($\sim 1$p)
\label{sec:physics-neut}

Although atom neutrality is commonly accepted, it raises the fundamental question of charge quantization in the framework of the SM~\cite{Foot1993,Foot1994}, and therefore relies mainly on experimental observations. Several measurements of the electrical neutrality of matter have been performed using different laboratory approaches. All experimental evidence to date is consistent with atoms being electrically neutral, i.e., there is an exact matching between the charge of the electron ($q_e$) and the proton ($q_p$), and the neutron charge ($q_n$) is zero. The best limits for the electron-proton charge asymmetry $(q_p+q_e)/q_e$ and the residual neutron charge $q_n/q_e$ are near $10^{-21}$ \cite{Unnikrishnan2004,DurstbergerRennhofer2011,Bressi2011}. Most of the methods used so far are measurements based on macroscopic dilute or bulk ensembles that suffer from difficulties in the modelling of systematic effects related to spurious charging effects or the inhomogeneity of electric fields.

Matter-wave interferometers with a macroscopic separation between interferometer arms allow one to shape electromagnetic and gravitational potentials~\cite{Gillot2013,Gillot2014,Overstreet2022}, opening the way to new measurements in fundamental physics based on geometrical phase shifts. In particular, a new test of atom neutrality with an atom interferometer based on the scalar Aharonov-Bohm effect has been proposed \cite{Champenois2001,Arvanitaki2008}. The scalar Aharonov-Bohm effect~\cite{Aharonov1959} appears when opposite electric potentials $\pm V$ are applied on the two interferometer arms during a time $\tau$. The electric potentials are turned on when the atoms are inside the electrode assembly where the electric field is vanishing. If the atom carries a non-zero electric charge $\delta q$, a phase shift proportional to $\delta q$ is induced: $\Delta \phi = \delta q V \tau / \hbar$. Therefore, one can infer a limit on the atom neutrality and the charge per nucleon from the uncertainty in the Aharonov-Bohm phase shift. Besides, by performing the measurements on the two atomic species, one can place independent bounds on the neutron charge $q_n$ and on the electron-proton charge asymmetry $q_e+q_p$. Then, assuming charge conservation in neutron $\beta$-decay ($n \rightarrow p + e^{-} + \overline{\nu}_e$), it is possible to infer a limit on the neutrino ($\overline{\nu}_e$) electric charge with the same accuracy as $q_p+q_e$ and $q_n$.

The Aharonov-Bohm method can potentially surpass by orders of magnitude the current bounds on atom neutrality. In particular, very large scale atom interferometers offer opportunities for efficient implementation of such tests, as they provide large separations between the interferometer arms leading to well-separated interaction zones and very long duration of the voltage pulse $\tau$. In addition, the gravitational antennas studied here anticipate an extremely high signal-to-noise ratio and a very low-noise environment~\cite{Abe2021}, opening up the prospect of electrical neutrality tests below $10^{-28} q_e$.

Astrophysical methods can also provide bounds on the neutrality of atoms \cite{Sengupta2000,Caprini2005} and neutrinos \cite{Sengupta1996,Raffelt1999}. However, these limits depend on specific assumptions and models. Therefore, the new bounds from atom interferometry could help to refine such astrophysical models, and might stimulate new studies in astrophysics. In addition, pushing the limits on atom neutrality is of great interest in particle physics, as the origin of the extreme fine-tuning between the charges of fundamental particles can be considered as a hint for new physics beyond the SM~\cite{Foot1993}. Indeed, some specific models propose de-quantization of the electric charge \cite{Witten1979,Arvanitaki2008}. We close by recalling the comment that most of the experiments {\it ``were done decades ago, and at the time were rather one-[person] shows. This is a pity in view of the effort invested in other searches beyond the standard model"} \cite{Dubbers2011}.

\section{Synergies of Cold Atom and Laser Interferometer Experiments}
\label{sec:GW}
%{\bf 5 pages guideline }
%\subsection*{Editors: Chiara Caprini, John Ellis}

\subsection{Introduction}
\label{sec:synergiesintroduction}

In this Section we summarise three aspects of the synergies between terrestrial atom and laser interferometer experiments. Both types of experiment must confront and mitigate the effects of the Earth's seismic activity. Laser interferometers are particularly vulnerable to vibrations of their mirrors, which can be mitigated by the design of their mounts, whereas the motions of clouds in atom interferometers are sensitive to fluctuations in the Earth's gravitational field as it vibrates, called Gravity Gradient Noise (GGN), which cannot be shielded. The synergies between the two types of experiment depend on the degree to which GGN effects can be minimised, allowing atom interferometers to explore frequency ranges that are inaccessible to laser interferometers and maximise their complementarity, as discussed in the first part of this Section. Prospective synergies in studies of astrophysical GWs are discussed in the second part of this Section.
These include the observations of inspiral stages of stellar-mass black holes that will be observed later by laser interferometers such as LIGO/Virgo/KAGRA, ET and CE. These will provide tests of gravity including constraints on the graviton mass and Lorentz violation~\cite{Ellis2020} as well as predicting the times and directions of subsequent mergers, thereby facilitating multimessenger observations. Furthermore, observing the long inspiral phases of stellar-mass black hole binaries also allows important tests of their, yet mysterious, formation processes and of their environments, e.g., through by detecting the eccentricities of their orbits~\cite{Nishizawa2017,Cardoso2021,Franciolini2022,Xuan2023} and/or of the peculiar velocities and accelerations of their centres of mass~\cite{Inayoshi2017,Toubiana2021,Sberna2022}. Atom interferometers could also observe the mergers of intermediate-mass black holes (IMBHs), which could probe the strong gravity regime as well as probe the hierarchical merger history of SMBHs. Other aspects of the complementarity between atom and laser interferometers
are discussed in the final part of this Section.

\subsection{Battling Gravity Gradient Noise}
%Leonardo Badurina}
\label{sec:Battling}

Due to their exquisite sensitivity to the propagation of atoms and changes in their structure, terrestrial very long-baseline atom gradiometers will be exceptionally powerful probes of gravitational waves (GW) in the unexplored ‘mid-frequency band’~\cite{Dimopoulos2009,Graham2013,Ellis2020,Badurina2021} and linearly-coupled scalar ultralight dark matter (ULDM) with mass between $\sim \SI{e-17}{eV}$ and $\sim \SI{e-11}{eV}$~\cite{Arvanitaki2018, Badurina2022}. 

The projected reach of these experiments is ultimately limited by fundamental noise sources.~\footnote{See Section~\ref{sec:Noise} for a more complete discussion of possible noise sources.}
For instance, vertical single-photon atom gradiometers such as AION~\cite{Badurina2020} and MAGIS-100~\cite{Abe2021} are designed to reach the atom shot-noise limit above $\sim\SI{1}{Hz}$, but would suffer from gravity gradient noise (GGN) at lower frequencies~\cite{Baker2012,Vetrano2013,Harms2013,Mitchell2022}. This type of phase noise arises as a result of mass density fluctuations of the ground and atmosphere~\cite{Harms2019}, which perturb the local gravitational potential around the atom clouds and imprint a noisy phase shift in a differential measurement. 

For the experimental configurations and frequency range of interest, the dominant source of GGN is expected to consist of ground density perturbations induced by horizontally-propagating seismic waves, in particular fundamental Rayleigh modes, that are confined near the Earth's surface by horizontal geological strata and are generated at strata interfaces, such as the Earth's surface~\cite{Hughes1998,Harms2019}. Representing the seismic field as an incoherent superposition of monochromatic plane waves propagating isotropically at the Earth's surface, following previous studies performed for LIGO~\cite{Hughes1998,Harms2019}, and using the Peterson low and high noise models (NLNM NHNM) driven by seismic-noise data from different terrestrial locations, Ref.~\cite{Badurina2023a} showed that GGN could significantly impact the projected reach of experiments with baselines $L\gtrsim\mathcal{O}(100\,~\mathrm{m})$ in the sub-Hz regime. 

\begin{table}
\begin{center}
\begin{tabular}{ %c | 
c c  c c c c c}
%\hline
\hline
%Design & 
$L$~[m] & $T$~[s] & $n$ & $\Delta z_\mathrm{max}$~[m] & $\delta \phi \, [1/\sqrt{\mathrm{Hz}}]$ & $T_\mathrm{int}$~[s] \\
\hline
%Advanced & 
$1000$ & 1.7 & $2500$ & $970$ & $10^{-5}$ & $10^8$ \\
\hline
%\hline
\end{tabular}
\end{center}
\caption{List of experimental parameters used for the computation of the sensitivity plots shown in this Section, which could be implemented in future vertical gradiometers, such as AION-km. For reference, $L$ is the length of the baseline, $T$ is the interrogation time, $4n-1$ is the total number of LMT kicks transferred during a single cycle, $\delta \phi$ is the shot noise-limited phase resolution,  $T_\mathrm{int}$ is the integration time, and $\Delta z_\mathrm{max}$ is the maximum gradiometer length given the choice of interferometer parameters. We also consider scenarios where $\Delta z$ is shorter than the maximum value. The set of geological parameters is taken from Ref.~\cite{Badurina2023a}.} 
\label{table:ExperimentalParameters}
\end{table}

\begin{figure*}
    \centering
    \includegraphics[width=.495\textwidth]{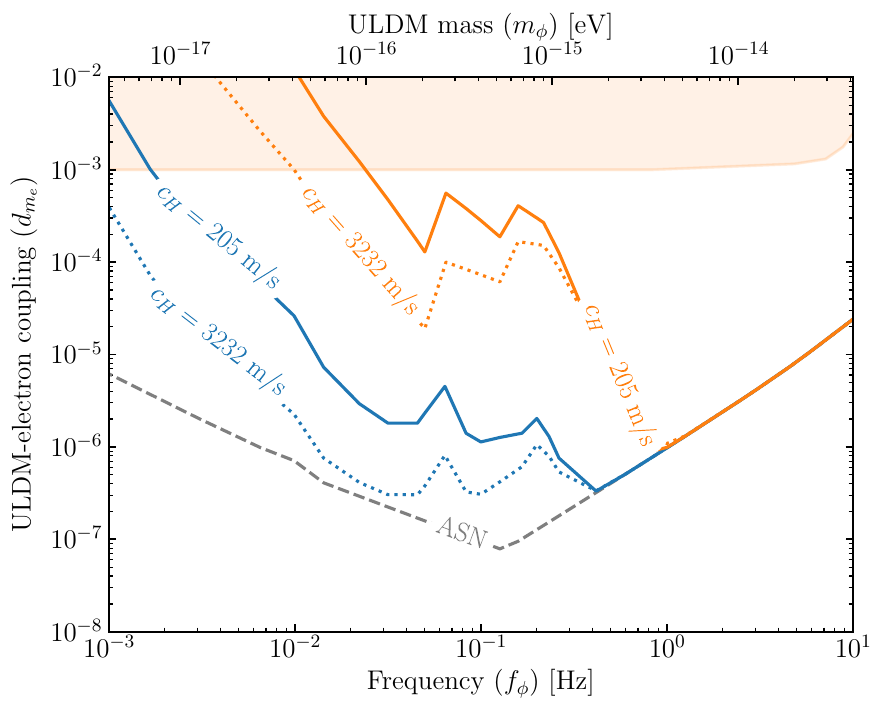}
    \includegraphics[width=.495\textwidth]{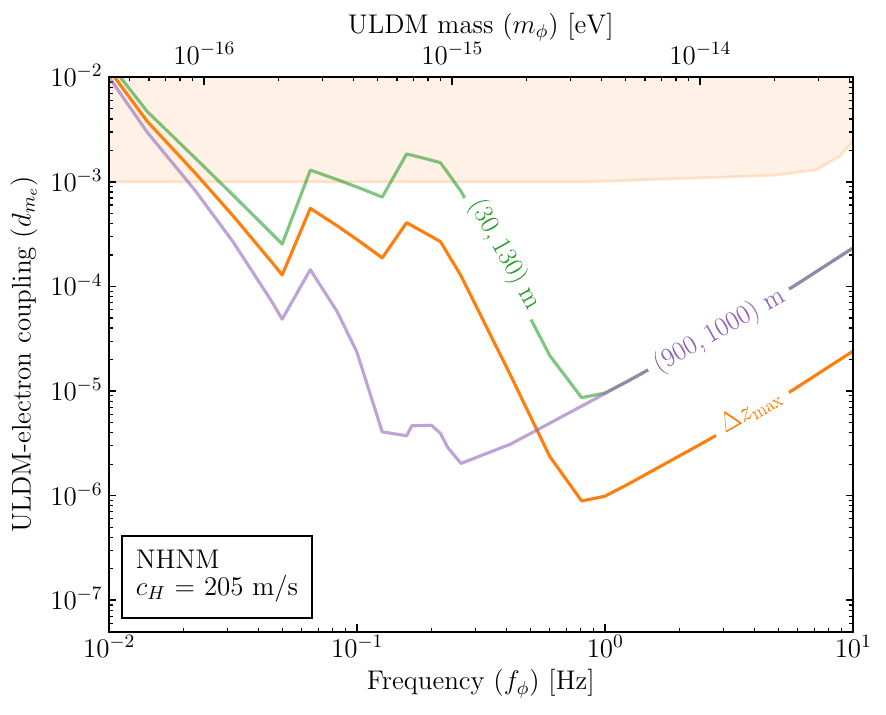} 
    \caption{
  Impact of GGN on the projected 95\% CL exclusion sensitivity to the ULDM-electron coupling of a single atom gradiometer with the %`advanced' 
  design parameters defined in Table~\ref{table:ExperimentalParameters}.
  \textit{Left panel}: comparison between the atom shot noise (ASN) (grey) and ASN-plus-GGN-limited sensitivities assuming that the GGN background is described by the Peterson NHNM (orange) or NLNM (blue). The solid and dotted lines are for Rayleigh wave velocities $c_H = 205\,\mathrm{m\,s}^{-1}$ and $c_H = 3232\,\mathrm{m\,s}^{-1}$, respectively. \textit{Right panel}: projected 95\% CL exclusion sensitivities for different values of $\Delta z$ and different atom interferometer positions, where we assume the NHNM and $c_H = 205\,\mathrm{m\,s}^{-1}$. We show exclusion curves for interferometers located towards the Earth's surface (green) and towards the bottom of the shaft (purple), assuming $\Delta z = 100$\,m but keeping all other experimental parameters unchanged. In both panels, the orange shaded region is excluded by MICROSCOPE~\cite{Touboul2022}.} 
\label{fig:single_gradiometer}
\end{figure*}

Several passive noise mitigation strategies could be implemented to recover large swathes of parameter space accessible to a purely shot-noise-limited device, especially in the crucial 0.1-10~Hz band. We focus here on ULDM searches, although the same applies in the context of GW searches with these large-scale quantum sensors, assuming the parameters listed in Table~\ref{table:ExperimentalParameters}. As illustrated in the left panel of Fig.~\ref{fig:single_gradiometer}, by choosing quiet sites (i.e., with GGN-limited sensitivity curves close to the NLNM curve), it may be possible to regain up to three orders of magnitude in sensitivity. 

It may also be possible to suppress GGN in the crucial mid-frequency band significantly by carefully choosing the experimental parameters, such as the vertical positions of a string of interferometers along the baseline. Since the GGN differential phase shift decays exponentially with depth~\cite{Badurina2023a}, experiments that are deep underground and far from sources of fundamental Rayleigh modes are expected to be more powerful probes of GW and ULDM. As shown in the right panel of Fig.~\ref{fig:single_gradiometer}, where we assume the NHNM, experiments that are deeper underground would be characterised by a sensitivity enhancement of up to two orders of magnitude in the sub-Hz regime, but would be less sensitive to a signal above 1~Hz. 

It would be advantageous to select sites whose geological composition supports Rayleigh waves with high horizontal propagation speed $c_H$. As shown in the left panel of Fig.~\ref{fig:single_gradiometer}, below 1 Hz where the decay length exceeds the separation between the interferometers, the GGN-limited background is suppressed by several orders of magnitude at high $c_H$~\cite{Badurina2023a}. This is because the GGN-induced gradiometer phase shift at low frequencies is proportional to $c_H$.

\begin{figure*}
    \centering
    \includegraphics[width=.495\textwidth, valign = t]{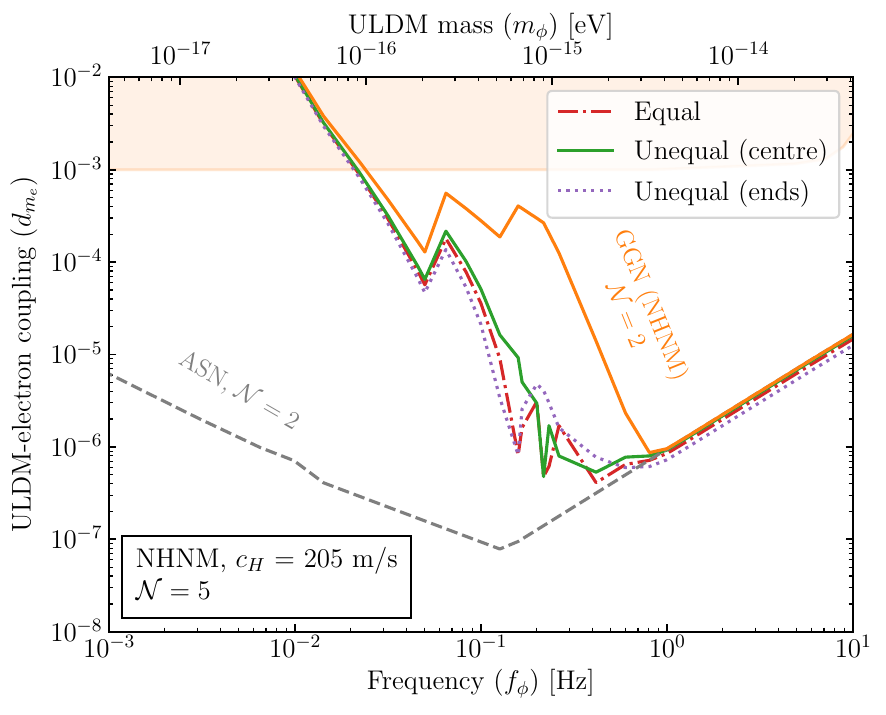}
    \includegraphics[width=.437\textwidth, valign = t]{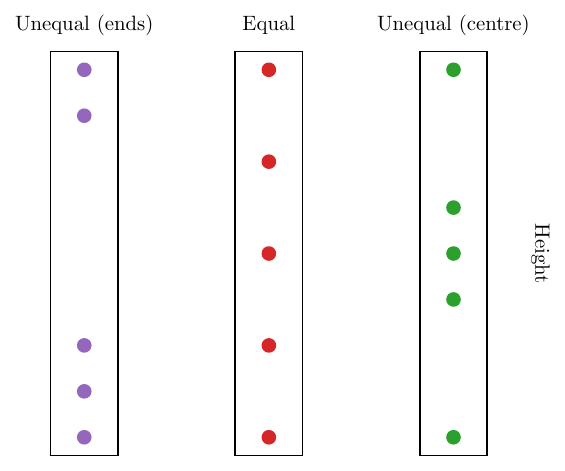} 
\caption{GGN mitigation using a multigradiometer configuration.
  \textit{Left panel}: projected 95\%~CL exclusion sensitivities for an atom multigradiometer with the
  experimental parameters listed in Table~\ref{table:ExperimentalParameters} and $\mathcal{N}=5$ interferometers, assuming that GGN is modelled by the NHNM.
The red dot-dashed, purple dotted and green solid lines show the atom multigradiometer exclusion curves for  equally-spaced, unequally-spaced (ends) and unequally-spaced (centre) configurations. The orange shaded region is excluded by MICROSCOPE~\cite{Touboul2022}.
For comparison, the grey and orange lines show the exclusion sensitivities for a single atom gradiometer ($\mathcal{N}=2$) with ASN-only and ASN-and-GGN backgrounds, respectively.
\textit{Right panel}: schematic representations of the three interferometer configurations with $\mathcal{N}=5$. The purple dots show the positions of the interferometers in the `unequal spacing (ends)' configuration, the red dots show their positions in the `equal spacing' configurations, and the green dots show the `unequal spacing (centre)' configuration.} 
\label{fig:multigradiometer}
\end{figure*}

In geological media where Rayleigh modes propagate with a low horizontal speed, it may instead be advantageous to employ a network of $\mathcal{N}\geq 3$ atom interferometers along the same vertical baseline, i.e., a multigradiometer configuration~\cite{Badurina2023a}.
%, which resembles the ELGAR experiment's proposed design~\cite{Canuel:2019abg}. 
Since the GGN and ULDM signals scale differently with the gradiometer length, this design would facilitate the mitigation of GGN by up to two orders of magnitude in the mid-frequency band for different spatial configurations. This is clearly visible in Fig.~\ref{fig:multigradiometer}, which shows the sensitivity reach in scalar ULDM parameter space of several illustrative configurations employing $\mathcal{N} = 5$ interferometers and assuming the NHNM.  

%In conclusion, terrestrial very-long baseline atom gradiometers offer the very exciting prospect of probing unexplored regions of GW and ULDM parameter space. However, GGN would severely limit reach of these experiments in the sub-Hz regime, where experiments are expected to have peak shot noise-limited sensitivity. 
These results show that more detailed and site-specific modelling will be needed to improve our understanding of the projected reach of these experiments in well-motivated parts of parameter space. For example, it would be desirable to assess whether specific sites have GGN levels closer to the NHNM or the NLNM, and the speeds of Rayleigh waves should be measured. Furthermore, in order to model correctly the GGN at realistic sites, the model presented here should be extended to anisotropic environments with different geological strata, which would give rise to a much richer spectrum of Rayleigh modes~\cite{Hughes1998,Harms2019}. In addition, beyond the simple noise mitigation strategies presented here, active GGN mitigation techniques could also be implemented. For example, it may be possible to model the GGN phase shift imprinted onto the atoms by using an array of seismic sensors, whose data would then be subtracted from the interferometer's data stream through Wiener filtering~\cite{Coughlin2014}.

The analysis summarised here should be interpreted as a first stepping stone towards understanding how to characterise and ultimately suppress the GGN background within the context of terrestrial very-long baseline atom gradiometers.~\footnote{A discussion of possible mitigation strategies can be found in Section~\ref{sec:Mitigation}.}

\subsection{Tests of gravity and formation of supermassive black holes}
%: Ville Vaskonen}
\label{sec:SMBHs}

\begin{figure}
\centering
\includegraphics[width=0.56\textwidth]{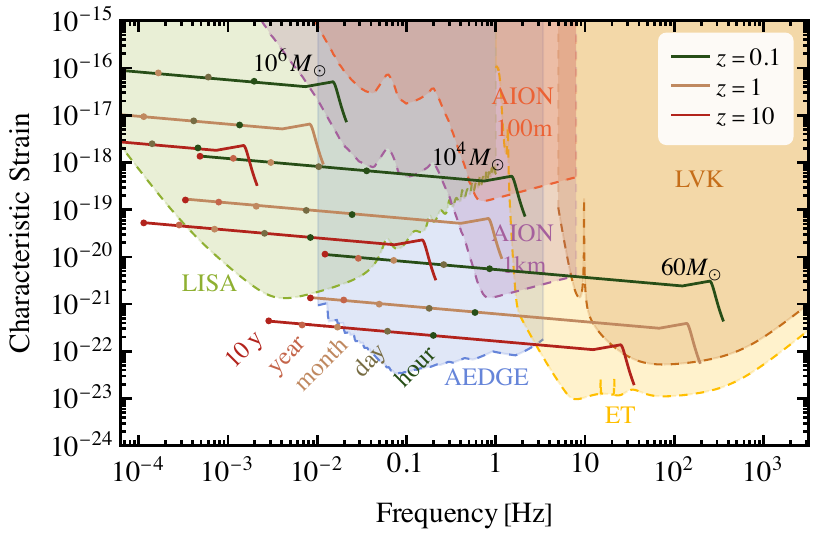}
\caption{The GW strain sensitivities and benchmark signals from BH binaries of different masses at different redshifts. The coloured dots indicate the times before mergers at which inspirals could be measured.}
\label{fig:BHstrains}
\end{figure}

Atom interferometers have the potential to bridge the frequency gap between LISA and ground-based laser interferometers, as shown in Fig.~\ref{fig:BHstrains}. As seen there, this frequency range is optimal for probes of inspiralling stellar-mass BH binaries and observing directly the mergers of IMBH binaries. 

As seen in Fig.~\ref{fig:BHstrains}, a space-based atom interferometer such as AEDGE could probe the inspirals of stellar mass BH binaries for several months. Depending on the level at which GGN can be suppressed, this may be possible also with very-long-baseline terrestrial atom interferometers. Over such a long observation time, the reorientation and movement of the detector with respect to the GW source become relevant. Because of these effects, the sky location of the binary can be measured very accurately~\cite{Graham2018a}.~\footnote{For comparison, the LIGO/Virgo network sees only the last seconds of the signals, limiting its sky-localisation accuracy without additional detectors. LISA will have similar accuracy to AION/AEDGE, but for heavier binaries.} In addition, as demonstrated in Table~\ref{table:BHparameters}, the long observation time enables a very accurate measurement of the binary chirp mass that controls how the frequency of the signal changes with time. Towards the end of the inspiral, hours or minutes before the merger and ringdown stages of the event, the GW signal leaves the sensitive frequency window of the atom interferometers. Since the time of the merger can be accurately predicted from their observations, together with the sky location, the atom interferometers provide an excellent early warning system for measurements of the GW signal from the binary merger as well as for searches for possible electromagnetic counterparts. 

\begin{table}
\centering
\begin{tabular}{ p{2.cm} | p{1.8cm} p{1.1cm} p{1.8cm} p{1.5cm} p{1.2cm} p{1.2cm}} \hline%\hline
 & $\sigma_{\mathcal{M}_c}/\mathcal{M}_c$ & $\sigma_{t_c}/{\rm s}$ & $\sigma_{D_L}/D_L$ & $\sigma_\theta/{\rm deg}$ & $\sigma_\phi/{\rm deg}$ \\ \hline
AION 1\,km & $2.2\times 10^{-5}$ & $12$ & $0.7$ & $2.7$ & $2.7$ \\
AEDGE & $3.6\times 10^{-7}$ & $0.38$ & $0.03$ & $0.15$ & $0.077$ \\
\hline%\hline
\end{tabular}
\caption{Estimated accuracies of AION 1\,km and AEDGE measurements of the chirp mass $\mathcal{M}_c$, the coalescence time $t_c$, the luminosity distance $D_L$ and the sky location, specified by the angles $\theta$ and $\phi$, assuming a GW150914-like benchmark source. From Ref.~\cite{Ellis2020}.}
\label{table:BHparameters}
\end{table}

\begin{figure}
\centering
\includegraphics[width=0.65\textwidth]{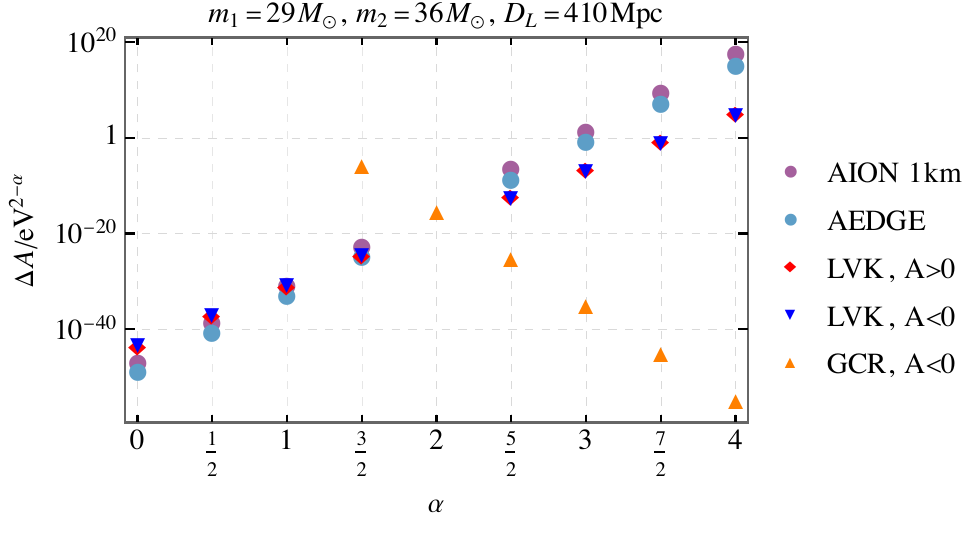}
\caption{Prospective sensitivities to modified GW dispersion relations of AION 1\,km and AEDGE, compared with the constraints from LVK and gravitational Cherenkov radiation. Figure from~\cite{Ellis2020}.}
\label{fig:beyondGR}
\end{figure}

The long observation time of the inspiral phase enables also searches for modifications of general relativity that could alter the GW signal. For example, a modified GW dispersion relation $E^2 = p^2 + A p^\alpha$ would introduce an extra phase for the GW signal, $\Psi(f) \to \Psi(f) + \delta \Psi(f)$, whose frequency dependence is determined by the magnitude $A$ and the parameter $\alpha$, $\delta \Psi \propto A f^{\alpha-1}$. Modifications with $\alpha < 1$ change the phase more at small frequencies and are easier to probe from the inspiral phase than from the merger. Therefore, as illustrated in Fig.~\ref{fig:beyondGR}, the measurements of the inspiral signals of stellar mass BH binaries with atom interferometers will complement the measurements done with the ground-based laser interferometers, providing a better probe of modified dispersion relations with $\alpha<1$. In addition, the low-frequency sensitivity of AI gravitational-wave detectors is beneficial for probing extra radiation channels of gravitational waves, in particular the possibility of dipolar radiation~\cite{Zhao2021}, which is ultimately related to the strong equivalence principle.

\begin{figure}
\centering
\includegraphics[width=0.4\textwidth]{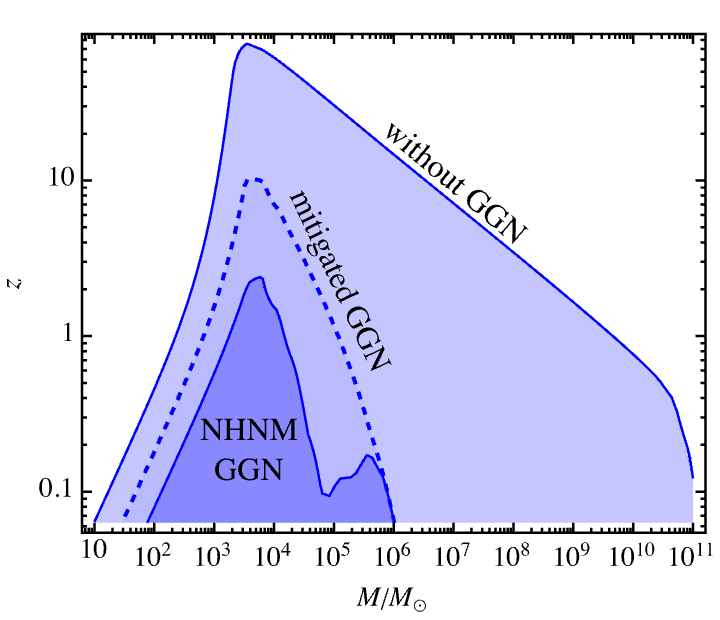}
\includegraphics[width=0.51\textwidth]{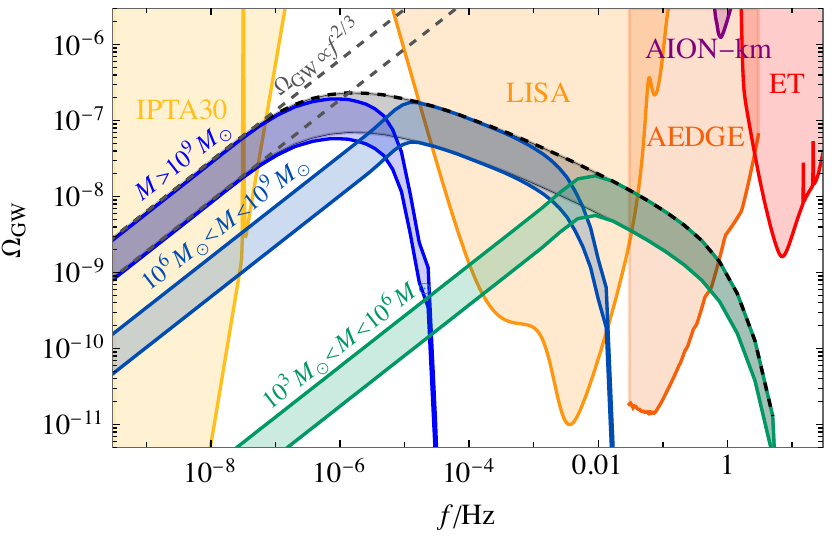}
\caption{Left panel: The sensitivities of AION 1\,km to GWs from equal mass BH binaries of total mass $M$ at redshift $z$, calculated assuming a level of GGN close to the NHNM and assuming that Rayleigh waves propagate with a speed of $205$~m/s. The contours compare estimates made assuming either no mitigation of GGN, or the level of suppression discussed in the previous Subsection, or complete suppression/mitigation of GGN. Right panel: The mean GW energy density spectrum from massive BH mergers compared with the sensitivities of the indicated experiments. The coloured bands correspond to different BH mass bands and are obtained assuming a constant merger efficiency factor $0.3 < p_{\rm BH} < 1$, following~\cite{Ellis2023}: plot adapted from~\cite{Ellis2023a}.}
\label{fig:IMBH}
\end{figure}

The population of BHs with masses intermediate between those whose mergers have been observed by LIGO and Virgo and those known to be present in the centres of galaxies has not been explored to the same extent, though insight into their potential importance for the assembly of SMBHs is now being provided by JWST observations of $10^7$ to  $10^8$ solar-mass black holes 
at redshifts up to $z=10$~\cite{Maiolino2023}. Very large terrestrial atom interferometers offer prospects for detecting IMBH mergers that are inaccessible to laser interferometers, as seen in Fig.~\ref{fig:BHstrains}. We see there that the early inspiral stages of IMBH binaries could be measured by LISA~\cite{AmaroSeoane2017} and TianQin~\cite{TorresOrjuela2023}, enabling the timings and sky locations of subsequent IMBH mergers to be predicted and facilitating multimessenger studies in combination with atom inteferometer observations, complementing the observations of heavier binaries that will be made with LISA and providing opportunities to probe strong gravity in a novel regime. However, the prospects for observing IMBH mergers depend on the extent to which GGN can be suppressed or mitigated. The left panel of Fig.~\ref{fig:IMBH} displays the IMBH merger sensitivities at the SNR = 8 level in three GGN scenarios: the NHNM with $c_H = 205$~m/s and no mitigation as in Fig.~\ref{fig:multigradiometer}, the same scenario with 5 gradiometers as also seen in Fig.~\ref{fig:multigradiometer}, and assuming total mitigation or suppression of the GGN. We see that in these scenarios AION 1\,km could detect $\mathcal{O}(10^4 M_\odot)$ binaries up to redshift $z \approx 2, 10$ and 70, respectively. Since the first astrophysical BHs are thought to form at $z\lesssim 20$, we would not expect to observe mergers of IMBHs beyond this redshift unless they are primordial, which would be of major interest.

The detection of IMBH mergers would enable studies of models for SMBH formation. In the absence of primordial SMBHs, they could have been assembled hierarchically from seeds such as BHs formed by the deaths of the first stars or by the direct collapses of gas clouds (for a review, see, e.g., Ref.~\cite{Greene2012}), combined with accretion. Astrophysical seeds would have been much lighter than the SMBHs we detect in galactic centres today, with masses ranging from $\mathcal{O}(100 \, M_\odot)$ in the stellar scenario to $\mathcal{O}(10^4 \, M_\odot)$ in the gas-cloud scenario. The hierarchical merging of BHs originating from such seeds provides a potential probe of the SMBH formation through GW observations. Atom interferometers could probe the mergers of the seeds, potentially distinguishing between the seed scenarios.

The existence of SMBH binaries is currently being probed by pulsar timing arrays (PTAs), and first detections of a GW background signal that may come from such binaries have recently been reported~\cite{Arzoumanian2020,Goncharov2021,Chen2021,Antoniadis2022}. The right panel of Fig.~\ref{fig:IMBH} shows the mean energy of the GW background from BH binaries estimated by extrapolating the PTA results to lower BH masses and higher frequencies,  assuming for simplicity a constant merger efficiency $0.3 < p_{\rm BH} < 1$~\cite{Ellis2023a,Ellis2023} and neglecting a possible additional contribution from low-mass seeds. Most of the GW signal expected in the LISA and atom interferometer frequency ranges is expected to be due to resolvable binaries, and it has been estimated that a space-based atom interferometer such as AEDGE could observe $\mathcal{O}(100) - \mathcal{O}(1000)$ IMBH mergers per year, while a terrestrial detector such as AION-km might observe as many as $\mathcal{O}(10)$ IMBH mergers per year~\cite{Ellis2023a,Ellis2023}, see also~\cite{Ellis2023b}.

The results summarised here show that terrestrial very-long-baseline atom interferometers could complement laser interferometers by exploring an intermediate frequency range where measurements could be used to probe general relativity, provide `early warnings' of future mergers of stellar mass BHs, and observe for the first time mergers of IMBHs. We note finally that the sensitivity to GWs from BH mergers would be closer to the ``without GGN" curve in Fig.~\ref{fig:IMBH} at a site where the GGN level is closer to the NLNM and/or the speed of Rayleigh waves is higher, emphasising the importance of future site studies (see the discussion in Section~\ref{sec:Battling}).

\subsection{New Physics with Cold Atom and Laser Interferometers}
%: Diego Blas}
\label{sec:NewPhysics}

The spectacular technological advances in atom and laser interferometers are having a significant impact on defining the current landscape of searches 
of new physics. This brief contribution describes some of the `new physics' cases that have not been discussed in the previous Subsections, emphasising how large atom interferometers and laser space interferometers can give complementary information on these problems. Finally, another direction that is gaining momentum among the community of physicists interested in precision   measurements and fundamental physics is also mentioned briefly: the search for ultra-high-frequency gravitational waves.

The LISA mission~ \cite{AmaroSeoane2017} is expected to be launched in the mid 2030s. The final configuration and specifications are to be decided, but LISA is expected to be able to
detect gravitational waves in a band  $\approx 10^{-4}-1$\, Hz, which has an enormous physics potential. In addition to the anticipated astrophysical signals, one of the best studied fundamental physics signals that could be measured by LISA is a stochastic background of GWs from cosmological sources~\cite{Auclair2022}. The physics of inflation, the existence and evolution of primordial black holes, the existence and dynamics of topological defects such as cosmic strings or the possibility of first-order phase transitions in the early Universe, especially at the electroweak scale, will be intensively explored by LISA. In all these cases it will be essential to make complementary observations in other frequency bands to establish the nature of any possible signal. Large atom interferometers offer a unique opportunity for exploring the frequencies between LISA and Earth-based laser interferometers, see, e.g., Fig.~~\ref{fig:CSexpansion} and~\cite{Badurina2020}.

However, the topics above are far from being a complete list of the fundamental physics horizons that LISA will explore, see, e.g., \cite{Arun2022,AmaroSeoane2022}. In order to assess the main scientific outputs of the LISA mission, and to plan and develop the work needed to ensure their delivery, the LISA Consortium has established a Science Investigation work Package (SIWP) within the LISA Science Group, which gives advice on how to optimise the LISA mission to maximise its scientific return. Furthermore, several Working Groups containing interested members of the scientific community also operate. Among these, the Cosmology Working Group and the Fundamental Physics Working Group are particularly focused on the ability of LISA to test ``new” physics and explore novel ideas. The efforts within these Working Groups aim at exploring the vast potential of GWs to test both the late and the early Universe, and thereby probe fundamental questions such as the nature of gravity and high-energy particle theory beyond the Standard Model, and yet mysterious phenomena such as inflation, dark energy and the late-time expansion of the Universe, dark matter, primordial phase transitions and the unification of forces, baryogengesis, and so on. In the following, we focus mainly on three directions for new physics: dark matter, tests of black holes and tests of general relativity. 

\begin{figure}
\centering
\includegraphics[width=0.6\textwidth]{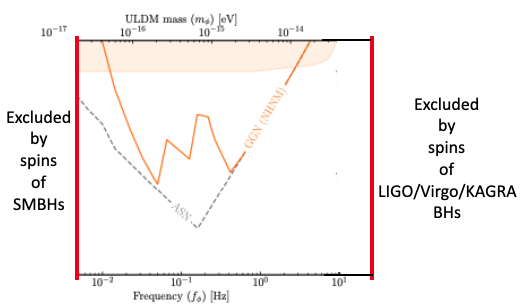}
\caption{Exclusions of weakly-interacting ultra-light bosonic fields from the
measured spins of SMBHs and LIGO/Virgo/KAGRA BHs compared with the prospective
sensitivity of a large atom interferometer, which could also exclude the
intermediate mass range by measuring spins of IMBHs. These constraints assume negligible bosonic self-interactions.}
\label{fig:IMBHDM}
\end{figure}

The dark matter studies are mostly devoted to two of the most significant ways in which it can impact the GW signals to be detected in LISA~\footnote{The possibility that part of the dark matter is made of primordial black holes may leave a signal in LISA, see, e.g., \cite{Arun2022}.}. The first is related to \emph{GWs from ultralight dark matter}, and is based on rotational superradiance, the phenomenon behind the fact that in the presence of ultra-light bosonic fields with mass $m_b$, a black hole of mass $M\sim 1/m_b$ (in $G=c=1$ units) with large spin will develop a non-spherical cloud of the bosonic field \cite{Brito2015}. This cloud  could have a mass that is a significant fraction of $M$, and would generate GWs as it rotates~\cite{Tsukada2021}. Calculations show that LISA may~\footnote{Many uncertainties fall into this number, in particular there are different models for the population of highly spinning black holes of different mass.} be sensitive to \emph{any} new boson with mass in the range $10^{-19}-10^{-17}$\,eV, while LIGO, Virgo and KAGRA are sensitive to masses in the range $10^{-13}-10^{-12}$\,eV \cite{Brito2017}. Such particles are currently being explored as dark matter candidates, as discussed in Sections~\ref{sec:physics-dm} and \ref{sec:Battling}, and atom interferometers can explore the mass range between the LISA, LIGO, Virgo and KAGRA ranges in two ways: by searching directly for ULDM as discussed in Subsection~\ref{sec:physics-dm} or by discovering mergers of IMBHs and measuring their spins, as illustrated in Fig.~\ref{fig:IMBHDM}. Note, however, that these constraints are are model-dependent, and those shown in  Fig.~\ref{fig:IMBHDM} assume negligible bosonic self-interactions.

The second way to explore dark matter is via its effects on GW emission from binary systems. The orbital motion of a binary BHs may be affected by dark matter, modifying the general relativity predictions for GWs emitted during inspiral and infall, for which different possibilities have been explored in the literature \cite{Barausse2014,Annulli2020,Cardoso2020,Coogan2022,AmaroSeoane2022}. The observability of this effect with LISA may require the density of dark matter in the binary to be much higher than that reached in minimal models, but it is nevertheless an interesting target for LISA and for large atom interferometers. 

\begin{figure}
\centering
\includegraphics[width=0.7\textwidth]{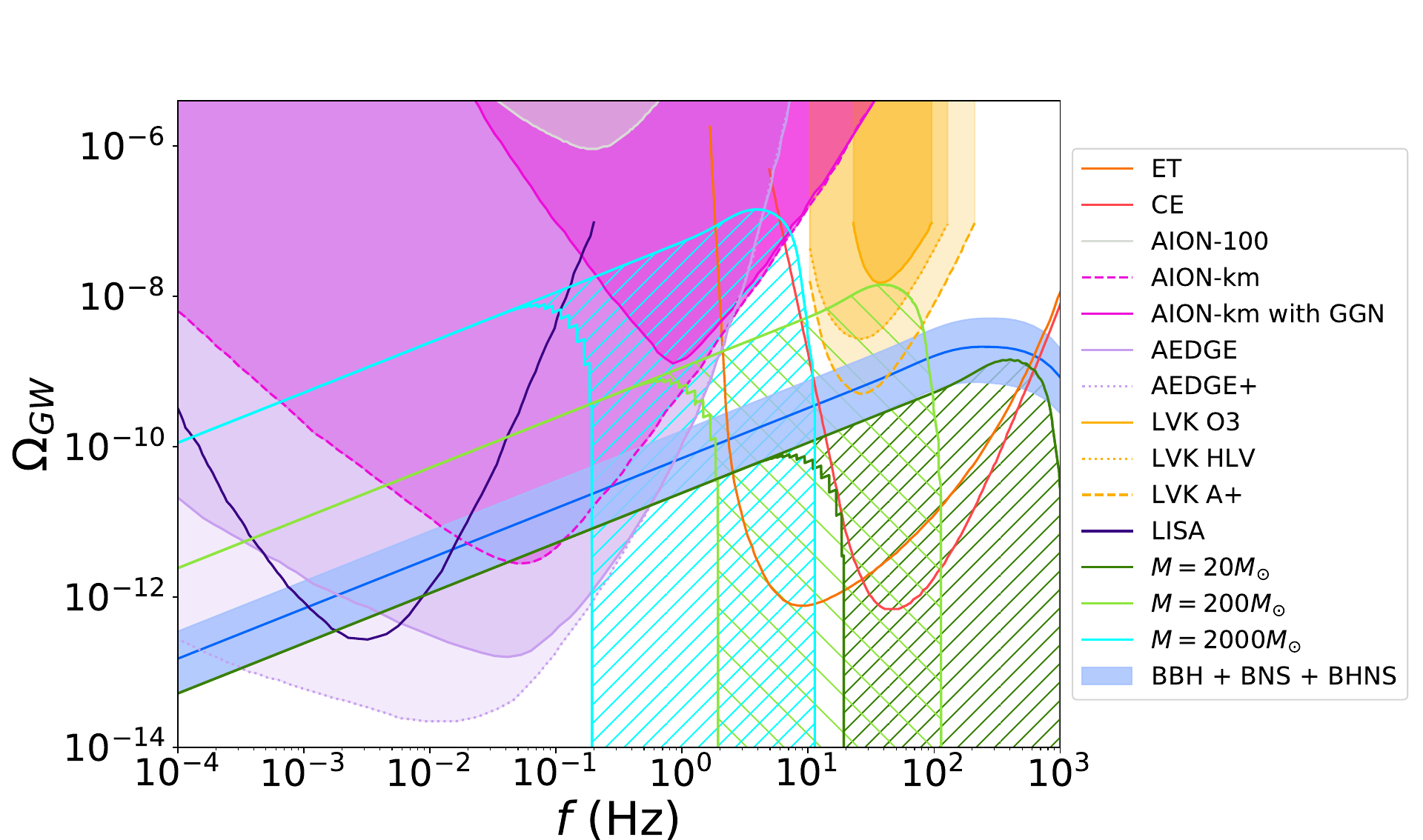}
\caption{Sensitivities of LVK, LISA and large atom interferometers to GWs
from mergers of ECOs weighing between 20 and 200 solar masses, compared
with the backgrounds from BH-BH and BH-neutron star binaries~\cite{Banks2023}.}
\label{fig:ECO}
\end{figure}

Regarding \emph{tests of black holes}, one possibility  is that (some) of the compact objects producing GWs may not be BHs, but other \emph{exotic compact objects} (ECOs). Different candidates are described in \cite{Cardoso2019}, and may correspond to new sectors of matter or to modifications of General Relativity (that may be inspired from quantum gravity or from agnostic modifications of the theory). The possible tests of this hypothesis from GWs are multiple and include new channels of emission, deviations from the unique properties and no-hair theorem of the Kerr metric (that uniquely describes a rotating black hole in General Relativity), changes in the tidal deformation and quasi-normal modes of the object, and even the presence of echoes of the original GW signal \cite{Arun2022}. 
The impact of the LISA mission duration on this possibility has been studied in \cite{AmaroSeoane2022}. See Fig.~\ref{fig:ECO} for results from a study of possible ECO signals in LIGO, LISA and atom interferometers, as well the background to be expected from the mergers of stellar mass BHs~\cite{Banks2023}.

As concerns \emph{tests of general relativity (GR)}, the approach followed by the SIWP is an agnostic one. The current and future probes of Lorentz-violating deviations from GR in the GW waveform~\cite{Yunes2009,AmaroSeoane2022} using laser interferometers can be found in~\cite{Yunes2016,Carson2020}, and the sensitivities of atom interferometers have been studied in~\cite{Ellis2020},
see Section~\ref{sec:GW}.

Some other new directions that  have been discussed within the AION Collaboration. One is the possible influence of dark matter (DM) scattering on the atoms~\footnote{A background that we know is present is the one produced by neutrinos of cosmic origin, whose detection remains a challenge \cite{Bauer2023}.}. Since the nature of DM is unknown, this search should consider all possible interactions to see if its scattering with the atoms may be substantial for some models. There may be detectable effects even at zero-momentum transfer that could fill a gap in searches for DM models below those explored by more traditional detectors, with possible  spectacular bounds \cite{Alonso2019,Du2022}. If one considers a possible transfer of momentum, the trajectories of the atomic samples may be modified, something that will also impact the patterns of interferometry, see, e.g., \cite{Domcke2017} for related ideas with LISA.

Another direction to explore with atom interferometers is the possibility of improving the measurement of the gravitational (Newton) constant $G_N$. This fundamental constant is not measured to high accuracy, and its value is the subject of some controversy \cite{Xue2020}. The measurement of $G_N$ would require a movable test mass located next to the interferometer, so that its effect on the atomic states may be measurable, see, e.g., \cite{Bertoldi2006,Fixler2007,DAmico2017}. The presence of such a moveable mass would also allow for probes of a possible fifth force, see, e.g., \cite{Jaffe2017,Sabulsky2019}.

It should also be noted that the dynamics of the \emph{lasers} operating in atom interferometers may also be impacted by new physics: some inspiration from  light interferometers can be found in \cite{Liu2019,Obata2018}.

For completeness, we mention another fundamental physics topic that is beyond the range of laser interferometers, namely the search for ultra-high frequency GWs (UHFGWs) \cite{Aggarwal2021}. It seems extremely hard to generate a substantial GW signal in this region with standard astrophysical processes, but this provides an opportunity, since there are primordial or non-standard processes that can do it. Detecting these high-frequency GWs requires small, laboratory-size experiments. The passage of a GW will slightly deform a solid, which was suggested long ago as a way to detect them \cite{Weber1960}. It will also generate photons out of a stationary magnetic field \cite{Gertsenshtein1962}. These effects will, for instance, affect the dynamics of a loaded cavity (they may generate mode mixing) \cite{Ballantini2005}. We have recently revisited these ideas and found them very promising for future searches of UHFGWs \cite{Berlin2022,Berlin2023} (see also \cite{Domcke2022,Bringmann2023}). There are still many opportunities for synergies between cutting-edge precision devices and UHFGWs, and we hope that this contribution triggers new ideas in this direction.

\subsection{Summary}
\label{sec:synergysummary}

The synergies between 
observations using atom interferometers and laser interferometers
depend on the complementarity between the frequency ranges they 
can cover. Atom interferometers are potentially more sensitive in
the mid-frequency band between terrestrial and space-borne laser
interferometers, but realizing this potential gain depends on the 
extent to which atom interferometers can overcome Gravity Gradient
Noise (GGN). Techniques
for mitigating GGN in atom interferometers were discussed in
Section~\ref{sec:Battling}, and the resulting synergistic tests
of gravity and probes of supermassive astrophysical black holes were
discussed in Section~\ref{sec:SMBHs}, while Section~\ref{sec:NewPhysics}
discussed possible synergies in searches for new physics. As examples
of the possible synergies, Fig.~\ref{fig:IMBH} illustrates the
potential sensitivity of a 1-km atom interferometer to the mergers of
intermediate-mass black holes that would have played key roles in the
assembly of supermassive black holes, Fig.~\ref{fig:IMBHDM} illustrates
the complementarity of atom interferometer searches for ultralight dark
matter to current constraints from measurements of black hole spins,
and Fig.~\ref{fig:ECO} illustrates the sensitivities of large atom
interferometers to exotic compact objects.

\section{Technology Overview} 
%($\sim 5$p)
\label{sec:tech}
%\subsection{Introduction (Jason Hogan, Wolf von Klitzing, Wolfgang Schleich) ($\sim 0.5$p)
\label{sec:tech-intro}

% The gains  achievable with TVLBAI are enormous.  However, reaching the sensitivity required, e.g.\,for gravitational wave detection, will require a number of technological advances, chiefly associated with the 

In this Section we summarize the status of the core atomic physics technologies needed to reach the target sensitivities for TVLBAI science.
As described below, reaching the sensitivity required for gravitational wave detection will require a number of technological advances, chiefly associated with atom optics and the reduction of the noise associated with atom detection.
Table~\ref{table:technologies} lists the current state-of-the-art for key performance metrics, along with the targets needed in each area to move towards a full-scale terrestrial gravitational wave detector.
The technology development path will involve improving the pulse efficiency of large momentum transfer (LMT) atom optics, developing atom sources with increased flux at low temperatures, and integrating spin squeezing to reduce atom shot noise.

The first part of this section (Sec. \ref{sec:tech-LargeM}) describes the use of LMT atom optics using additional laser pulses to enhance the beam separation and thus the sensitivity of atom interferometers.
Clock atom interferometry, in particular, is highlighted as a technique that takes advantage of narrow transitions commonly used in atomic clocks.
The use of these single photon transitions for atom optics enables improved common-mode suppression of laser frequency noise over long baselines and supports high pulse efficiencies, making it valuable for gravitational wave detection and dark matter searches.

The second part of this section (Sec. \ref{sec:tech-Diff}) focuses on LMT atom interferometers based on Bragg diffraction and Bloch oscillations.
It emphasizes the importance of understanding and controlling diffraction phases and inefficiencies in these processes, and references the recent developments in theoretically modelling and characterizing these effects in order to suppress them.
The final part of the section addresses the atom source technologies relevant to gravitational wave detectors based on atom interferometry.
It discusses the increased atomic flux requirements needed to reach the targeted strain sensitivity, and how this compares to the current state of the art.
Additionally, the control of external degrees of freedom of the atomic ensemble is highlighted as crucial for minimizing statistical and systematic errors.

\begin{table}[h]
\centering
\footnotesize
\caption{State-of-the art performance of key sensor technologies and their improvement targets for a full-scale terrestrial detector. The sensitivity enhancement is stated relative to current instruments.}
\vspace{3mm}
\label{table:technologies}
\begin{tabular}{lccccccc} 
 \toprule
   \multirow{2}{*}{Sensor Technology} & \multirow{2}{*}{State-of-the-art} & \multirow{2}{*}{Target} & \multirow{2}{*}{Enhancement}  \tabularnewline
  &   &  & \tabularnewline
  \midrule
 LMT atom optics & $10^2\, \hbar k$ & $10^4\, \hbar k$ & 100 \tabularnewline
 \quad \textit{Matter-wave lensing} & 50\,pK & 5\,pK & -- \tabularnewline
 \quad \textit{Laser Power} & 10\,W & 100\,W & -- \tabularnewline
 Spin squeezing & 20\,dB (Rb), 0\,dB (Sr) & 20\,dB (Sr) & 10 \tabularnewline
 Atom flux & $10^5$ atoms/s (Rb) & $10^7$ atoms/s (Sr)  & 10 \tabularnewline
 Baseline length & 10\,m & 1000\,m & 100 \tabularnewline
 \bottomrule
\end{tabular}
\end{table}

\subsection{Large Momentum Transfer Clock Atom Interferometry\label{sec:tech-LargeM}}

Atom interferometry makes use of the wave-like properties of matter that become evident at very low energy scales~\cite{Berman1997}. The concept is analogous to an optical interferometer that splits a coherent source of light into separate beams following different paths, which are then recombined and undergo interference. The interferometric measurement of the path length difference is the basis of current gravitational wave detectors such as Advanced LIGO~\cite{Aasi2015} and Virgo. Similarly, the wave-function of an atom can be split into a superposition of two states evolving along different paths. This is accomplished through absorption and stimulated emission of photons and the associated momentum transfers (recoils) between light and atoms~\cite{Kasevich1991}. The sensitivity of an atom interferometer can be enhanced by additional laser pulses, a technique called large momentum transfer (LMT) atom optics~\cite{McGuirk2000}. 
In a gravitational wave detector configuration, two clouds of atoms separated by a baseline are interrogated by common laser pulses. The differential phase signal measures the light travel time across the baseline and each additional pair of pulses adds another measurement of the baseline length.
This is analogous to the enhancement provided by Fabry-Perot cavities in Advanced LIGO and Virgo, where the laser beams in the interferometer arms are retro-reflected many times to enhance the signal and create a much longer effective baseline. Increasing the number of laser pulses in the atom interferometer through LMT atom optics is a key area of technology development on a par with increasing the physical detector baseline by building a larger instrument (see Table~\ref{table:technologies}).

\begin{figure}
\centering
\includegraphics[width=0.8\columnwidth]{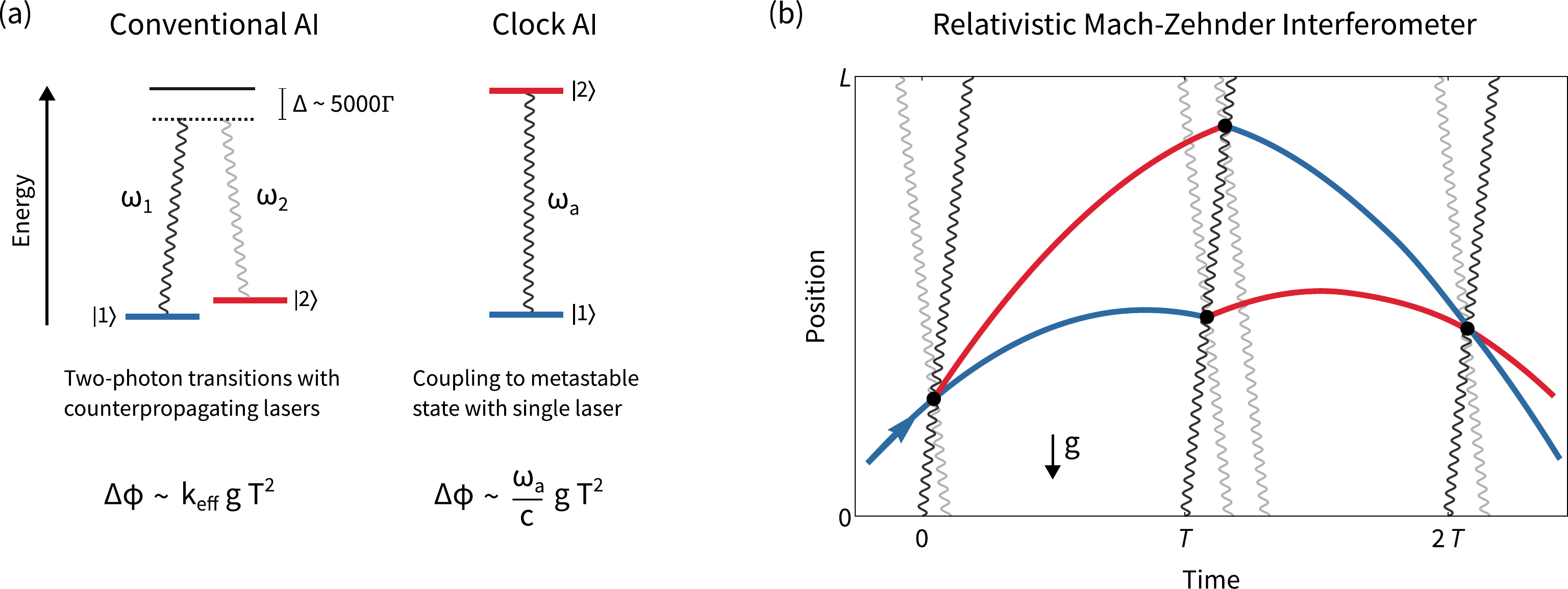}
\caption{\label{fig:clockai} (a) Comparison of the laser frequencies involved in conventional and clock atom optics as well as the leading order phase response of the associated interferometer. (b) Space-time diagram of a relativistic Mach-Zehnder interferometer using clock atom optics (dark lines) and conventional two-photon atom optics (dark and light lines). In a clock atom interferometer, the same laser pulse addresses the entire atomic superposition, imprinting the same laser phase and allowing for common-mode noise suppression.}
\end{figure}

It is desirable for the two atomic states used in an atom interferometer to be long-lived to minimize spontaneous decay, which diminishes the output signal. Conventionally, a set of counter-propagating laser beams is used to couple either two electronic ground states (Raman atom optics~\cite{Kasevich1991}) or two momentum states of the same electronic ground state (Bragg atom optics~\cite{Giltner1995}) via a far-detuned excited state. However, one can also resonantly drive transitions between an electronic ground state and a metastable state using a single laser beam~\cite{Graham2013,Hu2017}. Such a clock atom interferometer makes use of the narrow transitions commonly used in atomic clocks~\cite{Ushijima2015,Bloom2014,Marti2018}. Here, the phase response is proportional to the transition frequency $\omega_a$ instead of the effective wave vector $k_{\textbf{eff}}$ (frequency difference) of two lasers (see Figure \ref{fig:clockai}). Thus, clock atom interferometers support common-mode suppression of laser frequency noise, a dominant noise source for long-baseline interferometers~\cite{Graham2013}. Due to the long lifetime of the metastable state, spontaneous decay can be substantially suppressed compared to two-photon atom optics on a broad optical transition involving a far-detuned excited state. Therefore, clock atom interferometers can support much higher pulse efficiencies and many thousands of sequential laser pulses. They enable the substantial sensitivity enhancement that is crucial for reaching the ambitious sensitivity targets for gravitational wave detection and dark matter searches with very-long-baseline sensors~\cite{Abe2021}.

\begin{figure}
\centering
\includegraphics[width=0.5\columnwidth]{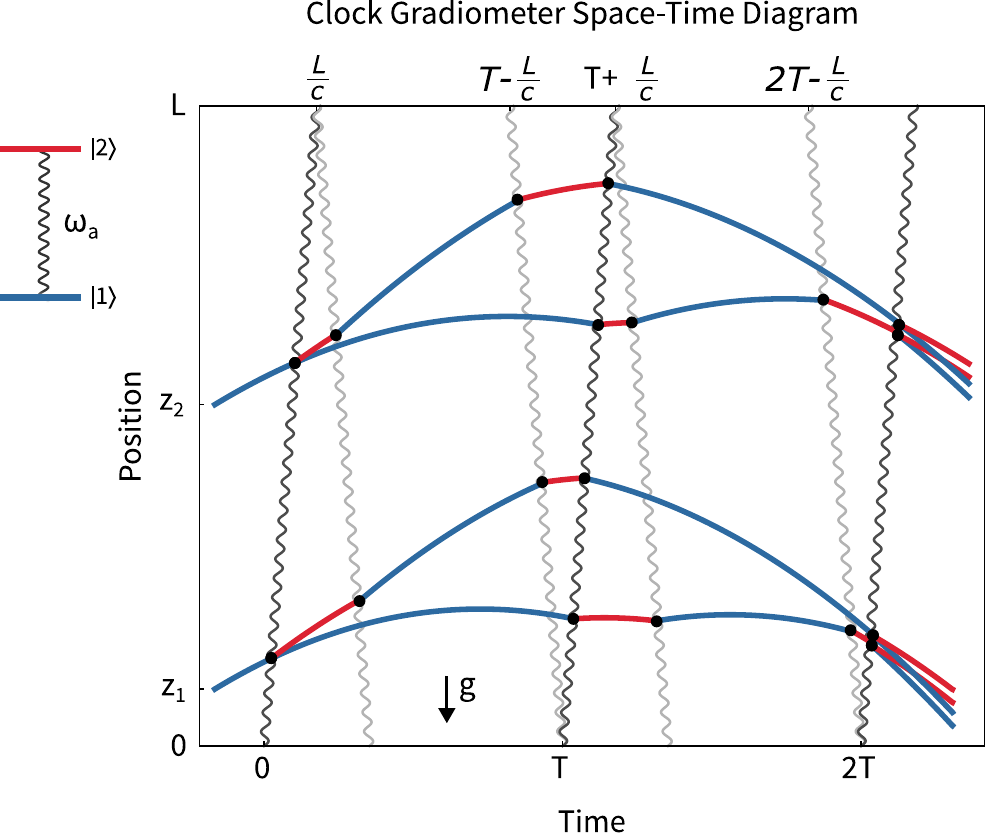}
    \caption{\label{fig:clockgradiometer} Space-time diagram of the interferometer trajectories based on single-photon transitions between ground (blue) and excited (red) states driven by laser pulses from both directions (dark and light gray). The pulse sequence shown here features an additional series of pulses (light gray) traveling in the opposite direction to illustrate the implementation of LMT atom optics (here $n=2$).}
\end{figure}

Figure \ref{fig:clockgradiometer} illustrates a sequence of light pulses generating a pair of clock atom interferometers, one on each end of the baseline. The timing of the atomic transitions, and thus the time the atoms spend in a superposition of the ground and excited states, depends on the light travel time across the baseline~\cite{Graham2013}. The resulting differential interferometer phase $\Delta\phi$ is then proportional to the baseline length $L$:
\begin{equation*}
    \Delta\phi \,\propto\,n\;\omega_a \, L/c,
\end{equation*}
where $\hbar\, \omega_a$ is the energy splitting of the clock transition, and $n$ the order of LMT enhancement. As a result, the differential phase measurement between the two atom interferometers is sensitive to variations in both the baseline $L$ and the clock frequency $\omega_a$ that arise during the light-pulse sequence. A passing gravitational wave modulates the baseline length, while coupling to an ultralight dark matter field can cause a modulation in the clock frequency. Thus, a differential clock atom interferometer combines the prospects for both gravitational wave detection and dark matter searches into a single detector design, and both science signals are measured concurrently~\cite{Abe2021}. In both cases, additional laser pulses linearly enhance the output signal and thus the sensitivity of the detector.

Proof-of-principle experiments with LMT clock atom interferometry have demonstrated enhanced interferometer and gradiometer sequences~\cite{Rudolph2020}, and a record momentum separation of over $600\,\hbar k$ using 1200 sequential laser pulses~\cite{Wilkason2022,Wilkason2022a}. These results have been achieved on an intermediately narrow transition with moderate lifetime and strong Rabi coupling. Ultra-narrow clock transitions can support even higher pulse efficiencies but necessitate longer pulse durations and colder atoms. Quantum degenerate ensembles and matter-wave lensing techniques are required to make full use of these efficiency gains. Ultimately, the LMT enhancement in any terrestrial clock atom interferometer is limited by the available laser power at the target wavelength and the free fall time of the atom i.e.\ the size of the instrument.

\subsection{Diffraction phases and losses in large-momentum-transfer atom interferometers} 
%(Klemens Hammerer)($\sim 1.5$p)
\label{sec:tech-Diff}

Large momentum transfer (LMT) atom interferometers take advantage of the improved scaling of their sensitivity with the momentum separation of the coherent superposition of matter waves. In Very-Long-Baseline Atom Interferometry, momentum transfers of hundreds and thousands of photon recoils will be realised to achieve the sensitivity needed to detect gravitational waves or test physics beyond the Standard Model. For LMT interferometer realisations, a number of mechanisms are currently being investigated, based on one- or two-photon processes, the latter of which can induce inelastic Raman transitions or elastic scattering processes. The highest metrological sensitivity is currently achieved by means of elastic transitions. This is accomplished in the form of (i) Bragg diffraction or (ii) Bloch oscillations of atoms in optical lattices. In both processes, a very high efficiency per transmitted photon pulse $\hbar k$ can be achieved. Bragg diffraction and Bloch oscillations underlie in particular the atom interferometric measurements of the fine structure constant \cite{Parker2018,Morel2020} and are also envisaged for the architecture of gravitational wave detectors \cite{Canuel2020,Canuel2020a}. Limitations in atomic interferometers based on these processes are currently determined by systematic effects due to so-called diffraction phases associated with atom-light interactions, as well as losses and inefficiencies of the beam splitters and mirrors. A key to future applications of atom interferometers with large scale factors, requiring the transmission of hundreds or thousands of photon pulses, will therefore be a detailed understanding and accurate control of diffraction phases and inefficiencies of elastic scattering processes. To this end, important achievements have been made recently~\cite{Siemss2020,Siemss2022,Fitzek2023,Rahman2023}, providing a promising basis for the further development of Very-Long-Baseline Atom Interferometry.

With respect to Bragg diffraction, a model for efficient beam splitting and mirror pulses was developed in \cite{Siemss2020} that covers the so-called quasi-Bragg regime. This regime describes pulses whose duration and intensity lie between the parameters of the deep-Bragg regime of very long, weak pulses and the Raman-Nath regime of very short, intense pulses that have been treated in the literature to date. The quasi-Bragg regime is characterised by a compromise between suppression of the velocity filter due to Doppler detunings and suppression of scattering to undesirable momentum orders. In \cite{Siemss2020}, it was shown that Bragg pulses with Gaussian intensity modulations achieve efficient beam splitters and mirrors exactly when the underlying quantum dynamics corresponds to an adiabatic process in the Bloch states of the optical lattice. With this insight, a number of important properties of Bragg operations can be quantitatively and, in some cases, even analytically characterised in a simple way. For example, the pulse area condition for beam splitters or mirror pulses can be formulated as a condition on energetic phases of Bloch states, and losses in parasitic scattering orders can be understood as non-adiabatic Landau-Zener transitions. All this was combined in \cite{Siemss2020} in the form of a scattering matrix that describes individual beam splitter or mirror operations with high accuracy, accounting correctly for losses and phases.

Based on this, the description of complete interferometer sequences was developed in \cite{Siemss2022}. Using the example of a Mach-Zehnder interferometer, the effects of Landau-Zener transitions into parasitic scattering orders were treated in detail. In a measurement of the relative atom number in the two main output channels of the final beam splitter, two essential effects come to light: First, parasitic output channels are inevitably also populated in a phase-dependent manner in the final beam splitter, so that the total number of atoms in the main output channels itself becomes phase-dependent. If the absolute number of atoms measured in an output channel is now related to this phase-dependent total number, a subtle and complicated phase dependence of the interferometer results. Secondly, this is further complicated because populations of parasitic momentum states in the first beam splitter can lead to further, parasitic paths closing in the last beam splitter in addition to the two main paths of the interferometer. This leads to intricate phase dependencies due to additional Mach-Zehnder and Ramsey-Borde interferometers. For generic Bragg pulses, this results in an overall interferometer signal that formally corresponds to an infinite Fourier series in the metrological phase (and its harmonics). Even though the description developed in \cite{Siemss2020,Siemss2022} correctly predicts the amplitudes and phases of this Fourier series in principle, this model cannot serve as a signal template due to its complexity. However, the detailed microscopic model from \cite{Siemss2020} allows the Bragg pulses to be designed in an optimal way so that parasitic interferometer paths are strongly suppressed, which greatly simplifies the signal shape. In \cite{Siemss2022}, it was shown that such suppression can achieve accuracy in the micro-radian range. Furthermore, with the analytical modelling of the transfer matrix of a Bragg interferometer, it was also possible to determine the fundamental bounds on the accuracy of such an interferometer in the form of the Cramer-Rao and the quantum Cramer-Rao bounds and thus to determine parameter ranges of optimal accuracy.

Remarkably, an analogous description is possible for LMT operations based on Bloch oscillations. In the literature, Bloch oscillations have long been considered as adiabatic processes in Bloch states of the optical lattice. Based on this view, losses were explained as Landau-Zener transitions at the avoided crossings at the edge of the Brillouin zone. However, such a description breaks down in the regime of LMT operations, which require very deep lattices in the range of tens of recoil energies and very fast accelerations in the range of hundreds of $\text{m}/\text{s}^2$. In this regime, the Bloch bands are effectively flat and the assumption of avoided crossing becomes untenable. However, it can be shown that in this regime an efficient description can be given in terms of Wannier-Stark eigenstates of the accelerated lattice rather than by Bloch states of the lattice at rest \cite{Fitzek2023}. Bloch oscillations can thus be interpreted as adiabatic dynamics in the Wannier-Stark spectrum, which in turn enables an efficient calculation of diffraction phases and losses. Especially for the losses from the accelerated lattice, this description shows complex dependencies on the lattice parameters due to tunnel resonances, which can thus also be chosen in an optimal way \cite{Fitzek2023,Rahman2023}. Another important consequence that can be drawn from this model concerns the necessary intensity stabilisation between interferometer arms to ensure a desired phase stability. For example, for the ELGAR gravitational wave detector \cite{Canuel2020, Canuel2020a} with sensitivity in the microradian range, a relative intensity stabilisation of $10^{-9}$ must be achieved. 

The above results show that there is considerable room for optimisation of elastic scattering processes for LMT atom interferometers. This concerns the microscopic modelling of the scattering processes as well as the description of full interferometer sequences and their signal shapes. All of these are important and essential steps towards the coming generations of atom interferometers and their applications in the field of gravitational wave astronomy and the search for new physics.

\subsection{Atom source technologies}
%(Christian Schubert)($\sim 1.5$p)
\label{sec:tech-source}

Proposals for gravitational wave detectors utilising atom interferometry on Earth target strain sensitivities in the range of $10^{-21}\,/\sqrt{\mathrm{Hz}}$ to $10^{-23}\,/\sqrt{\mathrm{Hz}}$~\cite{Dimopoulos2008a, Chaibi2016, Schubert2019, Canuel2020, Abe2021}.
They assume large momentum transfer processes for beam splitting, large baselines separating the atom interferometers, and a low phase noise down to  $1\,\mu\mathrm{rad}/\sqrt{\mathrm{Hz}}$.
Considering an ideal contrast of the interference signal, this corresponds to the quantum projection noise limit for a flux of $10^{12}\,\mathrm{atoms/s}$.
In principle, squeezing could partially alleviate the requirement on the atomic flux.
A derived requirement is the control of the external degrees of freedom of the atomic ensemble to minimise statistical and systematic errors, as well as assuring high beam splitting efficiency by preparing atom ensembles with very low residual expansion rates.
Finally, the cycle time affects the detectors bandwidth and has to be tuned accordingly~\cite{Dimopoulos2008a,Debs2011,LouchetChauvet2011,Szigeti2012,Loriani2019,Loriani2020,Canuel2020,Abe2021,Gebbe2021}.
	
Common choices for atom interferometers are alkali atoms, notably Rb and Cs~\cite{Gauguet2009,Stockton2011,Hu2013,Berg2015,Biedermann2015,Hamilton2015,Chiow2016,Freier2015,Gillot2016,Bidel2018,Janvier2022, Gautier2022}.
In the context of gravitational wave detection with atoms, Sr offers beneficial features, especially the possibility to drive a single-photon transition for frequency noise suppression~\cite{Graham2013,Yu2009}.
We now recall briefly published results about the creation of cold atomic ensembles, without claiming completeness.  
	
Usually, the preparation of the atoms starts with capturing and cooling in a 3D magneto-optical trap (MOT) (e.g. at 780\,nm for Rb) which may be loaded by a 2D magneto-optical trap or a Zeeman slower, and subsequent molasses cooling.	
Current state-of-the-art experiments are capable of providing molasses-cooled $^{87}$Rb ensembles with \SI{1e9}\,\si{atoms\per\second} and temperatures down to \SI{2}{\micro\kelvin} \cite{Beaufils2022,Rudolph2015}.

Further reduction of expansion rates is possible by evaporative cooling in magnetic or optical dipole traps (e.g., at 1064\,nm) down to the generation of Bose-Einstein condensates (BEC) and by additionally applying a delta-kick collimation (DKC) step\,\cite{Ammann1997}.
Evaporative cooling implies losses, especially for fast cycles.
An experiment utilising a cloverleaf style Ioffe-Pritchard trap generated BECs with $10^7$ $^{87}$Rb atoms in 1\,min~\cite{Streed2006}.
Using a time-orbiting potential with DKC via a pulsed magnetic trap potential enabled the production of $4\times10^6$ $^{87}$Rb atoms at an effective temperature of 50\,nK or $10^5$ $^{87}$Rb atoms at an effective temperature of 3\,nK~\cite{Dickerson2013}.
Rapid $^{87}$Rb BEC production was demonstrated with an atom chip, realising $^{87}$Rb BECs with $10^5$\,atoms ($4\times10^4$, $4\times10^5$) in 1\,s (850\,ms, 1.6\,s)~\cite{Rudolph2015} and subsequent DKC to an effective temperature of \SI{38}{pK}~\cite{Deppner2021}.
Control over the mean position and velocity has been shown on the \SI{<100}{\nano\metre} and \SI{100}{\micro\metre\per\second} scale using the Cold Atom Lab setup aboard the International Space Station, reaching effective temperatures after delta-kick collimation of \SI{100}{\pico\kelvin}~\cite{Corgier2018, Gaaloul2022}.
	
Optical dipole traps enabled the generation of $^{87}$Rb BECs with $2\times10^6$\,atoms at 50\,nK in about 10\,s~\cite{Hardman2016}.
Utilising painted potentials instead of static beam paths lead to $^{174}$Yb BECs with $10^5$\,atoms ($5\times10^4$, $1.2\times10^6$) in 1.8\,s (1.6\,s, 15\,s)~\cite{Roy2016}.
Moreover, evaporation with optical dipole traps in microgravity was demonstrated~\cite{Condon2019, Vogt2020} with  $4\times 10^4$ $^{87}$Rb atoms at 35\,nK in 13.5\,s~\cite{Condon2019}.
Analogous to magnetic trap potentials, optical dipole trap potentials can be exploited for delta-kick collimation.
Here, an experiment demonstrated the reduction to an effective temperature of 50\,pK~\cite{Kovachy2015}.
The combination of prematurely stopped evaporation followed by rapid decompression was studied to increase the effective atomic flux, leading to $4\times 10^5$ $^{87}$Rb atoms at an effective temperature in 2D of 3.2(0.6)\,nK~\cite{Albers2022}.
Furthermore, low-field Feshbach resonances in an optical dipole trap can be exploited to tune the rethermalisation rate, enabling the rapid production of $^{39}$K BECs with $5.8\times 10^4$\,atoms after \SI{850}{\milli\second}\,\cite{Herbst2022}.
	
Laser cooling of Sr isotopes typically relies on a two-color or dual stage magneto-optical trap at 461\,nm and 689\,nm, loaded by a Zeeman slower, leading to $10^7$ $^{88}$Sr atoms at \SIrange{1}{3}{\micro\kelvin} or down to \SI{400}{\nano\kelvin} after transfer to a red single frequency MOT at 689\,nm~\cite{Hu2019,Poli2014,Rudolph2020,Stellmer2014}.
Similarly, the red frequency MOT enabled the cooling of $10^7$ $^{84}$Sr atoms or $10^7$ $^{87}$Sr atoms to \SI{800}{\nano\kelvin}~\cite{Stellmer2014}.
	
Using a crossed optical dipole trap, $^{84}$Sr BECs were produced with $10^5$ atoms in 2\,s or $10^7$ atoms for 10\,s of evaporation after accumulating atoms for 40\,s to compensate for the low natural abundance~\cite{Stellmer2014}.
BECs of $^{86}$Sr or $^{86}$Sr featured $10^4$ atoms.
Implementing a continuous laser cooling of atoms in a reservoir dipole trap, a transparency beam rendering the atoms transparent for the cooling light, and a small dimple dipole trap led to the creation of $^{84}$Sr BECs with $10^5$ \textit{in parallel} to the laser cooling~\cite{Stellmer2014}.
This procedure was extended to the continuous creation of $^{84}$Sr BECs by continuously loading the reservoir, demonstrating an average BEC atom number of $7.4(2.3)\times 10^3$ with an estimated steady-state gain of $2.4(5)\times 10^5$\,atoms/s into the BEC~\cite{Chen2022}.
	
Delta-kick collimating $^{84}$Sr or $^{87}$Sr with a pulsed optical dipole trap potential to effective temperatures of 10\,pK in 3D is in theory possible, but is constrained in the case of the fermionic isotope  by the pre-DKC expansion time and size at lens (1.2\,cm), according to Ref.~\cite{Loriani2019}.
	
Other pathways for cooling atoms beyond the molasses stage are Raman sideband cooling in an optical lattice, leading, e.g., to $10^6$ atoms at \SI{0.4}{\micro\kelvin} in a Cs fountain experiment~\cite{Chiow2009,Estey2015}, or BEC generation by Raman cooling of $^{87}$Rb atoms in an optical dipole trap, demonstrating a 7\,\% condensate fraction of $2.5\times 10^4$ atoms after multiple cooling stages starting with $10^5$ atoms in a dipole trap~\cite{Urvoy2019}.
	
Squeezing may in future enhance the capability of atom interferometers.
Theoretical proposals predicted a noise of 20\,dB beyond the standard quantum limit (SQL) on momentum states for Sr atom interferometry~\cite{Salvi2018}, or up to 30\,dB beyond the SQL for a BEC with $10^6$\,atoms via delta-kick squeezing~\cite{Corgier2021}.
Experimental results showed a noise of 3.7(0.2)\,dB below the quantum projection noise for a Ramsey interferometer with 3.6\,ms Ramsey time and $2.4\times 10^5$\,atoms~\cite{Malia2020}, a squeezing parameter of -1.9(0.7)\,dB (conditional -3.1(0.8)\,dB) between two momentum modes for 9300\,atoms~\cite{Anders2021}, and an entanglement enhancement of 1.7(0.5)\,dB below the SQL in a Mach-Zehnder-like  atom interferometer with 660(17) $^{87}$Rb atoms and 0.7\,ms pulse separation time~\cite{Greve2022}.

\subsection{Summary} \label{sec:tech-concl}
%(Jason Hogan, Wolf von Klitzing, Wolfgang Schleich)($\sim 0.5$p) 
%\todo{Wolf: @Jason @Wolfgang summary of LMT and diffraction phases missing. @John: expert input needed here.} 

Large momentum transfer beam splitters increase the sensitivity of atom interferometers by opening up larger enclosed areas.  
Similarly, it is desirable to use long-lived states to allow longer interrogation times without spontaneous emission losses. 
Momentum transfer of up to $1200 \hbar k$ has been achieved. 
Clock atom interferometers combine long lived-states with highly efficient beam splitters. 
Another method to achieve LMT is via Bragg diffraction and Bloch oscillations.  
The beam splitters, however, can introduce their own phase shifts, which has been modelled in detail. 
From this analysis it has become clear that there is still considerable room for optimization, which will be guided by detailed microscopic modelling.

The signal-to-noise ratio of the atom interferometer will ultimately depend on the number of atoms available per time unit. 
Well-collimated thermal ensembles with  $\sim 10^5$ atoms can currently be generated within 1\,s, whereas larger BECs of up to $10^7$ atoms require a much longer preparation time (1\,min).
Improving the signal using squeezing is promising, but the transfer to large atom numbers, long time scales and large momentum transfer has yet to be demonstrated.
Even though scalability is not obvious, the BEC creation has some potential for optimisation towards larger atom numbers, e.g., increasing the initial atom number \cite{Feng2023}, vacuum quality, transfer efficiency from molasses to initial trap, and adjusting the trapping volume to avoid detrimental density regimes~\cite{Rudolph2015, Stellmer2014}.
Future source setups for atom interferometric gravitational wave detectors will require a trade-off between flux and required control to suppress error contributions, and their scope will be the subject of further dedicated research.

\section{Vertical Long-Baseline Detector Options}
\label{sec:vertical}
%{\bf 5 pages guideline, Editors: Dennis Schlippert, Tim Kovachy, Ulrich Schneider}

\subsection{Introduction}
%(Dennis, Tim, Ulrich): $\sim$0.5 pg}
\label{sec:VerticalLBLdetectoroptions}
We summarize in this Section the technological requirements for long-baseline vertical atom interferometers and discuss associated research and development needs.  We also provide an overview of four ongoing vertical long-baseline atom interferometry projects: the Hannover VLBAI atom interferometer (Section~\ref{Sec:VLBAI}), the Wuhan 10\;m Atom Interferometer (Section~\ref{Sec:Wuhan10m}), the AION Project in the UK (Section~\ref{Sec:AION}), and the MAGIS-100 atom interferometer under construction at Fermi National Accelerator Laboratory (Section~\ref{Sec:MAGIS100}).

Vertical configurations offer both opportunities and challenges.  A central motivation for orienting the atom interferometer vertically is that the trajectories of the freely falling atom clouds are collinear with the atom optics laser, so that the atoms remain nominally centered in the laser beam even for long interrogation times (some deviations from this idealized scenario arise due to deflections of the atom trajectories from Coriolis forces, as discussed in greater detail below).  Many of the applications of long-baseline atom interferometry involve differential measurements between two or more interferometers spaced over the baseline (gradiometer configurations).  Importantly, these differential measurements can suppress the impact of laser phase noise.  For atom optics based on multi-photon transitions using counter-propagating laser beams (e.g., Bragg transitions), this common mode cancellation of laser noise becomes inadequate for long baselines.  To circumvent this issue, horizontal long-baseline detectors can employ two orthogonal arms (see Section~\ref{sec:horizontal}).  Laser noise cancels between the two arms, while scientific signals of interest (e.g., from gravitational waves) are preserved.  By contrast, vertical detectors have only one baseline direction available, requiring suppression of laser noise over a single baseline. This requirement favors single-photon atom optics involving resonant driving of narrow clock transitions \cite{Graham2013, Yu2009}.  In addition to their beneficial features for laser noise suppression, single-photon atom optics offer an opportunity for dramatic improvements in large momentum transfers due to reduced spontaneous emission losses (see Section~\ref{sec:tech-LargeM}).  However, driving the associated weak clock transitions places demanding requirements on laser power.  The development of higher power laser sources is therefore important for maximizing the sensitivity of vertical long-baseline detectors.  Development of high-flux and spin-squeezed atom sources are also key to enhancing sensitivity, as discussed in Section~\ref{sec:tech-intro}. Moreover, scaling noise mitigation techniques and experimental infrastructure to longer baselines is nontrivial and in some cases requires new approaches. Sections \ref{Sec:verticalSensitivity} and \ref{Sec:vertricalTechnology} elaborate further on the challenges of vertical long-baseline atom interferometry and strategies for overcoming them.   

% \textbf{General considerations: vertical VLBAI in sections 8.1 and 8.2 below (Ulrich and Tim) $\sim$1.5 pg total}

% \begin{itemize}
%      \item scaling from 10\,m to 100\,m challenges
%      \item scaling to km baselines
% \end{itemize}

\subsection{Sensitivity scaling}
%(Ulrich)}
\label{Sec:verticalSensitivity}

The sensitivity of an atom interferometer can be divided conceptually into two parts.
The \textit{intrinsic sensitivity} governs the transposition from the targeted physical effect into a phase shift $\varphi$, for instance the conversion from characteristic strain of a gravitational wave at a particular frequency into the measurable $\varphi$ for a given interferometer configuration.
On the other hand, the \textit{readout sensitivity} of the detector is the minimum phase shift $\varphi$ accrued in the interferometer that can be read out. This is typically given in terms of the phase noise $\Delta\varphi$.

For both the search for ultralight dark matter and for gravitational waves, the relevant intrinsic sensitivities of an atomic gradiometer scale as $(C n \Delta r)^{-1}$~\cite{Badurina2021,Badurina2022}, where $C$ denotes the interferometer contrast, $n$ the order of LMT, and $\Delta r$ the separation between the interferometers. The scaling with $\Delta r$ drives the need to move to the km scale and the two remaining terms set the development goals for the laser and LMT development.

The readout sensitivity is limited by the atomic shot noise (quantum projection noise) and scales as $N_{\rm atom}^{-1/2}$. As discussed in Section~\ref{sec:tech-source}, this sets the development goals to increase the atomic flux by  a factor of 100 and to develop 20~dB squeezing for strontium. The underlying assumption is that all other noise sources are smaller than these strict requirements, driving limits on magnetic field noise, control of laser wavefront aberrations, camera stability and others.
In the following Sections we will look at those requirements and associated strategies that are specific to vertical long-baseline detectors in more detail.

\subsection{Technological challenges and methods }
%(Ulrich and Tim)}
\label{Sec:vertricalTechnology}

\subsubsection{Possible limits for LMT and laser power requirements}
%(Ulrich)}
The achievable order $n$ of LMT will typically be limited by the stringent pulse fidelities required to retain high contrasts $C$ at large $n$. As discussed in Section~\ref{sec:tech-LargeM}, the need to avoid spontaneous emission during long LMT pulse sequences mandates the use of very narrow transitions, such as the clock transition at $698\,$nm in fermionic $^{87}$Sr. For these transitions, high laser intensities are required to reach the high Rabi frequencies needed to drive several $10^4$ $\pi$-pulses within a sequence duration that is ${\cal O}(1)$~s. At the same time, the need to maintain high contrast $C$ requires the use of large beam waists to ensure homogeneous Rabi frequencies over the cloud, and this combination ultimately requires high laser powers for the narrow-linewidth interferometer lasers that drive the need for dedicated laser R\&D. 
The use of these narrow transitions furthermore sets stringent requirements on the cloud temperature, as the pulse fidelity is sensitive to Doppler shifts.

% \subsubsection{Improving cold atom flux (Ulrich)}
% Already discussed in sec 7

\subsubsection{Mitigating noise associated with source kinematics}
%(Tim)}
\label{Sec:MitigatingKinematics}

As atom interferometers have become more sensitive, it has become increasingly important to mitigate noise and systematic errors associated with the initial kinematics of the atom cloud.  One source of error arises from the coupling of cloud kinematics to gravity gradients.  For example, in the presence of a vertical gravity gradient, the gravitational acceleration experienced by an atom will vary depending on its vertical position and is thereby affected by the initial position and velocity of the atom.  Fluctuations or uncertainties in atom cloud kinematics will thus lead to errors in the interferometer phase shift.  A gravity gradient compensation scheme can be introduced to cancel these errors.  In a three-zone Mach-Zehnder interferometer (initial beam splitter sequence, mirror sequence, final beam splitter sequence), the laser frequency is shifted for the mirror sequence.  The frequency shift is chosen to produce a phase shift with a dependence on initial atom position and velocity that is equal and opposite to the corresponding kinematic-dependent phase shift arising from the gravity gradient, nulling the response of the total interferometer phase shift to initial atom kinematics \cite{Roura2017}. This scheme has been successfully demonstrated in multiple apparatuses \cite{Overstreet2018,DAmico2017}.

\begin{figure}[htp]
\centering
\includegraphics[width= \textwidth]{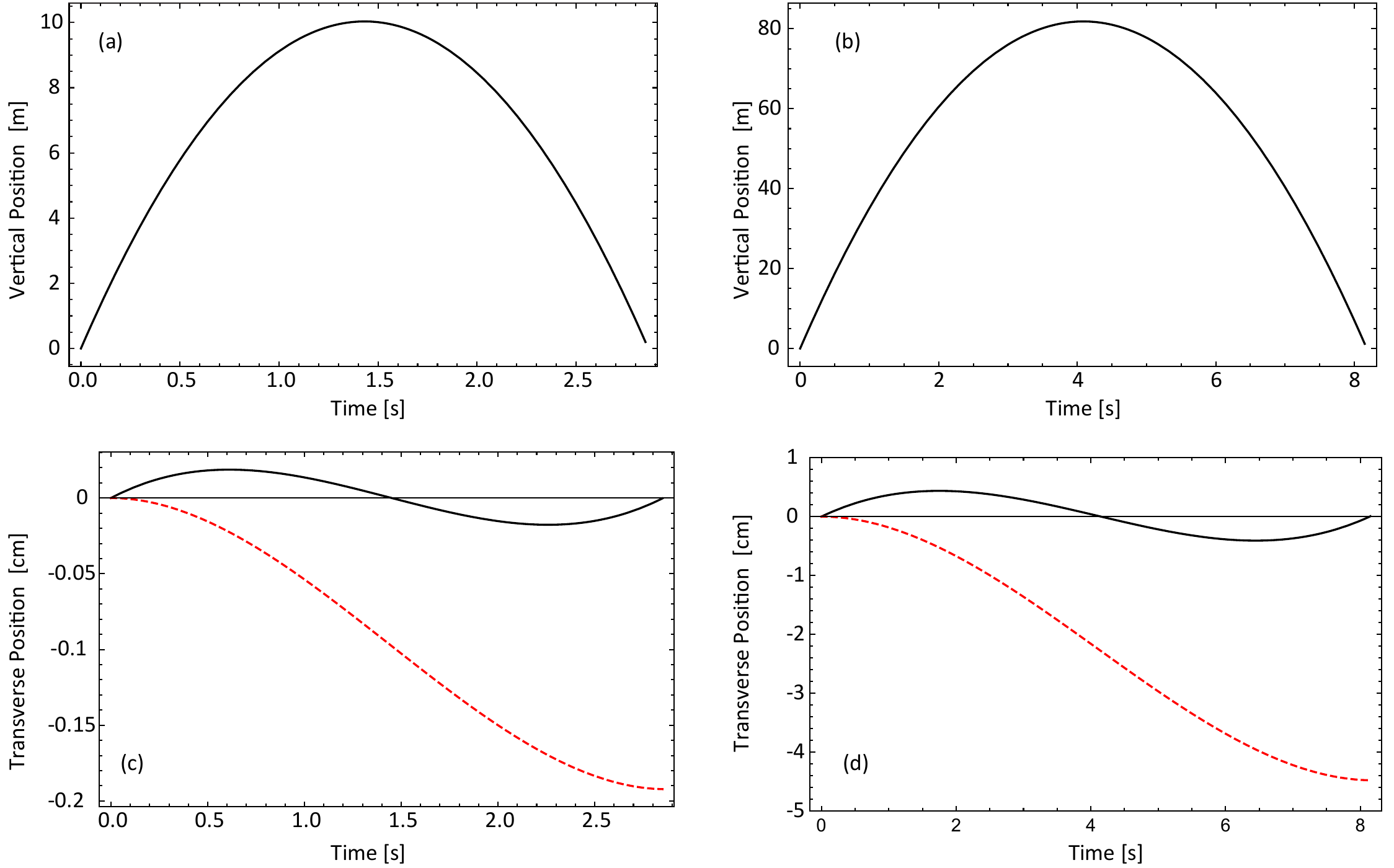}
\caption{Transverse deflections of atomic trajectories due to Coriolis forces.  (a) and (b) show trajectories in the vertical dimension for 10\;m and 80\;m launch heights, respectively. (c) Transverse trajectory deflections for a 10\;m launch height.  Dashed red curve: Purely vertical launch.  Solid black curve: Launch angle adjusted by $5 \times 10^{-5}$\;rad to minimize transverse deflections.  (d) Transverse trajectory deflections for a 80\;m launch height.  Dashed red curve: Purely vertical launch.  Solid black curve: Launch angle adjusted by $1.4 \times 10^{-4}$\;rad to minimize transverse deflections.}
\label{fig:CoriolisDeflections}
\end{figure}

There are several challenges in applying gravity gradient compensation to long vertical baseline detectors.  First, for the multi-second free fall times available in these detectors, the required frequency shift will be on the order of hundreds of MHz to GHz.  For Bragg transitions using a broad atomic transition, this frequency shift will often be small compared to the detuning of the lasers from the excited state, so that the effect of the frequency shift on the effective Rabi frequency will be modest.  However, frequency shifts of this scale are much larger than the widths of the narrow transitions used for single-photon atom optics.  Therefore, it is currently not clear that gravity gradient compensation can be applied to single-photon atom optics.  An additional challenge for long baseline gradiometers is that the gravity gradient will tend to vary over the baseline, so that needed frequency shifts for interferometers on either end of the baseline will be different.  It may be possible in some cases to adjust the local gravity gradient with local masses, but the corresponding infrastructure requirements would be nontrivial.  An alternative approach which circumvents these challenges is to suppress gravity gradient effects using multi-loop atom interferometers \cite{Dubetsky2006, Abe2021}. It should be noted that the sensitivity of multi-loop interferometers is reduced for signals at frequencies below the looping frequency (which corresponds to the inverse of the time between successive loops) \cite{Graham2016}.

Coriolis forces arising from the coupling of the atom velocity to Earth's rotation are another source of error that must be mitigated.  In apparatuses up to the 10\;m scale, rotation compensation schemes have been successfully applied by counter-rotating the angle of the atom optics laser beams against the Earth's rotation using tip-tilt mirrors \cite{Dickerson2013,Lan2012}.  Scaling this approach to 100\;m baselines and beyond offers new challenges.  For baseline length $L$ and laser beam rotation angle $\theta_r$, rotation compensation will cause a transverse beam deflection of $\sim L \theta_r$ over the baseline.  For a 10\;m baseline and an interferometer interrogation time of several seconds, this deflection will be on the mm scale, which is small compared to typical cm-scale laser beam widths \cite{Dickerson2013}. However, beam deflections grow to the cm scale for 100\;m baselines.  In addition to laser beam deflections arising from rotation compensation, transverse Coriolis deflections of the vertically launched atomic trajectories are also important to consider (see Fig. \ref{fig:CoriolisDeflections}). These deflections can be minimized, but not eliminated, by launching the atoms at a slight angle from vertical so that the launched atoms have a small transverse initial velocity (a capability that is foreseen for the MAGIS-100 experiment)~\cite{Abe2021}.  For example, for an 80\;m tall launch (which could be used, e.g., for dual species measurements with Bragg transitions), residual atom trajectory deflections will reach approximately 4\;mm. 

For these reasons, new approaches are needed for keeping the atom cloud centered in the atom optics laser beam for baselines of 100\;m and longer.  MAGIS-100, for example, will implement a laser beam pointing control scheme with two tip-tilt mirrors -- one at the top of the baseline to direct the laser beam downward, and a second at the bottom of the baseline to retro-reflect the beam \cite{Abe2021}.  The upper tip-tilt mirror will be placed before a magnifying telescope.  The location of the pivot point around which the laser beam rotates can be adjusted by tuning the distance between this tip-tilt mirror and the first telescope lens, providing an important degree of freedom for maintaining overlap between the laser beam and the atoms.  For instance, for the example of the tall launch described above, the atoms can be centered in the laser beam for each interaction zone (first beam splitter, mirror, final beam splitter) by placing the pivot point at the height of the beam splitter interactions (with the first and final beam splitters specified to occur at the same location) and setting the initial angle of the laser beam so that the atoms are centered in the beam for the mirror sequence.  As a further tool, MAGIS-100 will also include mirrors on precision translation stages (before the telescope) to enable dynamical adjustment of the lateral position of the laser beam.

It remains an open question whether rotation compensation will be practical for future km-scale detectors.  Beam deflections would grow to the 10-cm scale, requiring a larger diameter vacuum pipe and optics.  Multi-loop interferometers offer an alternative means to mitigate Coriolis phase shifts \cite{Dubetsky2006,Abe2021} and -- as discussed above -- may be required anyway to mitigate gravity gradients for interferometers using single-photon atom optics.

\subsubsection{Tall launch/drop opportunities and challenges}
%(Tim)}

Long-baseline vertical detectors offer an opportunity for tall atom launch or drop heights.  The resulting increase in free-fall time can in many applications boost sensitivity \cite{Dickerson2013}.  To take full advantage of this capability, the following challenges must be overcome.  As mentioned in Section~\ref{Sec:MitigatingKinematics}, taller launch or drop heights lead to increased transverse deflections of the atomic trajectories by Coriolis forces, which can be mitigated to some extent by launching the atoms at a slight angle.  Also, higher laser power is needed for taller launches.  Finally, if only one interferometer is operated at a time, the longer interrogation times afforded by tall launches or drops will cause a corresponding reduction in sensor bandwidth.  Multiplexed interferometer sequences could allow the lost bandwidth to be recovered (see Section~\ref{Sec:Multiplexing} for further details).

\subsubsection{Large-scale magnetic shielding}
%(Tim)}

Magnetic shielding over a long baseline poses fabrication difficulties.  For magnetic shields with large length-to-diameter ratios, small gaps between segments in a magnetic shield can lead to large magnetic flux leakage into the nominally shielded region. While a single-piece, welded magnetic shield has been realized at the 10\;m scale to overcome this problem \cite{Dickerson2012}, this approach is not scalable to longer baselines.  A modular, multi-piece design that minimizes flux leakage is therefore required.  Fortunately, such a design has already been developed and demonstrated for the Hannover VLBAI interferometer \cite{Wodey2020} (see Section~\ref{Sec:VLBAI}).  The MAGIS-100 magnetic shield design \cite{Abe2021} was adapted from the Hannover design.

\subsubsection{Impact of gravity gradient noise \& mitigation}
%(Tim)}

For terrestrial long-baseline atom interferometers, Newtonian gravity gradient noise (GGN) is expected to limit sensitivity to dark matter and gravitational waves for sub-Hz frequencies.  GGN can arise both from seismic and atmospheric effects.  For vertical detectors, preliminary studies have investigated the use of multigradiometer configurations (an array of multiple atom interferometers along the baseline) to distinguish the exponential decay of Rayleigh-wave-induced seismic GGN from the linear dependence of a gravitational wave or dark matter signal with depth \cite{Mitchell2022,Badurina2023a}.  Further investigations involving more realistic geophysical models (e.g., density layers of the Earth) are needed. Detailed studies of the impact of atmospheric GGN on long-baseline vertical detectors are also needed.  GGN and associated mitigation strategies are discussed in furter detail in Section~\ref{sec:Battling}.

\subsubsection{Vertical shaft infrastructure, access, installation}
%(Tim)}

Long-baseline vertical detectors require a significant amount of infrastructure. Electrical power for atom sources, vacuum pumps, and other equipment along the length of the baseline must be provided.  The atom sources will need to be accessed during installation and for periodic maintenance, requiring a system for safely lowering personnel along the baseline and platforms from which personnel can work.  Also, installation will require a large crane or an equivalent solution.  The approach MAGIS-100 takes to address these challenges is detailed in Section~\ref{Sec:MAGIS100}, and this issue was also addressed for a vertical 100-m device in an LHC access shaft at CERN~\cite{Arduini2023}, see Section~\ref{sec:CERN}. 

\subsubsection{Multiplexing}
%(Ulrich)}
\label{Sec:Multiplexing}
One advantage of the purely collinear geometry of vertical detectors is the possibility of multiplexing interferometer sequences. Here, the narrow linewidth of the clock transition and the associated sensitivity to Doppler shifts is beneficial, as it means that the shifts occurring during LMT sequences can be used to address separately clouds in different stages of the interferometer sequence by varying the frequency of the interferometer laser.  This enables higher effective repetition rates that can reduce aliasing effects at constant total interrogation time. In addition, velocity-selective launching, where different vertical velocity classes of the same cloud are launched separately can lead to reduced Doppler shifts during a given interferometer sequence and can increase  contrast for the same value of $n$. More modelling is needed to understand the trade-offs between this potentially higher interferometer contrast of velocity-selective schemes that minimize Doppler shifts and the higher effective repetition rate, on the one hand, and the reduced atom number per shot on the other.

\subsubsection{Phase shear readout, Referencing cameras, optical monitoring}
%(Ulrich)}
 
 In phase shear readout~\cite{Sugarbaker2013}, the interferometer phase is mapped on the spatial position of density fringes in space, which are then observed using fluorescence imaging. Assuming a realistic fringe spacing of $d =1000\,\mu$m, a phase shift of $\varphi=10^{-4}\,\pi$ translates into a spatial shift of the fringes of $\delta x=100\,$nm. 
Reliably detecting these spatial shifts translates directly into the required stability criteria for the imaging system, namely a required position stability of $<$~100~nm of the optical axis. 
In a 100~m detector in differential operation, such as AION-100 or MAGIS-100, this corresponds to a short-term stability requirement of the relative positions of the imaging systems of a pair of interferometers that are 50m apart of 100nm or less with respect to each other.

\subsection{The VLBAI Atom Interferometer in Hannover}
%(Dennis) $\sim$1 pg}
\label{Sec:VLBAI}

% \textit{Introduction -- }
The Very Long Baseline Atom Interferometry (VLBAI) facility~\cite{schlippert2020matter} enables ground-based rubidium and ytterbium atom interferometry with large scale factors. Utilizing a \SI{10}{m} free-fall distance, VLBAI has the potential for shot-noise-limited gravimetry with instabilities below \SI{1e-9}{m/s^2} at \SI{1}{s} and offers absolute measurements rivalling state-of-the-art superconducting gravimeters. Operation with ytterbium is key for the development and testing of methods for gravitational wave detection, e.g., with respect to high-flux atomic sources~\cite{Wodey2021} or interferometry using narrow-linewidth optical transitions~\cite{Rudolph2020,Hu2017}. Utilizing rubidium and ytterbium simultaneously, it will be able to test the universality of free fall at an accuracy level below \SI{1e-13}~~\cite{Asenbaum2020,Hartwig2015,Schlippert2014} and the inherently large scale factors and related spatial superposition states will enable tests of quantum mechanics on macroscopic scales~\cite{Kovachy2015a}. This summary presents an overview of the VLBAI facility, highlighting its key features and its fundamental science prospects.

% \textit{Design -- }
The VLBAI facility, which is currently being commissioned in Hannover, comprises three main components:

1. The core of the facility consists of a \SI{10}{m} long, vertical ultra-high vacuum tube with a \SI{10}{cm} inner diameter. To mitigate the impact of stray magnetic fields, a high-performance octagonal dual-layer magnetic shield encloses the tube, reducing magnetic-field gradients to values below $\SI{2.5}{nT/m}$, corresponding to a \SI{6e-14}{\meter\per\second\squared} acceleration bias for operation with rubidium~\cite{Wodey2020,Lezeik2022}. In addition, by means of relative gravimetry, the uncertainty contribution originating from non-trivial gravity gradients along the baseline can be quantified and modelled~\cite{Schilling2020,Lezeik2022}. Temperature probes along the baseline help detect and correct errors caused by temperature gradients.

2. Dual-species sources of quantum-degenerate ensembles of various ytterbium isotopes as well as \textsuperscript{87}Rb will be installed at both ends of the vacuum tube. By combining magnetic and optical trapping techniques together with all-optical matter-wave collimation~\cite{Albers2022}, the facility targets atom fluxes on the order of $10^6$ atoms/s, with temperatures in the picokelvin regime~\cite{Deppner2021,Kovachy2015}.

3. A seismic attenuation system (SAS) is utilized to suspend a retroreflection mirror that serves as the atom interferometer's inertial reference. The SAS employs 6-degrees-of-freedom active stabilization through electromagnetic actuation using signals from onboard seismometers and novel opto-mechanical devices~\cite{Richardson2020}.

VLBAI employs rubidium as a standard choice for inertial sensing due to its well-established source and laser technology. In addition, the heavy lanthanide ytterbium is utilized, providing advantages such as strong cooling forces on the singlet line, narrow intercombination transitions, and minimal magnetic susceptibility for improved control over systematic effects and environmental decoherence. Furthermore, the internal composition of the species enhances sensitivity for the exploration of new physics beyond the Standard Model~\cite{Hartwig2015}.

% \textit{Performance Estimation -- }
The VLBAI facility aims to surpass the current state of the art in absolute gravimetry by mitigating external noise sources through the seismic attenuation system and magnetic shielding. In a simplified drop configuration with \SI{2e5}{} atoms per cycle, \SI{3}{\second} preparation time, and a \SI{800}{\milli\second} free-evolution time, the shot-noise-limited sensitivity is estimated to be \SI{1.7}{\nano\meter\per\second\squared} at \SI{1}{\second}. By adopting an advanced configuration that launches \SI{1e6}{} atoms per cycle, resulting in a \SI{2.8}{\second} free-evolution time and utilizing four-photon atom optics, the shot-noise-limited acceleration sensitivity improves to \SI{40}{\pico\meter\per\second\squared} at \SI{1}{\second}. In a gradiometric configuration with a \SI{5}{\meter} baseline and \SI{1e5}{} atoms per cycle, the shot-noise-limited sensitivity is \SI{5e-10}{\per\second\squared} at \SI{1}{\second}. Furthermore, a simultaneous-comparison measurement involving \textsuperscript{87}Rb and \textsuperscript{170}Yb allows for testing the universality of free fall by determining the E\"otv\"os ratio with a precision better than one part in $10^{13}$~\cite{Hartwig2015}.

% \textit{Fundamental Physics -- }
Apart from its metrological applications through highly sensitive absolute gravimetry (cf. Section \ref{Sec:geodres}), the VLBAI facility acts as a platform for interferometry with large scale factors as are essential for gravitational-wave detection. It also provides an opportunity to test the equivalence principle, to investigate fundamental decoherence mechanisms and the limitations of quantum mechanics at a macroscopic scale. Finally, putting optical clocks into spatial superposition in so-called quantum clock interferometry will give rise to novel measurements of the gravitational redshift~\cite{Loriani2019a,Ufrecht2020a,Roura2021,DiPumpo2023}.

\subsection{The Wuhan 10~m Atom Interferometer}
%(Lin Zhou) $\sim$1 pg}
\label{Sec:Wuhan10m}

\begin{figure}[htp]
\centering
\includegraphics[width=0.6\textwidth]{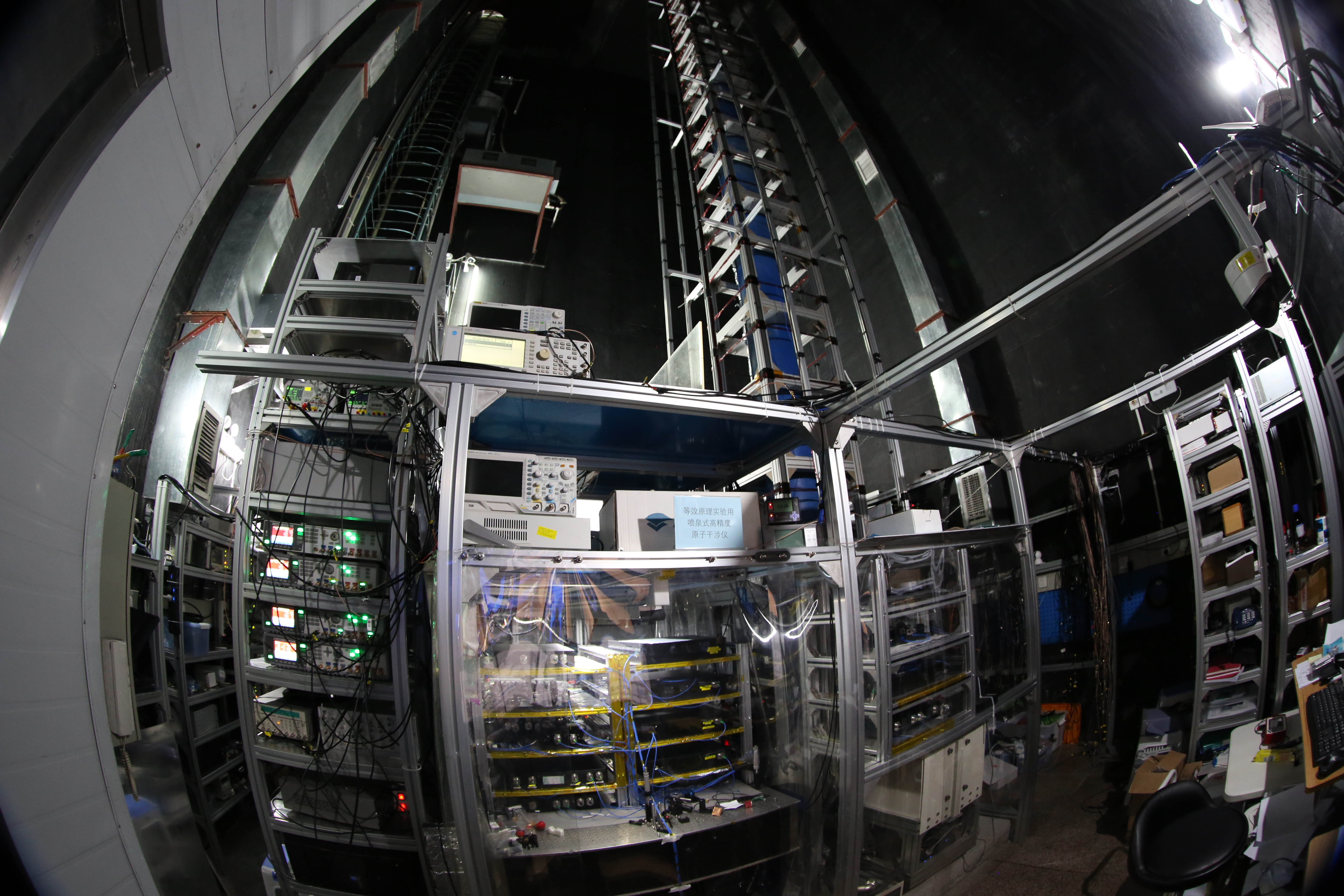}
\caption{The Wuhan 10 m Atom Interferometer~\cite{Zhou2011}.}
\label{fig01_Wuhan 10m.JPG}
\end{figure}

The equivalence principle (EP) is one of the key ingredients in general relativity. Quantum tests of the EP with atoms provide an important way to examine the applicable scope of the current physical framework, as well as to discover new physics. The design and appplications of the Wuhan 10-m atom interferometer shown in Fig.~\ref{fig01_Wuhan 10m.JPG} have been discussed in~\cite{Zhou2011}. The dual-species atom interferometer was used to test the EP with an accuracy of $3 \times 10^{-8}$ in 2015~\cite{Zhou2015}. The detector has also been used to perform a joint mass-energy test of the EP using the 4WDR-e method, and the mass and internal energy violation parameters were both constrained to the $10^{-10}$ level~\cite{Zhou2021}. The latest experimental progress of the Wuhan 10-meter atom interferometer has been to use the large momentum transfer technique with the 4WDR method to realize an 8-photon recoil atom interferometer. Interference fringes obtained after a free evolution time of $2T = 2.6\,\mathrm{s}$ have been observed, which is the longest duration realized in the laboratory, and the corresponding precision for gravity measurement is $4.5 \times 10^{-11}$ g per shot. The resolution of differential measurements with the dual-species atom interferometer is $2.5 \times 10^{-11}$ at 7168 s~\cite{Zhou2022}.

\subsection{The AION Project in the UK}
%(Charles Baynham) $\sim$1 pg}
\label{Sec:AION}

The Atom Interferometer Observatory and Network (AION)~\cite{AION} is a proposed research infrastructure allowing studies of dark matter and gravitational waves (GWs) from cosmological and astrophysical sources in the theoretically relevant but currently inaccessible mid-frequency band. It will develop and demonstrate the necessary deployable and scalable quantum technology by constructing and operating 10~m- and 100~m-scale instruments, paving the way for a future km-scale facility.
Stage 1 of the AION project, funded via the UK Science and Technology Facilities Council (STFC) Quantum Technologies for Fundamental Physics (QTFP) programme and involved institutions, has successfully delivered a new UK collaborative network in cold-atom interferometry, and by 2024 the current phase of the project will have made key steps towards the construction and deployment of a first 10~m prototype instrument. The long-term AION programme comprises:
\begin{itemize}
\item  Stage 1 (10~m): Construction of a first full interferometer system, providing proof-of-principle of the basic technology, along with evidence of scalability from lab-based to purpose-built infrastructure. As seen in Fig.~\ref{fig:Oxford}, the system will be deployed in a dedicated low-vibration basement in the Oxford Physics Department with an adjacent laser laboratory, and its performance fully characterised. Stage 1 will deliver an improvement in the sensitivity to the coupling of ultra-light dark matter (ULDM) to the electron by up to a factor of 3 for a mass $\sim 10^{-15}$ eV, but is not expected to have significant sensitivity to GWs. 

\begin{figure}
    \centering
    \hspace{1cm}
    \includegraphics[width=7cm]{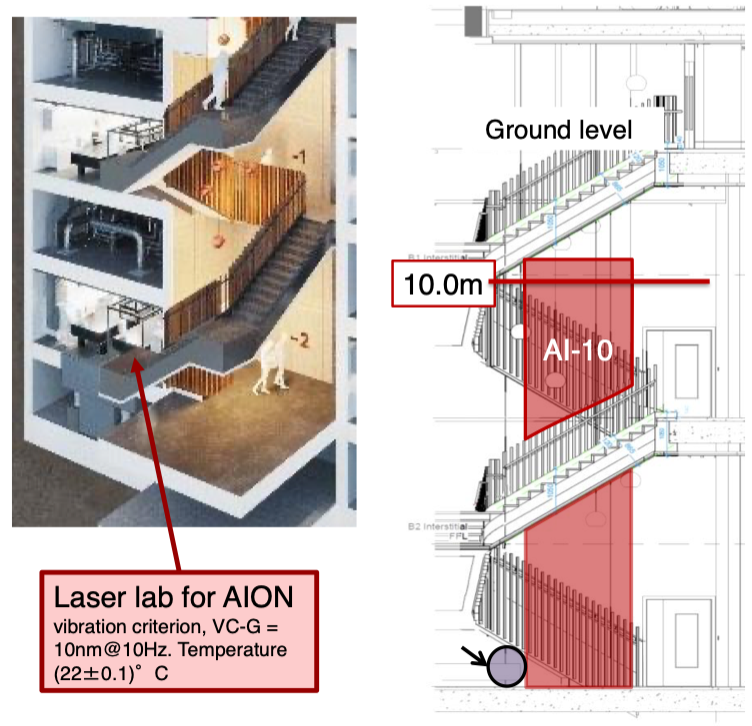}
    \caption{\it Layout of the AION-10 atom interferometer in the basement of the Oxford Physics Department.}
    \label{fig:Oxford}
\end{figure}

\item Stage 2 (100~m): Construction is expected to start in 2027, with operation from 2030 to search for both DM and GWs over an operational period of several years. The expanded Boulby laboratory in the UK is under study as a potential site for the detector (see Section~\ref{sec:Boulby} for more information). This stage will address fundamental challenges for construction and operation of a large detector ``in the field''. Stage 2 aims at an improvement in sensitivity to, e.g., the ULDM-electron coupling by up to more than one and a half orders of magnitude over a wide range of ULDM masses. It will also have unique sensitivity to GWs in a frequency band around 1 Hz, being sensitive, e.g., to phase transitions in the early universe and mergers of intermediate-mass black holes weighing $10^4$ solar masses at a redshift $z\approx 1$.

\item Stage 3 (1~km): Construction would start in the mid-2030s, with a target of reaching ultimate terrestrial sensitivity for GW and DM observation by the end of the decade. This stage would constitute a new international infrastructure. As for stage 2, the expanded Boulby laboratory in the UK is under study as a potential UK site for the detector. 

\item Stage 4 ($\gtrsim 1000$~km): A mission proposal for an Atomic Experiment for Dark Matter and Gravity Exploration in Space (AEDGE)~\cite{ElNeaj2020} has been submitted to ESA within its Voyage 2050 programme. This would directly use AION technology and could be flying from 2045.

 \end{itemize}

The goals of AION Stage 2 explicitly include further development of the necessary national and international partnerships and collaborations, along with the development of the necessary technical and operational experience.
 Stage 3 will require cooperative international investment and strategy funding.
Therefore, the AION Collaboration strongly supports the formation of an international proto-collaboration to support very long-baseline AI for fundamental physics exploitation.

In the following, we provide a short summary of the status of the AION project in Stage 1, during which the AION programme has three major deliverables to complete:\\

A):  Design and construction of five Ultra-Cold Strontium Laboratories in Birmingham, Cambridge, Imperial, Oxford and RAL;\\

B):  Establish a partnership with the MAGIS experiment in the US;\\

C): Design and Construction of a 10 m prototype detector to be built in the Beecroft building in Oxford.  \\

Milestones A) and B) have been completed, and the project is now mainly focusing on C), the design and construction of a 10 m prototype detector.

\begin{figure}[t]
    \centering
    \includegraphics[width = \textwidth]{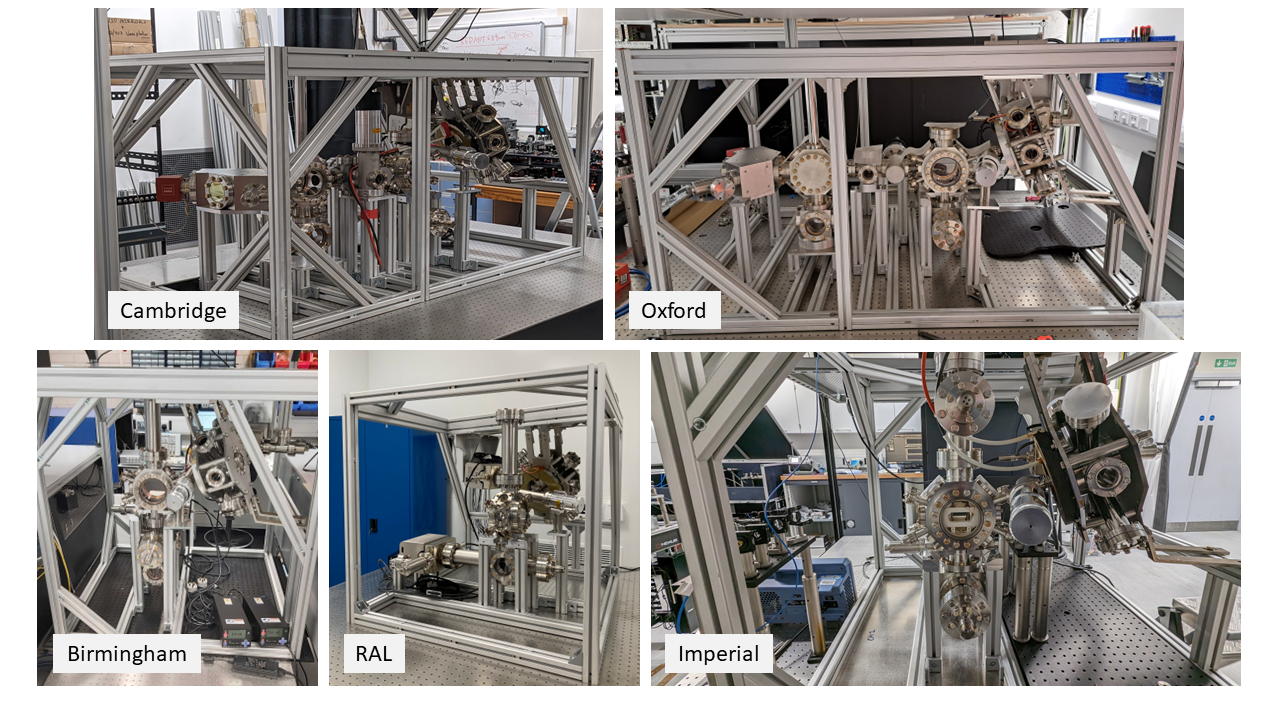}\\
    \includegraphics[width = 0.6\textwidth]{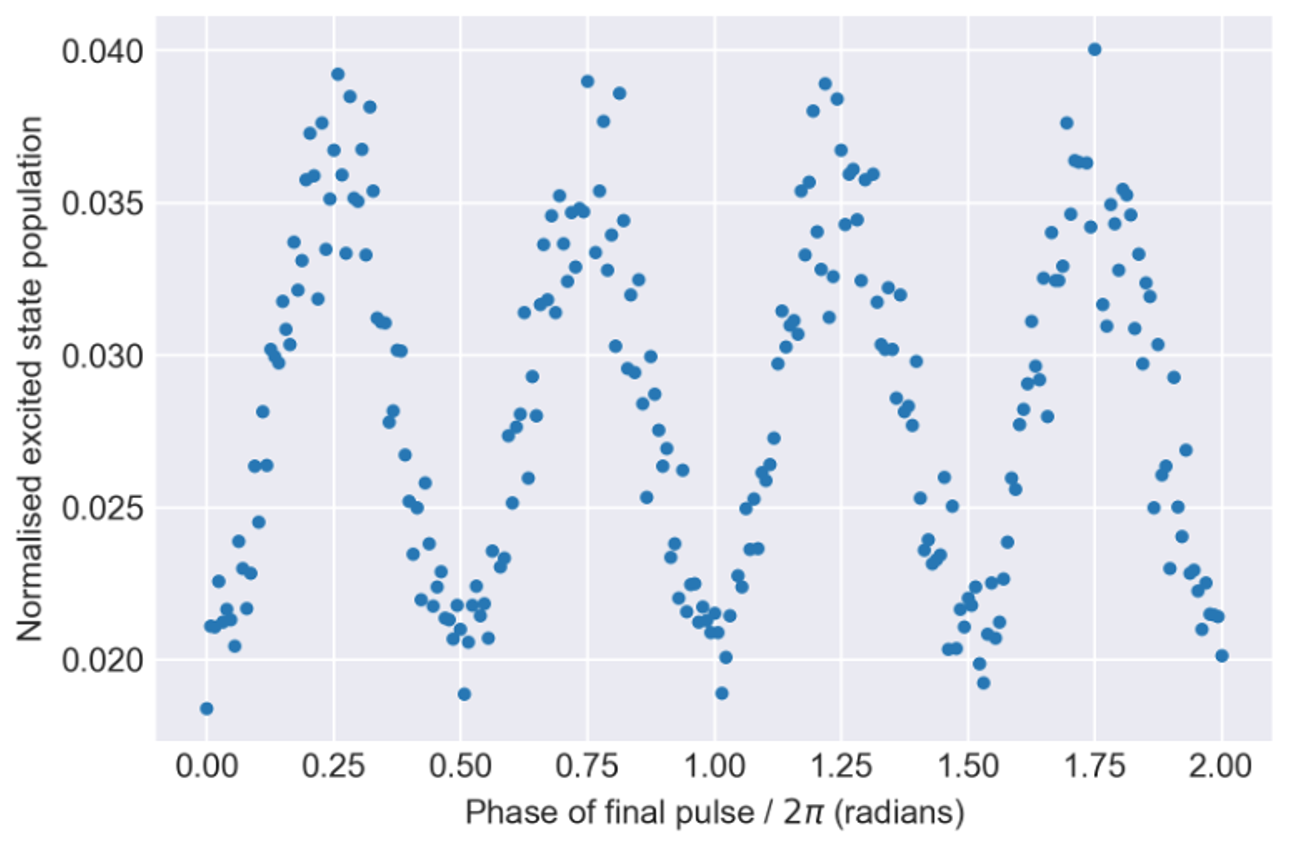}\\
    \caption{{\it Upper panels:} Photographs of the five AION sidearm systems, installed at their corresponding institutions~\cite{AION:2023fpx}. {\it Bottom panel:} Measurements at the University of Birmingham of the occupation levels of an excited strontium state following atom interferometry sequences in which the phase of the final laser pulse is varied, demonstrating interference fringes analogous to those in an optical Mach-Zehnder interferometer.}
    \label{fig:all_sidearms}
\end{figure}

The biggest challenge faced by AION in Stage 1 was the centralised design and production of the ultra-high-vacuum sidearm and laser-stabilisation systems for the AION Ultra-Cold Strontium Laboratories, as summarised in~\cite{AION:2023fpx}. Streamlining the design and production of the sidearm and laser stabilisation systems enabled the AION Collaboration to build and equip in parallel five state-of-the-art Ultra-Cold Strontium Laboratories within 24 months and observe atomic interference fringes - see Fig.~\ref{fig:all_sidearms} - by leveraging key expertise across the collaboration. This approach could serve as a model for the development and construction of other cold atom experiments, such as atomic clock experiments and neutral atom quantum computing systems, by establishing dedicated design and production units at national laboratories, such as Rutherford Appleton and Daresbury Laboratories in the UK.  

\subsection{The MAGIS-100 Atom Interferometer}
\label{Sec:MAGIS100}

\begin{figure}[htp]

    \centering 
    \includegraphics[width=.9\textwidth]{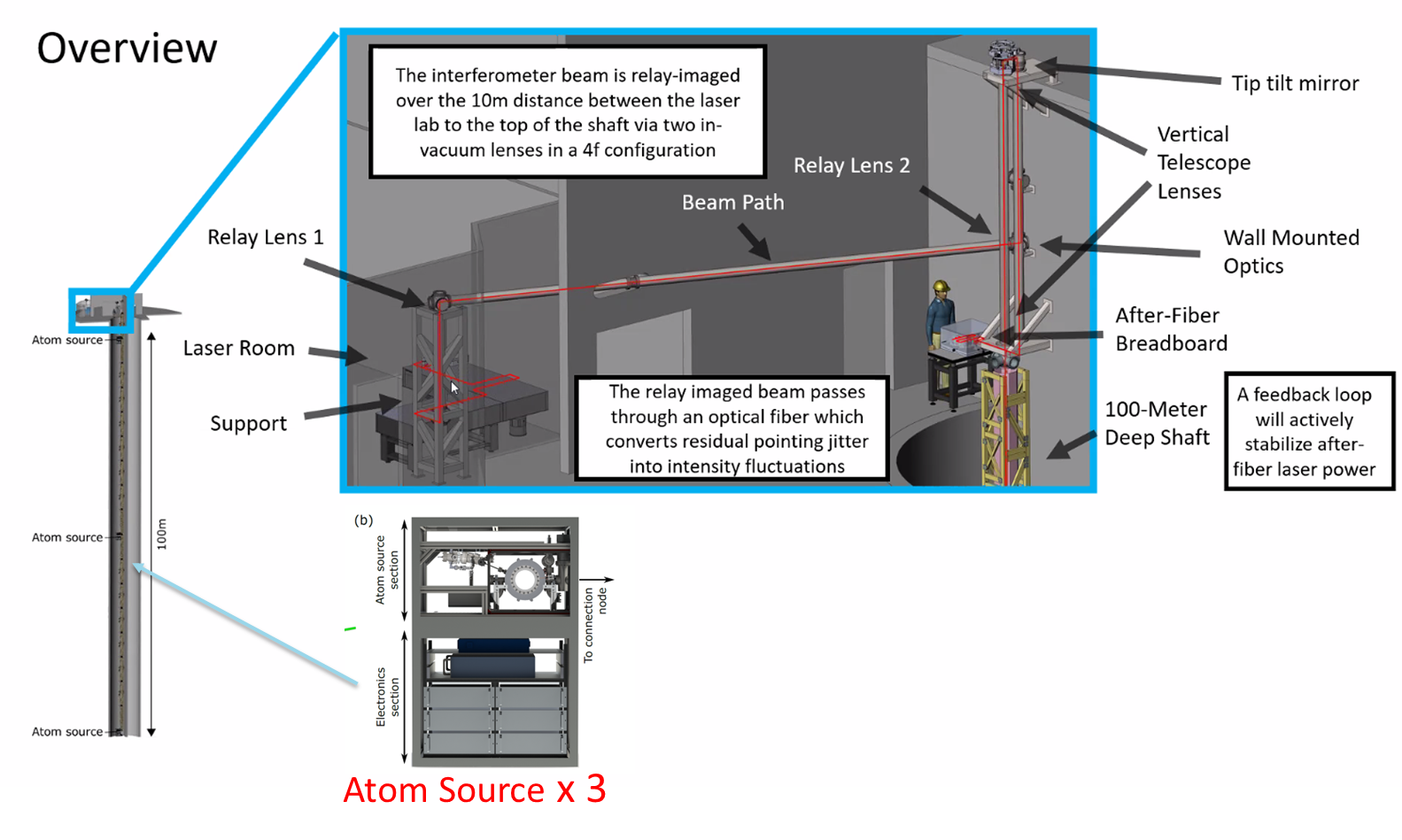 }
\caption{Basic layout of the MAGIS experiment.}
\label{fig:MAGIS_Layout}
    
\end{figure}

The MAGIS-100 experiment is Fermilab project E-1101 (FERMILAB-TM-2700-PPD, FERMILAB-CONF-23-430-ETD), which consists of a collaboration of 9 universities and national laboratories, and is funded by the US DOE QuantISED program, the Gordon and Betty Moore Foundation, the UK STFC, and the Kavli Foundation. To carry it out it is necessary to build and install at Fermilab a next-generation instrument that uses light-pulse atom interferometry to search for physics beyond the Standard Model.  MAGIS-100 will exploit the existing $\sim$100\;m vertical MINOS access shaft and will be an upgrade of the existing 10\;m scale experiment at Stanford with greatly increased sensitivity due to its increased length.

In the MAGIS detector design, dilute clouds of Ultra-Cold atoms at both ends and the middle of the baseline (the MINOS shaft, see Fig.~\ref{fig:MAGIS_Layout})  act as both inertial references and clocks.  Light from a laser propagates in a high-vacuum region between the two atom ensembles and interacts with the atoms, driving transitions between the ground and excited atomic clock levels. The accumulated phase difference between the states in the atom ensembles (interferometers) is sensitive to the time variation in the atomic energy levels, which would be caused by a light scalar dark matter field interacting with the atoms. Additionally, the timing of these transitions depends on the light travel time across the baseline, so a passing gravitational wave can result in a shift of the atomic state, with an associated interferometer phase difference. MAGIS will use Sr atoms, which are able to act as excellent clocks. The MAGIS detector concept is shown schematically in Fig.~\ref{fig:MAGIS_Layout}, with the upstairs laser lab (currently under construction), the above-ground laser transport for the interferometry laser beam, the vertical interferometry vacuum tubes, and the three strontium atom sources. The use of any two atom sources can produce a difference measurement. In addition, the use of three sources provides the opportunity to correct for gravity gradient noise, and serves as a general constraint for systematics control.  MAGIS-100 will also have the capability to perform co-located, dual-isotope Sr atom interferometry in an alternative dark matter search mode \cite{Abe2021}.

%\begin{figure}[htp]

%    \centering 
%    \includegraphics[width=.9\textwidth]{Fig_2_RP.png }
%\caption{Plan of the MAGIS-100 laser laboratory}
%\label{fig:MAGIS_Lasers}
%    
%\end{figure}

Construction at Fermilab provides many benefits to the experiment. The site provides a laboratory environment with the required power to operate the experiment, appropriate support for laser and vacuum systems, and safety oversight. Engineering for civil components such as shaft supports, vacuum systems, and electrical systems benefits from the existing Fermilab staff, who have extensive experience with equipment deployment at this scale. Controls, computing, and networking systems are readily available, and safety and interlocks controls are standardized. In addition, there are facilities for operations that will be exploited.

MAGIS-100 has a modular construction: 17 sections of approximately 5.7\;m length are assembled from the bottom of the shaft to the top. Specialized connections are found at the location of the three atom sources to allow for optical shuttle and launch of the atom clouds. The apparatus is surrounded by an octagonal, multi-layer mu-metal shield to provide a clean magnetic environment. Necessary optics (mirrors, mode cleaning, and retro-reflection) are located at the top and bottom of the experiment. Items needing maintenance are found at the connections between nodes, and are shielded by sliding couplers.

The sections are surrounded by an aluminum strongback frame for rigidity and protection from impacts. The beam tube components for the modules can be aligned using a 6-strut system to provide the necessary degrees of freedom. Ferromagnetic materials have been minimized in the design.

Installation will be aided by a personnel access system, which is currently planned to be a cage suspended from the 15-ton overhead bridge crane located in the MINOS building. It is likely that a supplementary lightweight gantry crane will be used to lower the modules and atom sources into position on their permanent, fixed, mounts. 

Most (16/22) of the MAGIS-100 lasers are located on the surface in a purpose-built laser laboratory. These constitute the high-power interferometry laser system \cite{DeRose2023} and the red lasers for each atom source, which are transmitted to the sources over fibers. Most of the laboratory lasers are frequency locked using an optical comb. The IR and blue lasers used at the atom source locations are installed in the shaft together with the sources.

MAGIS-100 will provide a powerful instrument for science. In addition, its atom sources and laser systems are modular and readily upgradable in the future. Full technical details can be found in \cite{Abe2021}.

\subsection{Summary}

In this Section, we have discussed the features and technological challenges of vertical long-baseline atom interferometry.  Key areas where further research is needed include the development of higher power lasers and robust atom optics techniques \cite{Butts2013,Dunning2014,Berg2015, Wilkason2022,Saywell2020,Goerz2023,Chen2023,Saywell2023,Louie2023} to enable interferometers with several $\times 10^4$ $\pi$-pulses on narrow clock transitions, investigations of the scalability of rotation and gravity gradient compensation techniques to km-scale baselines, demonstration of multiplexing techniques to allow multiple interferometers to be operated simultaneously, and designing systems for stabilizing and/or monitoring the relative positions of cameras used for phase shear readout of interferometers separated over a long baseline.  Moreover, further investigations into Newtonian gravity gradient noise mitigation strategies for vertical configurations will be essential. This Section has also provided an overview of four vertical long-baseline atom interferometry projects: the Hannover VLBAI atom interferometer, the Wuhan 10 m atom interferometer, the AION project in the UK, and MAGIS-100 at Fermilab, which will serve as testbeds for the necessary technology development for TVLBAI projects.

\section{Long-Baseline Detector Options (horizontal)}%\todo{Wolf: @Bejamin/Sven: Summary missing?}
\label{sec:horizontal}
%{\bf 5 pages guideline }
%\subsection{Editors:Benjamin Canuel, Sven Abend, Mingsheng Zhan}
\subsection{Introduction}
Horizontal gravitational wave detectors have some different features from vertical ones, making them more similar to laser-based gravitational wave detectors such as LIGO and Virgo. The first key feature is that beam splitting is decoupled from the atomic motion due to free fall. The second is the possibility to build a two-arm detector. Unlike vertical detectors, where single photon beam splitters are used to prevent laser phase noise entering the signal, a two-arm configuration can be used for the same purpose, but with two conventional photon beam splitters. This enables the use of alkaline elements for the atomic source, which have been used in a variety of previous atom interferometry experiments, so that the key technologies (i.e., high flux, large momentum transfer, long interferometry times, delta-kick collimation or squeezing) are more mature. Two long-baseline horizontal experiments are under construction currently in France and China, namely MIGA and ZAIGA, which are presented in this Section and their current status and challenges discussed. In subsequent parts of this Section, possible strategies for horizontal geometries are introduced, which realize multi-loop configurations originally introduced for space-based detectors. Using a distinct relaunch strategy in combination with delta-kick collimated Bose-Einstein condensates and symmetric horizontal beam splitters the necessary parameters for the next generation of detectors seem plausible. Tackling the issue of laser front phase aberrations and the need for increasing laser powers, cavity-based diffraction can be used in horizontal geometries, due to the decoupling of atomic motion and beam splitting, making it an ideal candidate for implementation.

\subsection{Large-scale atom interferometry with MIGA }
%($\sim 1$p)}
\label{sec:MIGA}

The MIGA~\cite{Canuel2018} large-scale gravity antenna is a demonstrator for low-frequency GW detection based on atom interferometry~\cite{GWHandbook2021}. This project is carried out by a consortium that gathers 17 leading French laboratories and companies in atomic physics, metrology, optics, geoscience and gravitational physics. The geometry of MIGA consists of a 150 m array of three Rb atom interferometers manipulated by cavity-enhanced Bragg pulses to provide simultaneous measurements of strain and inertial forces sensed in an optical cavity.

\begin{figure}[htp]
\centering
\includegraphics[width=0.9\textwidth]{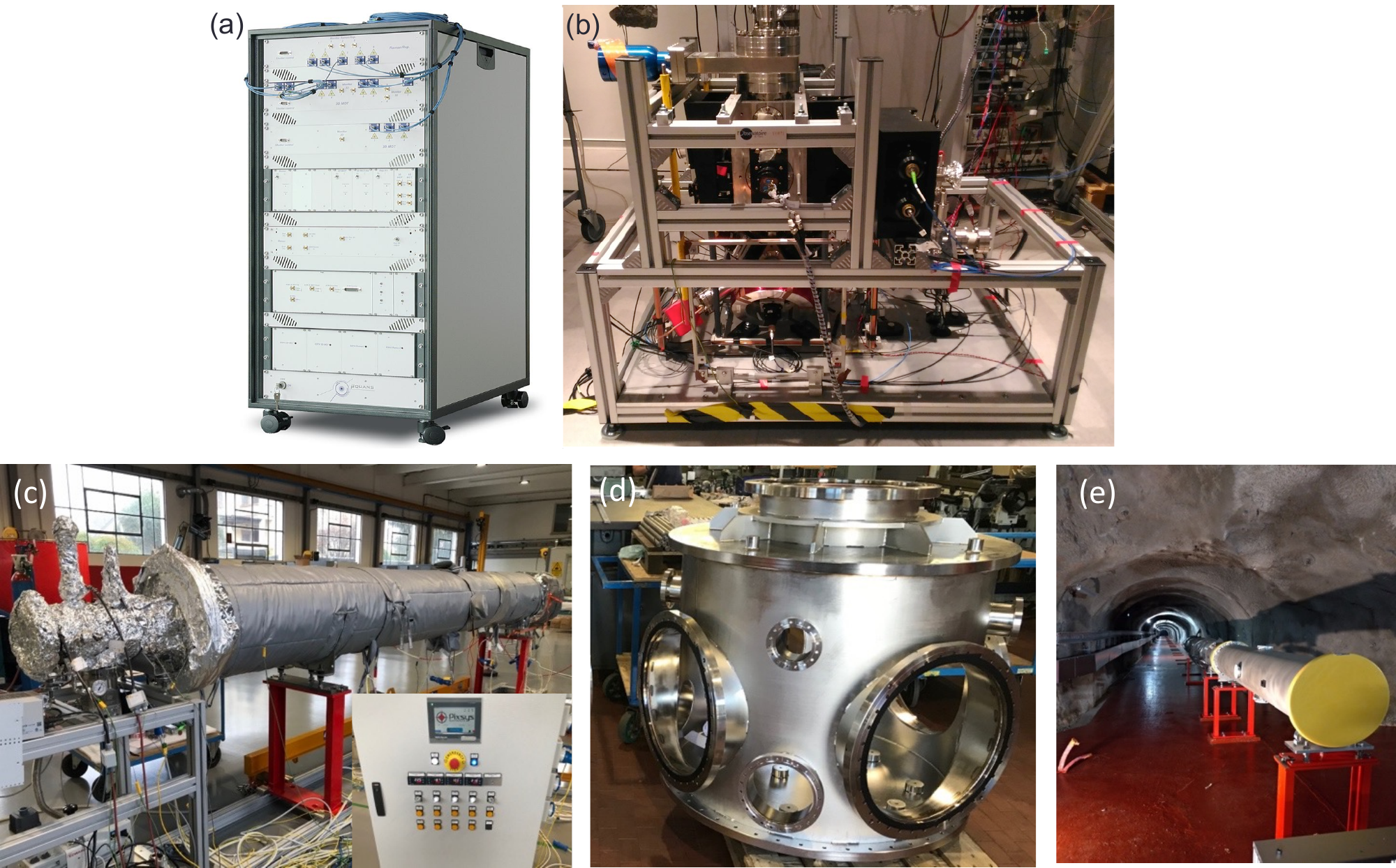}
\caption{(a) Fibre laser system developed by the Muquans company~\cite{Sabulsky2020}, (b) cold $^{87}$Rb atom source~\cite{Beaufils2022}, (c) standard 6 m long section under vacuum test, (d) vacuum tower in production at SAES Parma (Italy), (e)  MIGA gallery within the Laboratoire Souterrain {\`a} Bas Bruit (LSBB) and installation of the first sections of the vacuum vessel~\cite{Canuel2022}.}
\label{fig:Miga_status}
\end{figure}

This underground antenna is located in the LSBB laboratory~\cite{Gaffet2009}. This facility is a former command centre of the military nuclear force on the “plateau d’Albion” that has been converted into a unique underground research platform. Located 500~m deep inside the mountains of the pays d’Apt, this facility is ideally located away from major anthropogenic disturbances and benefits from very low background noise, in particular  seismic~\cite{Canuel2018,Rosat2018} and magnetic noise sources~\cite{Henry2016}, which are major disturbances usually preventing atom interferometer experiments from reaching their ultimate performances.

Targeting an initial strain sensitivity of $2 \cdot 10^{-13}/\sqrt{\mathrm{Hz}}$ at 2~Hz, MIGA will be a demonstrator for a new generation of GW detectors based on atom interferometry (AI) that could open a new frequency band for the observation of GWs. The instrument will be used to study advanced measurement strategies and atom manipulation techniques to push further the sensitivity of large-scale AI experiments.  MIGA will also provide precise measurements of the local gravity sensed by an AI network to study the impact of Gravity Gradient Noise (GGN) and possible mitigation strategies. According to current models, GGN is a central issue for low-frequency GW detection, and instruments relying on single atom gradiometer geometries would be strongly limited in their performances in a large portion of their sensitivity window.

An upgraded version of the antenna, using the Large Momentum Transfer (LMT) generated by $2\times 100$ photon transitions and an improved detection noise of $0.1$~mrad/$\sqrt{\mbox{Hz}}$ would make it possible to study the different components of GGN over different baselines and obtain a better modelling of the space-time correlation properties of GGN~\cite{Junca2019}. These studies are crucial to develop strategies for mitigating GGN based on spatial averaging using AI sensor networks~\cite{Chaibi2016}.

The infrastructure work at LSBB has been completed and two new perpendicular galleries of 150~m were bored to host the initial antenna and a possible evolution towards a 2D instrument~\cite{Canuel2022}. The Rb atom sources were also produced and fully characterized~\cite{Sabulsky2020,Beaufils2022} in collaboration with the SYRTE laboratory (see Fig.~\ref{fig:Miga_status} (a), (b)). The vacuum vessel was produced, tested~\cite{Sabulsky2022} and delivered to LSBB and MIGA is now currently in its final assembling and commissioning stage at the LSBB (see Fig.~\ref{fig:Miga_status} (c), (d), (e)).

\subsection{Very-large-scale atom interferometry with ELGAR}
%($\sim 1$p)}
\label{sec:ELGAR}

\begin{figure}[htp]
\centering
\includegraphics[width=0.9\textwidth]{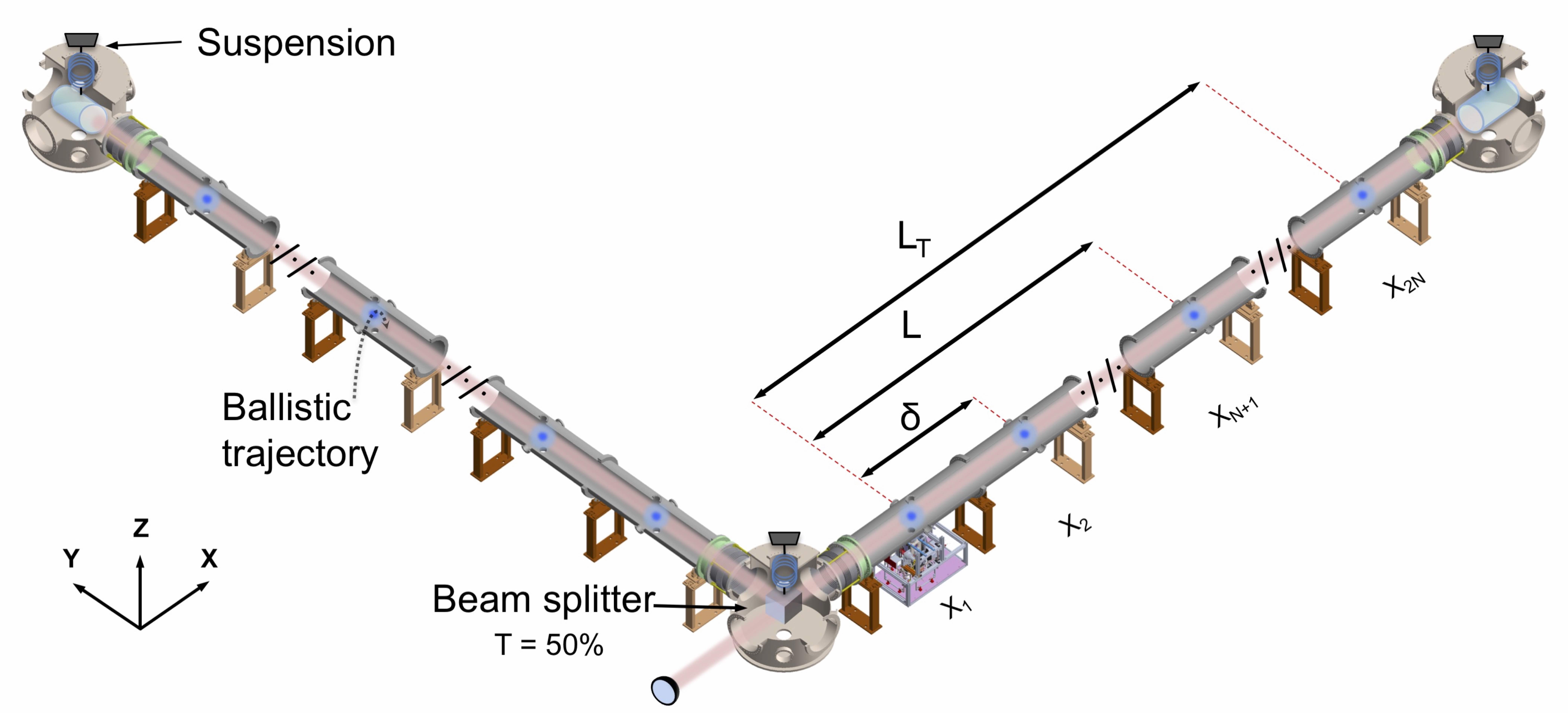}
\caption{Geometry of ELGAR, based on a distributed 2D array of gradiometers with baseline $L= 16.3$~km with a total baseline $L_T=$ 32.1~km. Taken from~\cite{Canuel2020}.}
\label{fig:ELGAR}
\end{figure}

This expected outcomes of MIGA will pave the way for the development of future research infrastructures based on quantum technologies such as ELGAR (European Laboratory for Gravitation and Atom interferometry Research)~\cite{Canuel2020}. Based on an ${\cal O}(10)$~km horizontal geometry, the ELGAR initiative aims at the realization of a European underground infrastructure to study space-time and gravitation with the primary goal of detecting GWs in the band between 0.1 Hz and 10 Hz. This instrument would use a geometry based on an array of atom gradiometers to reduce the contribution of GGN (see Fig~\ref{fig:ELGAR}). 
A common ultra-stable laser interrogates two perpendicular arms of total baseline $L_T=$ 32.1~km. Each arm is composed of $N=80$ gradiometers with baselines $L=$ 16.3~km, spaced from each other by a distance $\delta=$ 200~m; these two last parameters have chosen taking into account the space-time correlation properties of GGN. The gravitational signal is then extracted from the difference of the mean of the gradiometer signals in each arm. This averaging method makes it possible to reduce the GGN by a factor that can be better than $1/\sqrt{N}$ in the detector bandwidth, and the 2D geometry limits the influence of the residual frequency noise of the interrogation laser. The site of the detector would be an underground facility at a depth of a few hundred metres, reducing further both seismic and atmospheric GGN~\cite{Creighton2008,Harms2013,Harms2019}.

A preliminary description of the antenna subsystems can be found in~\cite{Canuel2020a}. It would use $^{87}$Rb sources cooled down to 100 pK, launched on a vertical trajectory towards the common interrogation laser where they would be interrogated using a four pulse sequence ``$\pi/2$-$\pi$-$\pi$-$\pi/2$''. The atom manipulation protocol would use a combination of Bragg diffraction and Bloch oscillations~\cite{Gebbe2021} to reach a total number of $2n=$~1000 transferred photons during each interrogation pulse. Assuming an atom flux of $10^{12}$~atoms/s and an integration time of $4T=800$~ms, the shot noise limited strain sensitivity of the detector would be $3.3 \times 10^{-22}/\sqrt{\rm{Hz}}$ at its optimal detection frequency of 1.7~Hz. The bandwidth of ELGAR would therefore complement that of the future space-based detector LISA and third-generation ground-based optical detectors such as ET~\cite{Punturo2010} or the Cosmic Explorer~\cite{Abbott2017,Reitze2019}. A simultaneous operation of these instruments would open multiband GW detection over an extended band, ranging from 0.1~mHz up to 10~kHz.
The ELGAR infrastructure would focus on GW observation in the band around 1~Hz, but would also have extended applications in various domains including geology, fundamental physics, gravitation and general relativity.

\subsection{Status of the ZAIGA project}
%($\sim 1$p)}

\begin{figure}
    \centering
    \includegraphics[width=0.6\textwidth]{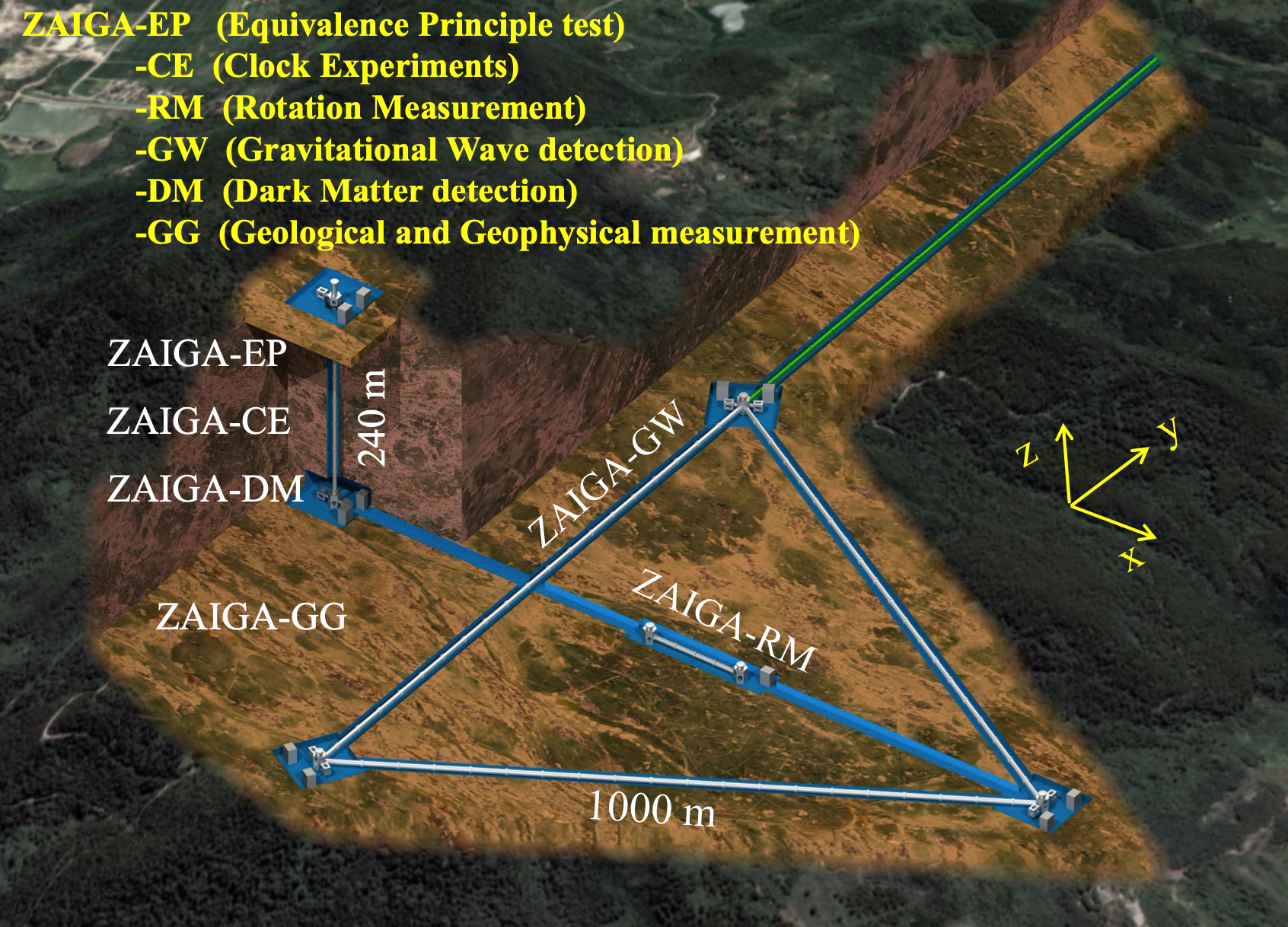}
    \caption{\it Layout of the ZAIGA laboratory near Wuhan, China for a range of experiments using atom interferometry~\cite{Zhan2019}.}
    \label{fig:ZAIGA}
\end{figure}

ZAIGA is a facility for underground laser-driven atom interferometer experiments near Wuhan, China~\cite{Zhan2019} that is illustrated in Fig.~\ref{fig:ZAIGA}. %[https://doi.org/10.1142/S0218271819400054].
Its first stage is already funded and under construction, with completion scheduled for 2027. It includes a 240-meter vertical shaft equipped with an atom fountain and atomic clocks, which targets probes of quantum mechanics, tests of the Einstein Equivalence Principle, measurements of the gravitational redshift using atomic clocks and searches for ULDM and GWs, as well as a horizontal gallery of length $> 1$~km that will host an experiment to measure rotation and the Lense-Thirring effect, and probe general relativity.
The second phase, planned for 2027 to 2035, will feature an equilateral triangle of horizontal 1-km galleries instrumented with atom interferometers that will provide enhanced sensitivities to ULDM and GWs, and there is provision for extending one of the arms of the triangle to multiple kilometres in a subsequent stage. In addition to this comprehensive programme of fundamental physics, the ZAIGA laboratory will be the principal node in a network for environmental monitoring including measurements of earthquakes, changes in ground water, meteorology, geogravity and relativistic geodesy.

\subsection{Future perspectives for horizontal detectors: Scalable large momentum transfers and multi-loop geometries}

Horizontal beam splitting faces the challenge of an intrinsic symmetry, due to the vanishing Doppler shifts of atoms falling perpendicularly to the laser beam. The implementation of symmetric horizontal beam splitting has been studied for either Bragg~\cite{Ahlers2016} or Raman diffraction. Instead of using higher $n^\mathrm{th}$ order or sequential diffraction processes, which are mainly limited by atom loss caused by non-ideal beam splitter efficiencies, accelerated optical lattices may be used, where the atomic ensemble suffers significantly fewer losses from dephasing effects due to light shifts. 

\begin{figure}%[htbp]

    \begin{tabular}{@{}c@{}}
    \centering 
    \includegraphics[width=.45\textwidth]{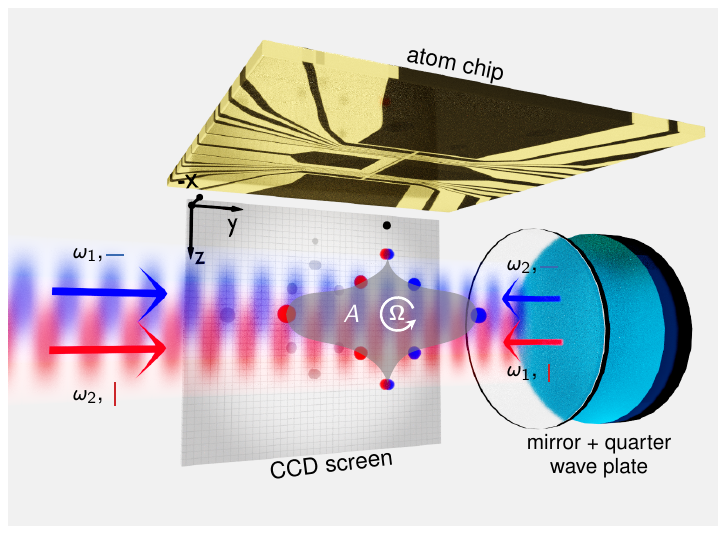}
    \end{tabular}%
    \vspace{\floatsep}
    \begin{tabular}{@{}c@{}}
    \centering 
    \includegraphics[width=.45\textwidth]{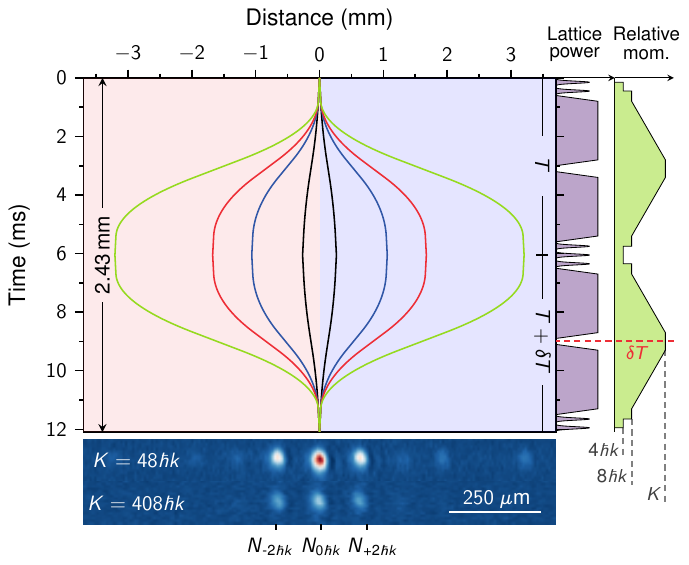}
    \end{tabular}
    
	\caption[Twinlattice scheme]{The twin lattice is formed by retroreflecting light at two frequencies with linear orthogonal polarization. A quarter-wave plate in front of the retroreflector alters the polarization to generate two counterpropagating lattices (indicated in red and blue). After release from the atom chip and state preparation, the BEC is symmetrically split and recombined by the lattices, driving double Bragg diffraction (DBD) and Bloch oscillations (BOs). In this way, the interferometer arms form a Sagnac loop enclosing an area $A$ (shaded in gray) for detecting rotations $\Omega$. The interferometer output ports are detected on a CCD chip by absorption imaging. Figure from \cite{Gebbe2021}.}
	\label{fig:twinlattice}
\end{figure}

The twin-lattice approach presented in \cite{Gebbe2021} is based on a combination of double Bragg beam splitters, which realize the symmetric momentum splitting of the initial atomic ensemble, with a simultaneous acceleration of both interferometer arms with the help of Bloch oscillations. 
The individual steps of the sequence are presented in the following and the overall scheme is shown in Fig.~\ref{fig:twinlattice}. The light fields of the twin lattice are aligned perpendicular to gravity. Additionally the retroreflection of $\omega_1$ and $\omega_2$ is realized with a combination of a quarter-wave plate and a separate mirror. The two applied frequencies are combined in a single beam with orthogonal linear polarization. This helps to suppress losses caused by parasitic standing waves as well as systematic effects like laser phase noise and wavefront distortions. Due to the beam being aligned horizontally there is a vanishing velocity in the beam splitter direction, which directly entails symmetric diffraction through a double Bragg or Raman process.

For the sequence itself a BEC is created with the help of the atom-chip trapping and cooling capabilities. The individual steps are shown in Fig.~\ref{fig:twinlattice} for different amounts of transferred momenta. 
After release delta-kick collimation is performed and followed by an adiabatic rapid passage to a non-magnetic state. The interferometer itself is created by an initial sequential double Bragg diffraction $\pi/2$-pulse and an acceleration via Bloch oscillations. 
This creates a maximum momentum separation between the two interferometer arms of up to $\Delta p=408\,\hbar k$. To mirror the momentum states for the second half of the interferometer the atomic ensembles are decelerated again to $\pm 4\,\hbar k$. The following three-pulse combination can be described as a sequential double Bragg $\pi$-pulse that inverts the atomic velocity to $\mp 4\,\hbar k$. 
Afterwards the ensembles are accelerated again to the same momentum separation $\Delta p$ as before and decelerated again, with the difference that the trajectories are now pointing towards each other to allow for a spatial overlap at the end of the sequence. The interference is closed by a final sequential double Bragg diffraction $\pi/2$-pulse. 
After a waiting time of $\tau_\mathrm{det}=9$ms the three output ports of the interferometer can be imaged using absorption detection. 

A folded triple-loop geometry can be realized as an alternative to a standard double-loop or butterfly geometry using this kind of symmetric beam splitters together with relaunches~\cite{Abend2016} at the intersections of the trajectories so that only a single laser link is required~\cite{Schubert2019}.
This enables scalability in $T$, consequently a broadband detection mode~\cite{Hogan2016}, and a resonant detection mode~\cite{Graham2016} by adding additional relaunches and beam splitting pulses.
In addition, the triple-loop geometry is robust against fluctuations of the mean position and velocity of the wave packet entering the interferometer.
The scheme requires a pointing stability of the relaunch vectors at the level of $\sim$\,prad, comparable to the requirement on the initial launch vector in a single-loop geometry~\cite{Schubert2019}.
Omitting other terms, the differential phase shift between two triple-loop interferometers in a gradiometric configuration caused by a gravitational wave is $8k_{\mathrm{eff}}hL\sin^4(\omega{T}/2)\left[ (7+8\cos(\omega{T}))/2 \right]$~\cite{Hogan2010}. 
As an option, this geometry can be implemented at a later stage for broadband and resonant detection modes.

\subsection{Multi-photon atom interferometry via cavity-enhanced Bragg Diffraction}

The target strain sensitivity for decihertz GW detectors based on atom interferometry requires the application of Large Momentum Transfer (LMT) atom optics. 
Modern atom optics techniques utilize stimulated Raman transitions \cite{Kasevich1991}, Bragg diffraction \cite{Giltner1995}, and Bloch oscillations \cite{Peik1997} or a combination of the three to realize LMT atom interferometry \cite{Cadoret2008,Muller2008,Jaffe2018,McGuirk2000,Gebbe2021,Muller2009,Pagel2020,Ahlers2016,Kovachy2012}. 
It appears to be feasible to generate $1000 \hbar k$ of momentum transfer by a combination of Bragg diffraction and Bloch oscillations~\cite{Gebbe2021}, but this imposes severe requirements on the laser power \cite{Muller2008}. 
Beam splitting of order $n$ requires a laser intensity scaling as $n^4$ for a fixed spontaneous emission rate \cite{Muller2008a}.

An optical cavity could provide the required power enhancement, while adding spatial filtering for the beam splitting, with added benefits for both sensitivity and accuracy with careful design of the device: 
the power enhancement increases linearly the scale factor of atomic inertial sensors via coherent multiphoton processes \cite{Leveque2009,Chiow2011,McDonald2013,Sabulsky2022a} utilizing a large intensity of the interrogation laser. 
The natural spatial filtering of cavity modes reduces wavefront distortion \cite{LouchetChauvet2011}, an effect that can eventually limit the accuracy of atom interferometers but grows quickly in LMT atom interferometry. 
Other notable effects on atom interferometers, such as diffraction phases generated by intensity and mode shape fluctuations, can be controlled within an optical cavity. 

At the limits of optical cavities, high finesse is required for efficient atom optics, but a long cavity is also needed for GW detection. However, this leads to narrow linewidth cavities and so a long temporal response \cite{DovaleAlvarez2017}. 
Such a response is known to degrade the temporal shape of the light pulses used to manipulate coherently  the matter waves \cite{Fang2018}. 
These effects are fundamental issues limiting the use of optical cavities for atom interferometry \cite{DovaleAlvarez2017}, but new techniques have been proposed to circumvent this problem using light-shift engineering \cite{Bertoldi2021} and intracavity frequency modulation of circulating pulses \cite{Nourshargh2021,Nourshargh2022}.
Optical cavities have proven useful for modern atomic physics, having been used to produce Bose-Einstein condensates in the strong-coupling regime \cite{Colombe2007} and to generate entanglement in particles of an atomic ensemble in order to beat the so-called standard quantum limit \cite{Cox2016,Hosten2016,Greve2022}. 
Recent work has shown that a matter-wave interferometer operating as a gravimeter can be realized using a stable optical cavity \cite{Hamilton2015}.
Hence optical cavity manipulation for enhancing light-matter interactions is a promising pathway for mediating LMT atom optics in future experiments. 

Furthermore, the realization of a GW detector based on atom interferometry can further benefit from optical cavities in mediation of the light-matter interaction.
Interrogation of atom interferometers functioning as array of atom gradiometers \cite{Chaibi2016,Canuel2018,Canuel2020} requires that the atom optics be driven by the same oscillator. 
This atom gradiometer, inside an optical Fabry-P\'{e}rot Michelson interferometer, forms a hybrid device that would be the start of a new generation of instrument where both atomic and optical readout are used simultaneously to provide an extended GW detection band into lower frequencies. 
A 2D geometry naturally lends itself to reducing efficiently the impact of frequency noise on the atomic readout, as is already the case for its optical counterpart \cite{Harry2010,Acernese2014}.
Optical cavity mediation of LMT atom optics is a promising technological pathway for future high sensitivity large scale AI detectors. 

A dedicated experiment to explore cavity manipulation in the horizontal direction demonstrated an LMT atom interferometer using cavity-enhanced beamsplitters \cite{Riou2017,Mielec2020,Sabulsky2022a} in the framework of the MIGA project \cite{Canuel2018}.
It utilized a marginally stable cavity \cite{Arnaud1969} that allows for the propagation of large spatial modes. This device is generally useful for all atomic sources, trapped or on ballistic trajectories.
This configuration, developed for atom interferometry \cite{Riou2017, Mielec2020}, comprises two high reflectivity mirrors placed at the focal planes of a biconvex lens. 
This horizontal 80~cm cavity is coupled with a sub-Doppler cooled $^{87}$Rb atomic source developed for the MIGA project \cite{Beaufils2022} and used to operate a Mach-Zehnder atom interferometer using a momentum transfer of up to $2n = 8 \hbar k$ on a large resonating mode with a $1/e^2$ diameter of 4~mm \cite{Sabulsky2022a}.

This experiment \cite{Sabulsky2020,Beaufils2022,Sabulsky2022a} launches a cloud of atoms into the cavity on a ballistic trajectory where interference patterns are generated by the cavity not being perfectly horizontal and scanning the scale factor \cite{Sabulsky2022a}. 
The optical gain of the cavity ($38$) is sufficient to drive diffraction with mW level input from a low-noise source, while the finesse ($200$) results in a cavity temporal response of a few hundred ns, which is sufficient to avoid distorting the applied Gaussian pulses; diffraction up to $n = 4$ has been observed. 
Tilting the experiment at various angles, as observed with an external tilt measurement, induces small but measurable accelerations from gravity. 
By studying the cantilevering of fringes as well as spectroscopy, it was possible to conclude that the phase shift is inertial for the different diffraction orders.

This work paves the way towards increased sensitivity of matter wave interferometer experiments and hybridization with laser interferometers. 
In the short term, this method is applicable to various atom interferometer sources and geometries, like mobile, sub-Doppler cooled vertical gravimeters \cite{LouchetChauvet2011} and horizontal gyroscopes \cite{Dutta2016}. 
%\subsection{Conclusion ($\sim 0.5$p)}

\section{Site Options} %\todo{Wolf: Summary missing?}
\label{sec:site}

\subsection{Introduction}
%(\emph{Editors}: Jeremiah Mitchell, Sean Paling, Sergio Calatroni)}
\label{subsec:siteintro}

The focus of this Section is to elucidate some requirements and desiderata for potential future locations for a terrestrial-very-long-baseline atom interferometry experiment. This includes discussion of the presumed infrastructure requirements and environmental feasibility investigations for a large-scale international experiment. As a small survey of current site options globally, we include summaries of presentations from representatives of:
CERN in Geneva, the Boulby underground laboratory in the UK, the Sanford Underground Research Facility (SURF) in the USA and the Callio Lab in Finland. This Section also includes shorter discussions of other deep underground facilities that may be considered.

This document highlights future plans for scaling atom interferometry technology far beyond the confines of small, highly-controlled laboratories. There are various different physical configurations that a future experiment can employ, see the discussions in Sections~\ref{sec:vertical} and \ref{sec:horizontal}. Nevertheless, one may construct a set of common criteria that factor in driving forces from experimental constraints and specifications, and other limiting factors from systematic and environmental noise sources. Our criteria can be neatly split into: site infrastructure requirements and ambient environmental characteristics. Site infrastructure requirements capture experiment-specific needs that we as the experimenters can control, such as local temperature, electric, and magnetic field fluctuations, within constraints imposed by the cost and the knowledge how to mitigate such effects. On the other hand, ambient environmental characteristics encompass all of the location-dependent traits that we cannot directly control, such as seismic and atmospheric fields. These noise sources will require deeper understanding and measurement of each site's geological structure and historical fluctuations to devise stable long-term means of forecasting and mitigation.

Ideally, the experimental collaboration would construct the perfect site for a large-scale experiment by selecting the most isolated environment and building in all control measures required to match the specifications. This would include digging vertical shafts, for vertical atom interferometer configurations, or excavating long deep underground galleries, for horizontal layouts. Realistically the expense of this approach would be far too great if one would like to implement and compare various different techniques and configurations. Instead the pragmatic approach is to examine existing deep underground laboratories and explore their abilities to meet the experimental criteria. This is the approach we suggest here.

\subsection{Infrastructure considerations for long-baseline detectors}
%: Richard Hobson}
\label{subsec:siteinfra}

A long-baseline atom interferometer requires two large spaces, the laser laboratory and the baseline shaft or gallery, which must be connected to each other by a short laser link.
 
The laser laboratory should be a windowless room with a floor area of at least \SI{50}{m^2} and a height of roughly \SI{3}{m} with space for overhead gantries and cable supports. The laser laboratory is needed to house specialised instruments for the manipulation of cold atoms, e.g., lasers, optics, control electronics, and coil drivers. A typical layout (for the MAGIS experiment) is shown in Fig.~\ref{fig:MAGIS_Layout}. The room should have a high-specification heating, ventilation and air conditioning (HVAC) system capable of stabilising the temperature at approximately \SI{22}{\degreeCelsius} with fluctuations below $\pm$\SI{1}{\degreeCelsius}. The HVAC system should be able to handle a heat load of up to \SI{30}{kW} from instruments in the laboratory. Services such as processed cooling water, single-phase and three-phase mains outlets (approx. \SI{35}{kW} total), compressed dry air, and network ports must be available in the laser laboratory. The room must have appropriate engineering and access controls for laser safety, since the room will contain several class 3B and 4 lasers. Routine access for laser-trained users greater than 12 hours per day will be necessary, for instrument installation, maintenance and operation.
 
The main part of the detector will be housed in either a vertical baseline shaft or horizontal baseline gallery, depending on the choice of detector orientation. The longer the baseline, the better the detector sensitivity, up to a limit of a few kilometers. For vertical shafts, a target shaft length of \SI{100}{m} or more is currently being sought for the next-generation detectors similar in scale to MIGA and MAGIS-100. The baseline shaft or gallery will contain the baseline tube (a long, straight, ultra-high-vacuum tube) with between 2 to 10 atom sources attached at intervals along the baseline. Each atom source will require services such as processed cooling water, mains power (approx. \SI{10}{kW}), and various optical fibre and cable connections to the laser laboratory. For a vertical baseline, the atom sources will each occupy a volume of approximately $\SI{1}{m}\times\SI{1}{m}\times\SI{2}{m}$, extending as ``sidearms” from the vertical tube. For a horizontal baseline, a few-meter vertical UHV tube will be required above each atom source to enable long atom free-fall time. A more stable temperature and humidity environment in the baseline shaft or gallery would be beneficial, especially for the atom sources, though local enclosures for the atom sources, internally controlled to \SI{0.5}{\degreeCelsius} pk-pk fluctuations, could be a viable solution for less temperature-stable shafts. Routine and safe access to the atom sources will be required for over 12 hours per day for instrument installation, maintenance and operation.

\subsection{Environmental characterization for site selection}
%: Jeremiah Mitchell}
\label{subsec:siteenv}

Long-baseline atom interferometers are susceptible to various noise sources. Locally to the laser laboratory and atom sources, temperature fluctuations, spurious electric and magnetic fields, and vibrations can feed into the subsystems of the experiment causing noise in the laser light that interacts with the atoms or by directly modifying the atom structure of the Ultra-Cold atom ensembles. These effects have been briefly discussed in Section~\ref{Sec:verticalSensitivity}, see also~\cite{Abe2021} for more details.~\footnote{See Section~\ref{sec:Noise} for a more complete discussion of possible noise sources.} For very-long-baseline detectors the science signal we search for could vary along the extent of the baseline. This requires us also to consider ambient large-scale effects such as seismic and atmospheric density perturbations that cause noise in differential measurements of the phase shift of the atom interferometers along the baseline. These ambient effects, which have been referred to as Gravity Gradient Noise (GGN), become crucial terrestrial limitations in the lower frequency domain. Other large experiments including LIGO, Virgo and the LHC have also needed to explore these potential backgrounds~\cite{Abbott2017,Harms2019}, though these backgrounds are not yet limiting significantly their respective science cases. For atom interferometers the current generation of experiments including MIGA, MAGIS-100, ZAIGA, and the VLBAI plan to explore GGN experimentally as this will unlock the full science reach of such long-baseline atom interferometers~\cite{Canuel2018,Abe2021,Zhan2019}.

By measuring the local effects mentioned above we can establish what levels of passive shielding and active tracking will be required to meet experimental specifications for a given noise budget. This can be quickly done for a potential site with a survey of temperature and pressure using thermometers and pressure/humidity sensors, magnetic fields using flux-gate magnetometers, and arrays of accelerometers to characterize vibrations above \SI{1}{Hz}. It should be noted that a single set of measurements would only provide enough information to establish the DC, or static noise backgrounds. Time-dependent fluctuations of these fields can also lead to noise in the atom interferometer thus longer-duration surveys should be carried out. Once the amplitudes and fluctuations of these fields are known we can specify the levels for thermal enclosures, magnetic shields, and vibration isolation for the various subsystems of the experiment. For some references for expected noise levels in MAGIS-100, see~\cite{Abe2021}, and for measurements at the CERN site discussed below, see~\cite{Arduini2023}.

One should also examine the characteristics of the larger ambient environment surrounding the site. The effects of most interest for the ${\cal O}(1)$~Hz domain are seismic field fluctuations sourced by: pressure waves (P-waves), transverse waves (S-waves) and Rayleigh waves (surface waves), atmospheric fluctuations, sound waves reflected from the Earth's surface, and advected temperature gradients. The primary concern from these fluctuating noise sources is a density perturbation leading to a fluctuating gravitational field and time-dependent GGN coupled to the Ultra-Cold atom ensemble through its free-fall trajectory in the perturbed gravitational field. One mode of this coupling through seismic Rayleigh wave channels was discussed in Section~\ref{sec:Battling} and further details on investigations into seismic GGN can be found in references~\cite{Junca2019,Mitchell2022,Badurina2023a}. Atmospheric fluctuations couple in a similar manner and are currently being investigated. In order to characterize the site for these environmental backgrounds a seismic survey can be carried out to provide histograms of the power spectral density of ground motion. For vertical configurations synchronized borehole seismometers can be arranged in the shaft to profile the vertical dependence of the ground motion. Atmospheric effects can be best understood through large area microbarometer surveys, LIDAR measurements, or databases of historical meteorological data. With ambient noise levels collected the feasibility of the site can be assessed, an example of this was done for the CERN vertical shaft PX46~\cite{Arduini2023}. Mitigation for these low-frequency noise sources will most certainly require passive and active isolation with large auxiliary environmental monitoring systems and filtering techniques.

To maximize the sensitivity of a TVLBAI the proposed site should meet noise sensitivity limits informed by large sensor arrays and local surveys. Noisy environments may be suitable so long as the appropriate active monitoring systems and data filtering techniques are employed. From current investigations the most suitable method would be to locate the experiments deep-underground and site a location with very regular and predictable weather and seismic activity.

\subsection{Potential deep underground laboratories}

We now review studies that have been made of some prospective sites.

\subsubsection{CERN}
%: Sergio Calatroni}
\label{sec:CERN}

The Large Hadron Collider (LHC) at CERN is sited in an underground tunnel at a depth varying between roughly \SI{80}{m} and \SI{140}{m}, due to the changing altitude of the overburden terrain. Several vertical shafts connect the surface to the LHC tunnel for personnel access, for lowering and raising accelerator components, for utilities and control systems.  A feasibility study was conducted with the support of the Physics Beyond Colliders initiative at CERN and the PX46 access shaft was identified as the most suitable location for installing a \SI{100}{m} class vertical atom interferometer~\cite{Arduini2023}. 

\begin{figure}
    \centering
    \includegraphics[width=0.6\textwidth]{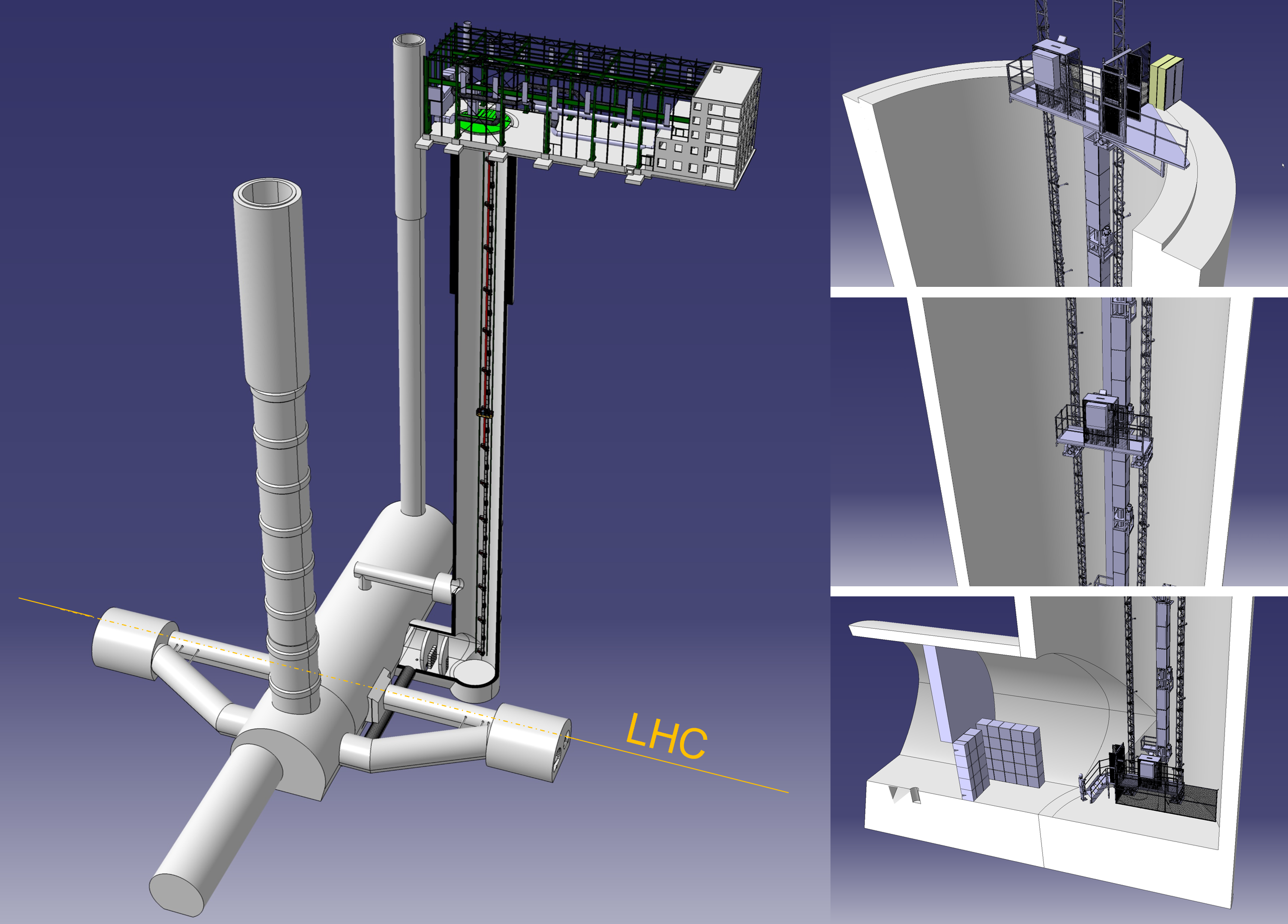}
    \caption{3D model of the underground civil infrastructure at Point 4 of the LHC. The vertical atom interferometer is in the PX46 shaft. There is concrete shielding in the gallery connecting to the main cavern. A fast and safety-proof elevator platform surrounds the experiment and is used for assembly, operation and escape in case of hazards~\cite{Arduini2023}.}
    \label{fig:CE:proposed} 
\end{figure}

This access shaft, shown in figure~\ref{fig:CE:proposed}, is towards the north of the LHC ring, is \SI{143}{m} deep and has a diameter of \SI{10.10}{m}, is fully lined with reinforced concrete and its surface access is inside a large open space technical building while at its bottom it is connected via a short gallery to a large side cavern where high-power radiofrequency (RF) equipment is located. The shaft is used mostly for lowering accelerator components and is otherwise free of any equipment. Following the approach discussed in the preceding sections, the study team analyzed the available technical infrastructure, the environmental aspects relevant for an atom interferometer, and any further requirements or limitations arising from being part of a large operating particle collider. No showstoppers have been identified. All details are discussed in the feasibility study report~\cite{Arduini2023}, which are summarized very briefly in the following.

The technical building where the PX46 shaft is housed in has all relevant general utilities such as electricity supply, cooling water, etc., with capacities exceeding the requirements set out in Section~\ref{subsec:siteinfra}. The floor space is in excess of \SI{500}{m^2}, which should be largely sufficient for housing the laser laboratory and any needed access control system.

An extensive experimental campaign has been conducted using seismometers and geophones as measurement probes showing that the site largely satisfies the requirements in terms of surface acceleration fluctuations in the range of \SI{100}{\milli\hertz} to \SI{100}{\hertz}, except for some disturbances at \SI{50}{\hertz} that would require some mitigation. The amplitude of the surface vertical displacement measurements at frequencies $< \SI{1}{\hertz}$ generally lies below the NHNM~\cite{Peterson1993}, albeit with some fluctuations.

From previous analysis and estimates, magnetic field fluctuations below $\delta B \leq \SI{100}{\pico \tesla/\sqrt{\hertz}}$ would be desirable~\cite{Abe2021,Arduini2023}. These constraints apply most strictly within the peak detector sensitivity band \SI{50}{\milli\hertz} to \SI{10}{\hertz}. Electromagnetic noise has been measured at PX46 with the use of fluxgate magnetometers at the lower frequencies and inductive pick-up coils at the higher frequencies, both at the surface and at the bottom of PX46, closer to the LHC active devices and components. The measured noise is largely within the specified limits. The ramping of the high-field dipole magnets of the LHC results in a small change of the DC background, within the acceptable limit of $\gtrsim\SI{50}{\nano\tesla}$.

The atom interferometer should be accessible at all times by the operators, and in view of this requirement the study team performed an in-depth assessment of the situation and identified any needed mitigations to guarantee the operators safety in any event during the LHC machine operation. The main potential concerns identified~\cite{Arduini2023} would be radiation due to LHC beam losses, helium release from cryogenic apparatus leading to oxygen deficiency hazards, and hazard due to fires potentially developing in the nearby LHC equipment. All these aspects have been studied in detail. Simulations of the dose to personnel in case of beam loss have been performed, leading to the identification of the need for a further shielding wall at the bottom of PX46, which is illustrated in Fig.~\ref{fig:CE:proposed}.  With a suitable set of doors and an access control system, this wall allows furthermore physical separation from the LHC environment, guaranteeing access to the interferometer at any time. Helium release and fire hazards have been found to be of no major concern, provided that a sufficient fast escape route is available for the operators. A fast elevator, able to operate in any conditions via redundant equipment, has been proposed and studied by a consultant company. This elevator may be used for the assembly of the interferometer itself, during routine operation and for escape, either from the top or from the bottom of the tunnel. The elevator surrounds the experiment itself, thereby leaving enough free space in the access shaft allowing routine lifting of LHC components if required.

Cost and schedule estimates have also been performed, and are discussed in the feasibility study \cite{Arduini2023}. If the site were selected by a collaboration for building an atom interferometer, and the proposal accepted by CERN management, the next step would be launching a detailed technical design of the facility, in order to start adaptation work as early as possible, compatibly with the LHC schedule, so as to make the site available to the experimental community.

\subsubsection{Boulby underground laboratory}
%: Sean Paling}
\label{sec:Boulby}

\begin{figure}
    \centering
    \includegraphics[width=.8\textwidth]{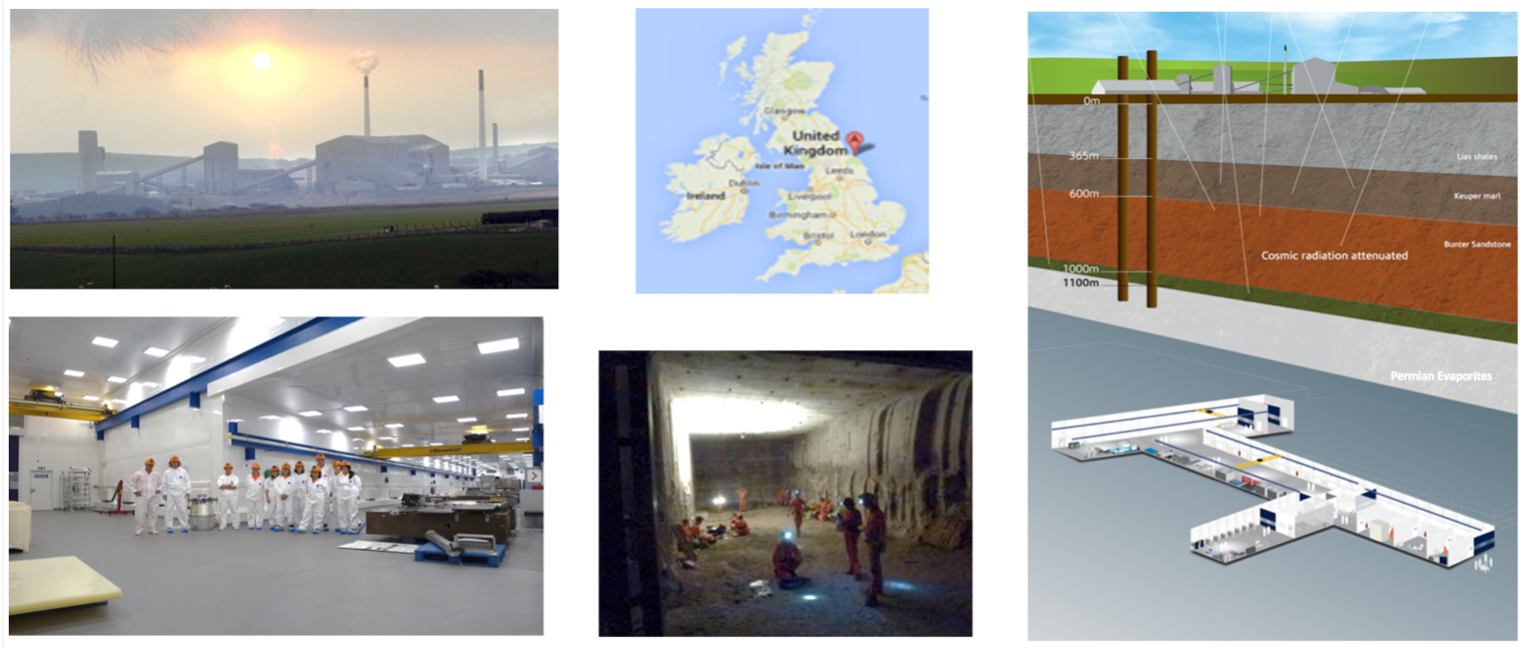}
    \caption{The Boulby underground laboratory, the UK's deep underground science facility operating in a working mine in the North-East of England.}
    \label{fig:boulby}
\end{figure}

The STFC Boulby underground laboratory is the UK’s deep underground science facility, and one of the few facilities in the world suited to projects and studies requiring an ultra-low background radiation environment and/or access to the deep underground environment. Boulby operates in a working polyhalite and salt mine in the North of England. At \SI{1100}{m} deep, with low-background surrounding rock, good support infrastructure, experienced facility staff and a strongly supportive host in the mine operators (ICL-UK), the Boulby facility is an ideal place for low-background and deep underground science studies, see Fig.~\ref{fig:boulby}.

The Boulby facility has been in operation since the late 1980s and has seen significant development in its infrastructure over this time. The current Boulby facility consists of its main \SI{4000}{m^3} underground laboratory which is fully supported with \SI{1}{Gb} internet capability, \SI{10}{T} lifting capacity, power, cooling and air filtration. The facility is operated as a class 10k clean room throughout and 1k in its Boulby UnderGround Screening (BUGS) material screening facility. Directly outside the facility is the Outside Experimentation Area (OEA), a \SI{3000}{m^3} space in a bare salt cavern which is provided for various science projects and teams requiring access to the deep underground environment and geology. The facility also has a \SI{400}{m^2} surface support facility for administration and science/visitor support. Finally, there is a \SI{300}{m^2} surface warehouse facility for incoming goods storage and handling. 

For over 3 decades the Boulby facility has hosted world-leading astro-particle physics studies, in particular searches for Dark Matter that resulted in the development of key detector techniques including the ZEPLIN two-phase Xenon detector system, which subsequently became the technique used by world-leading detectors such as the LUX-ZEPLIN (LZ) detector operating in the SURF facility in the USA. The current facility continues to host Dark Matter R\&D projects (the CYGNUS and NEWS-G detectors) and also operates a suite of high-sensitivity germanium detectors, XIA surface alpha counting systems, Rn emanation systems and an ICPMS system - all part of the BUGS facility providing world-class material screening capability for ultra-low background science projects. Beyond astro-particle physics and ultra-low background science studies, the Boulby facility has a rich and varied range of multidisciplinary science studies underway in the fields of geology, seismology, environmental carbon reduction, sustainable energy technologies, studies of life in extreme environments and technology development for planetary exploration, mining and more.

In the future there are plans underway to expand greatly the facilities and science underway at Boulby. The UK is currently looking to create a new $\sim$ \SI{30000}{m^3} underground facility close to the current laboratory to host major new particle and astro-particle physics studies (in addition to an expanded multidisciplinary science programme) from 2030 onwards. Aside from studies to be operated in the current and new facilities, Boulby is also inviting proposals from science projects interested in the wider infrastructures and environments provided by the Boulby mine. This includes projects interested in access to the shafts at Boulby, of which there are three: two \SI{1.1}{km} primary shafts for people access and rock/product removal, and one \SI{180}{m} shaft, which is a tailings shaft used for disposal of water from the main mine and its processing plant.

Boulby is keen to support the interest of the AION-100 and AION-km projects in utilising the shafts at the Boulby site. Preliminary studies have already been undertaken to assess the space needs and availability for these projects. The next key steps are to undertake a detailed study of the seismic and gravitational backgrounds at the facility, in addition to further exploring logistics needs and challenges. A programme of work to install instrumentation to assess backgrounds in the 3rd shaft at Boulby (the \SI{180}{m} tailings shaft) to assess its suitability for AION-100 is expected to begin this year.

\subsubsection{The Sanford Underground Research Facility}
%: Jaret Heise}

The Sanford Underground Research Facility (SURF) has been operating for more than 15 years as an international facility dedicated to advancing compelling multidisciplinary underground scientific research in rare-process physics, as well as offering research opportunities in other disciplines~\cite{Heise2022}. SURF laboratory facilities include a surface campus as well as a significant underground footprint consisting of more than \SI{15}{km} of accessible areas across seven main elevations. Enhanced services are available on the 4100-foot level, and especially the 4850-foot level, to support significant research needs and laboratory facilities. In particular, campuses at the 4850-foot level (\SI{1500}{m}, 4300~m.w.e.) host a range of significant physics experiments, including the LUX-ZEPLIN (LZ) dark matter experiment and the MAJORANA DEMONSTRATOR neutrinoless double-beta decay experiment. The CASPAR nuclear astrophysics accelerator recently completed the first phase of operation and is set to resume activities in 2024. SURF is also home to the Long-Baseline Neutrino Facility (LBNF) that will host the international Deep Underground Neutrino Experiment (DUNE). SURF offers an ultra-low background environment, low-background assay capabilities, and ultra-pure electroformed copper is produced at the facility. 

\begin{figure}
    \centering
    \includegraphics[width=.7\textwidth]{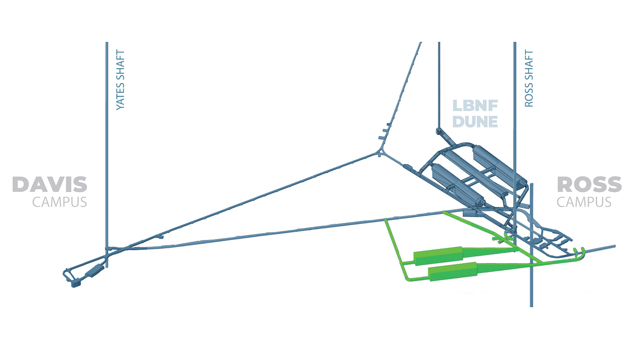}
    \caption{Current and proposed underground laboratory space at SURF, including up to two new caverns on the 4850-foot level (\SI{100}{m} L $\times$ \SI{20}{m} W $\times$ \SI{24}{m} H). There are more than \SI{15}{km} of accessible areas across seven main elevations as well as vertical options.}
    \label{fig:surf}
\end{figure}

SURF is preparing to increase underground laboratory space. Plans are advancing for construction of new large caverns (nominally \SI{100}{m} L $\times$ \SI{20}{m} W $\times$ \SI{24}{m} H) on the 4850-foot level (\SI{1500}{m}, 4200~m.w.e.), aligned with the timeline for next-generation experiments (approximately 2030). SURF has also performed an initial evaluation for creating a vertical facility. Candidate areas were identified based on preliminary requirements, with options for medium scale (\SI{100}{m}) and large scale (\SI{1000}{m}) vertical facilities, see Fig.~\ref{fig:surf}.

SURF is a dedicated research facility offering opportunities and space for diverse science, including new opportunities for vertical and horizontal tunnels to accommodate infrastructure associated with atom interferometry science.

\subsubsection{Callio Lab}
%Julia Puputti}

Callio Lab is a multidisciplinary underground research centre coordinated by the University of Oulu,  operating at the Pyhäsalmi mine in Finland~\cite{Joutsenvaara2021}.  Callio Lab was established in 2015 to continue the work started in the year 2000 by the Centre for Underground Physics in Pyhäsalmi (CUPP). The site has since hosted research and experiments in a wide range of fields, such as mine reuse, geothermal research, underground health and safety, circular economy and muography. The local research team offers coordination, cooperation, networking and facilitation, and is a founding member of the European Underground Laboratories Association. 

\begin{figure}
    \centering
    \includegraphics[width=0.45\textwidth]{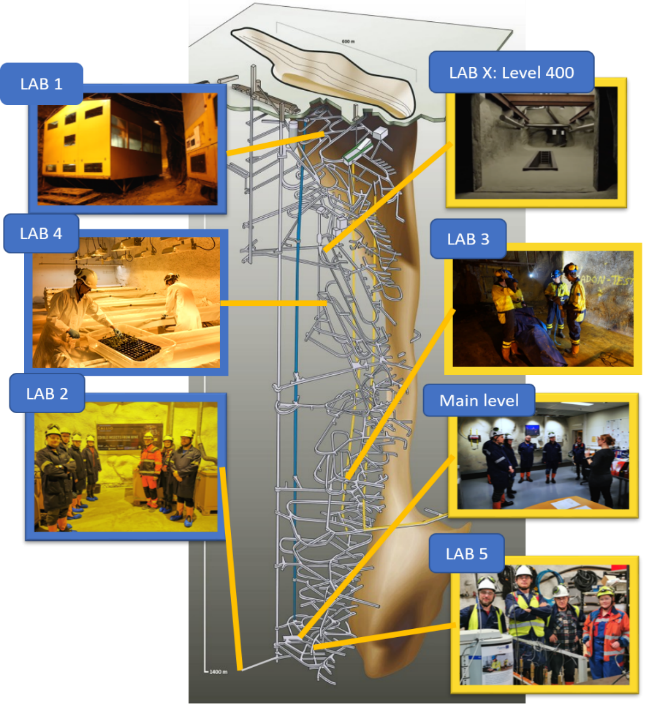}
    \caption{3D model of the Callio Lab tunnel network with insets of the various deep underground labs at the mining site.}
    \label{fig:callio}
\end{figure}

The Pyhäsalmi mine opened in 1962, producing copper, zinc and pyrite. On-site there are three major operators, the Pyhäsalmi Mine Ltd mining company, the post-mining activity coordinator Callio, and the University of Oulu Callio Lab. Underground mining ended in autumn 2022 and work to utilise the available mine infrastructure as a pumped hydro energy storage has begun. The establishment of a permanent reuse operator for the underground mine ensures the site will be maintained and scientific activities at the Pyhäsalmi mine can continue. The excavation and building of new laboratory spaces (horizontal and vertical) is also possible, as shown in Fig.~\ref{fig:callio}. 

Access to facilities and different levels is possible via the elevator shaft (3-minute ride to the main level) or the inclined tunnel. Driving from the surface to the bottom of the mine at \SI{1.43}{km} takes about 40 minutes. As mining has only recently ended, the available services and infrastructure are still in good condition. Electricity and fibre optic connections are readily available, the re-use sites are well characterised, and due to the long history of the mine and scientific activities, extensive historical data sets are available. The mine also has a microseismic monitoring network and preliminary analysis suggests that the already low seismicity has decreased significantly since the end of underground extraction. 

During the CUPP years, the site was a participant in two FP7 design studies for a next-generation neutrino observatory. Now with the certainty of the underground mine remaining open and accessible even after closure, the potential of Callio Lab as a site for large scale experiments endures.

\subsubsection{Other sites}

\paragraph{Low Noise Underground Laboratory: LSBB}
The Laboratoire Souterrain {\`a} Bas Bruit (LSBB, \url{https://lsbb.cnrs.fr/}) is an international underground research facility located in Rustrel-Pays-d'Apt, France~\cite{Gaffet2009}. It is a converted missile control centre featuring \SI{4}{km} of horizontal galleries dug into limestone. Of these galleries include \SI{250}{m} of anti-blast gallery at depth of \SI{280}{m} as well as two sets of approximately perpendicular galleries with lengths of $\SI{800}{m}\times\SI{1250}{m}$ and $\SI{150}{m}\times\SI{150}{m}$, respectively. The site is designed for low-noise academic and industry research and is the host location of the MIGA experiment utilizing the perpendicular \SI{150}{m} galleries for horizontal atom interferometry (see Section~\ref{sec:MIGA}).

The site also hosts experiments for 3D seismic observations, testing very high sensitivity gravity sensors, qualification of field-programmable-gate-arrays (FPGAs) and muon tomography, with plans for future expansion.

\paragraph{Canfranc Laboratory} 
%(\emph{Contributor}: Diego Blas)
The Canfranc Underground Laboratory (LSC, \url{https://lsc-canfranc.es/en/home-2/}) is a world-class deep underground laboratory with a suite of experiments that require very low environmental radiation levels on topics related to neutrino physics, dark matter and other phenomena.
It is located at Canfranc, a village in the Spanish Pyrenees that has
a train connection to the Spanish network and is easily accessible by car and truck. The LSC is 800 meters below ground, located between the Somport road tunnel and an old railway tunnel, which are about \SI{8}{km} long on the Spain-France border. The LSC is the second largest deep-underground laboratory in Europe, with a total area of about \SI{1250}{m^2} and a volume of about \SI{10000}{m^3}, and is easily accessible by road or through the old train tunnel. 

\begin{figure}
\centering
\includegraphics[width=0.9\textwidth]{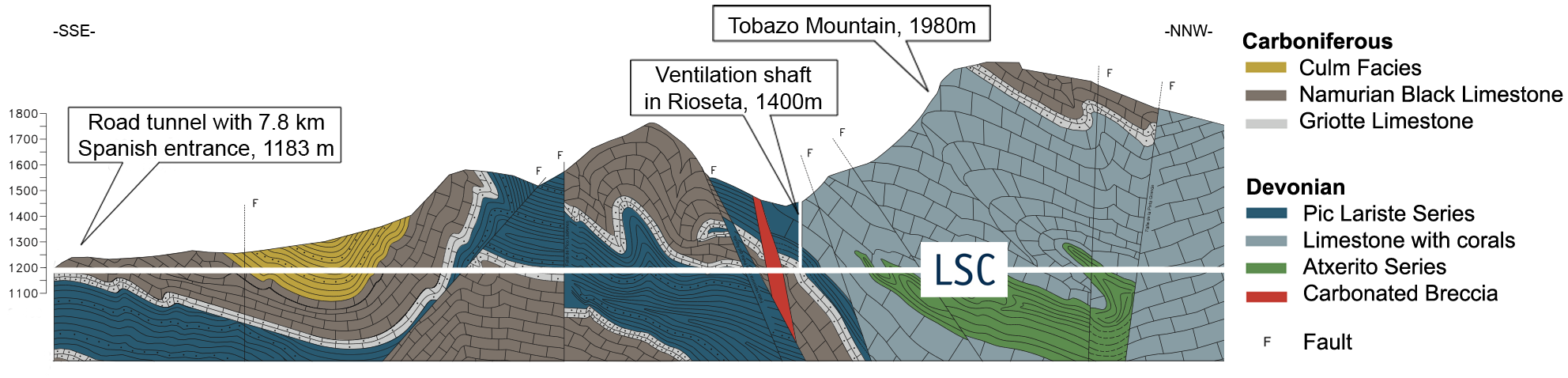}
\caption{A diagram of the LSC at Canfranc, showing the horizontal gallery and the vertical shaft used for ventilation~\cite{PerezPerez2022}.}
\label{fig:LSC}
\end{figure}

Close to the LSC there is a ventilation shaft with a length of a few hundred meters, including vertical sections of more than $\sim$\SI{100}{m}, as seen in Fig.~\ref{fig:LSC}. There are also horizontal tunnels that may be considered. The seismic noise (that impacts directly the GGN level) at LSC was measured while it was being considered as a potential site for a third-generation gravitational wave detector~\cite{Beker2012}, and it was found that LSC has a relatively low seismic noise level, close to the Peterson NLNM. This noise measurement includes the contribution from the traffic through the adjacent road
tunnel, implying that this should not be a concern for the performance of a TVLBAI experiment at LSC. More recent measurements can be found in~\cite{Bulik2018}. We also note that the remoteness of LSC suggests that the impact of anthropogenic electromagnetic noise should also be small.

To date there has not been more exploration of the feasibility of this site for future large interferometers, though contact has been established with the Director of the LSC (Carlos Pe\~na-Garay) who is very positive about exploring in depth this possibility. According to him, LSC has also base funding which could help to prepare the local infrastructure and technicians experts in several aspects of precision experiments.

%{\bf COMMENT BY D. BLAS: I DON'T KNOW IF WE HAVE RIGHTS TO REPRODUCE THESE IMAGES.}
%{\bf Comment from S. Calatroni: reference \cite{PerezPerez2022} is published from MDPI, they adhere to the STM permissions protocol, and as such the figure can be republished without further permission. See \url{https://www.stm-assoc.org/intellectual-property/permissions/permissions-guidelines/} }

%\begin{figure}
%\centering
%\includegraphics[width=0.7\textwidth]{figures-10/ggncanfranc.png}
%\caption{Seismic noise spectra at various sites. The shaded regions are bounded by the 90 and 10 percentiles while the solid curves represent the mode or most common PSD
%value in each frequency bin at different locations including the LSC in Spain. The dashed black curves indicate the new Peterson high and low
%noise models From \cite{2012JPhCS.363a2004B}.}
%\label{fig:ggnC}
%\end{figure}

\paragraph{AION Porta Alpina} 
%(\emph{Contributors}: Ellis, Buchmueller, Lombriser)

\begin{figure}
\centering
\includegraphics[width=0.9\textwidth]{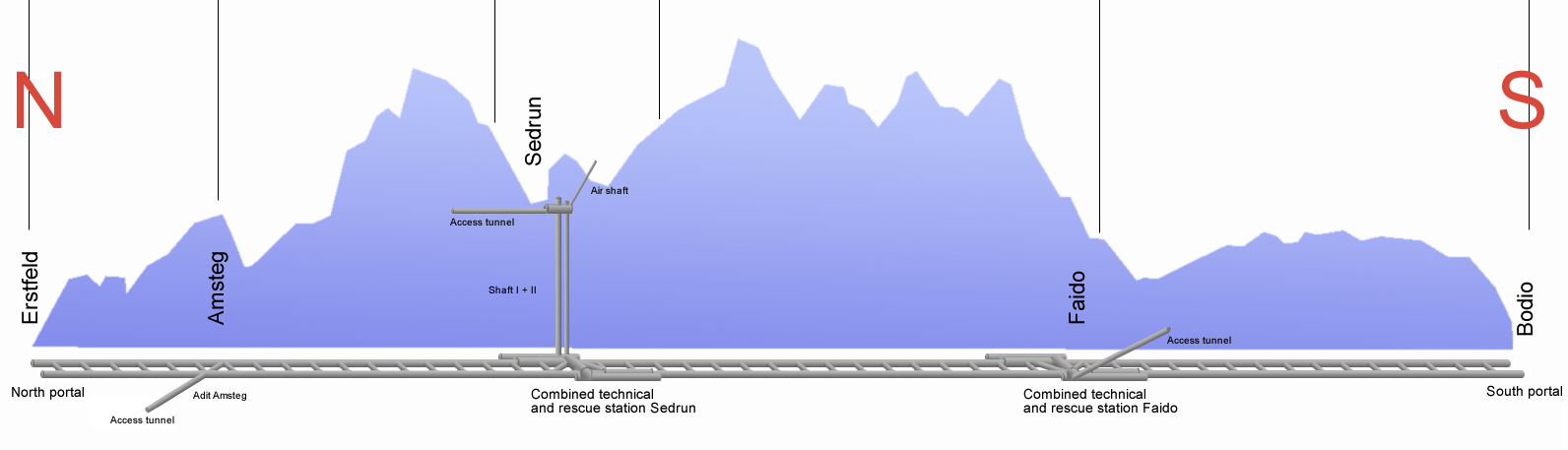}
\caption{A diagram of the Gotthard Base Tunnel running from North to South under the Swiss Alps, showing the horizontal gallery and the pair of 800-m vertical shafts that provide access from Sedrun to the site of the envisioned ``Porta Alpina'' underground railway station.}
\label{fig:Gotthard}
\end{figure}

The Gotthard Base Tunnel is a \SI{57}{km} long railway tunnel in Switzerland connecting the north and south sides of the Alps at a depth of ${\cal O}$(km). A pair of vertical access shafts from the village of Sedrun present a promising site for the deployment of a ${\cal O}$(km) atom interferometer. Built to expedite the construction of the base tunnel between 1999 and 2016 and to provide ventilation, the two \SI{800}{m} vertical shafts are located at the end of a \SI{1}{km} horizontal gallery with road access from Sedrun that extends into the mountain. With inner diameters of \SI{8.6}{m} and \SI{7}{m}, respectively, the shafts lead down to a multi-function station within the base tunnel, as seen in Fig.~\ref{fig:Gotthard}. They are currently used for maintenance access and could be used as an emergency exit. {\it Prima facie}, there is sufficient space in either shaft to accommodate a long-baseline atom interferometer. We note in addition that four caverns (\SI{38}{m}~$\times$~\SI{10}{m}~$\times$~\SI{5.5}{m} each) connected to the station have been constructed for the envisioned ``Porta Alpina'' underground passenger railway station, a project that was placed on hold in 2007 with new negotiations initiated in 2020 but deferred again in 2023. Influential national, cantonal, and municipal politicians in Switzerland have expressed interest and support for the vision of a ``Porta Alpina'' atom interferometer experiment. 
    
To date there has not been more exploration of the feasibility of this site for future large interferometers. An inspection of the site to prepare for a possible feasibility study is planned for the first quarter of 2024.
%Autumn 2023.

\paragraph{Mines in Sweden}

Some mines in Sweden have been considered as possible underground sites for a
proposed large underground water Cherenkov neutrino experiment using a beam from the European Spallation Source.
These include the Garpenberg mine, which has a 800~m deep shaft of around 4~m in diameter, and the Zinkgruvan mine. The latter does not currently have such a shaft, but one would be dug in the future for the extraction of rock debris from the excavation of the large detector caverns if the neutrino experiment is located there. 

\section{Supplementary Topics}
\label{sec:above}
%{\bf 5 pages guideline \\
%Each subsection around 0.5 - 1.5 pages}
%\subsection{Editors:Lisa Woerner}

\subsection{Introduction}
This Section outlines research objectives, technological necessities, synergies, and other possibilities for deploying long-baseline systems. 
As such, this chapter is an aggregation of separate topics, which are not necessarily connected to one another. 
Consequently, each Subsection outlines their individual impact on terrestrial systems or their research objective.

The importance of this Section lies in drawing attention beyond the obvious technology developments and research objectives outlined in the other Sections of this paper. 
Here, we span lessons learned from other terrestrial long-baseline systems, enhancements by usage of additional technologies, applications to geodetic research, usage for early warning systems, and synergies with further fundamental research.

\subsection{Noise Sources}
\label{sec:Noise}
%Here: Discussion of possible sources and how other long baseline systems, such as LIGO / VIRGO, have worked that in the past. 
% Jeremiah, Soumen, Jan, if you want to change the structure or responsibility, please do so! 
\subsubsection{Newtonian Noise Sources}
%Here: Jeremiah, I guess, this would be mainly yours
Newtonian noise (NN) sources can be summarized as density fluctuations that modulate the local gravitational potential of the Earth \citep{Saulson1984}. These in turn affect the accelerations of test masses moving through the background gravity field. Specific to the atom interferometry experiments being discussed here, the atomic centre-of-mass motion during an interferometric sequence acquires a quantum phase shift as the propagation trajectories of the quantum states sample the either static or dynamic variations of space-time. A general formalism for capturing these effects is through a continuity equation relating the gravity potential to the source density perturbation:
\begin{equation}
    \delta \phi_{g} = G \int_V \frac{\delta \rho (\mathbf{r},t)}{|\mathbf{r} - \mathbf{r}_0|} \mathrm{d}V.
    \label{eqn11_1}
\end{equation}
At leading order in a Mach-Zehnder atom interferometer this causes a phase shift
\begin{equation}
    \delta \phi = \mathbf{k}(\mathbf{g} + \delta\mathbf{g})T^2,
\end{equation}
where $\delta \mathbf{g} = -\nabla (\delta \phi_g)$.
As one would imagine there are many potential sources of density fluctuation. The density fluctuation term $\delta\rho(\mathbf{r})$ in Eq.~(\ref{eqn11_1}) for seismic NN can be expressed as $-\nabla(\rho_s(\mathbf{r})\chi(\mathbf{r},t))$, where $\rho_{s}(\mathbf{r})$ is the density of the soil surrounding the test mass, and $\chi(\mathbf{r},t)$ is the seismic displacement of the medium. Similarly, the density fluctuations due to changes in atmospheric pressure ($\delta \rho^\mathrm{P}$) or temperature ($\delta \rho^\mathrm{T}$) in the vicinity of the test masses can be expressed as $\delta \rho^{\mathrm{P}}(\mathbf{r},t) = \frac{\bar{\rho}_{a}}{\gamma \bar{p}_{a}}\delta p_a(\mathbf{r},t)$ and $\delta \rho^{\mathrm{T}}(\mathbf{r},t) = -\frac{\bar{\rho}_{a}}{\bar{T}_{a}}\delta T_{a}(\mathbf{r},t)$, where $p_a(\mathbf{r},t)$ and $T_a(\mathbf{r},t)$ are the air pressure and temperature; $\gamma$ is the adiabatic index; and $\bar{\rho}_a$, $\bar{p}_a$, and $\bar{T}_a$ are the average density, pressure, and temperature of the atmosphere, respectively. We can simplify matters by considering effects within the categories of:
\begin{itemize}
\setlength\itemsep{0em}
    \item Anthropogenic Sources
    \begin{itemize}
        \item Foot-fall
        \item Motors, pumps, fans
        \item Vehicles, lifts
    \end{itemize}
    \item Earth Generated
    \begin{itemize}
        \item Seismic
        \item Atmospheric
        \item Oceanic
    \end{itemize}
\end{itemize}

The dominance of anthropogenic sources of noise is observed worldwide for frequencies greater than 1 Hz \citep{BonnefoyClaudet2006}. A trough in global seismic noise model is observed in the band 1-2 Hz which marks the transition from natural to anthropogenic sources \citep{Peterson1993}. Most anthropogenic seismic sources tend to be sporadic, with the exception of motors and fans that generate monochromatic peaks in the seismic displacement spectrum, and are a well known source of noise for second-generation terrestrial GW detectors like the Advanced LIGO and the Virgo detectors \citep{Coward2005,Fiori2020}. Foot-fall is important in the frequency range of \SIrange[range-phrase=--]{1}{3}{Hz}, but can have unpredictable resonance effects at times of heavy foot traffic~\cite{Ekimov2006}. The gravitational effects of moving vehicles and lifts can be modelled by considering masses with constant acceleration or velocity that lead to predictable signals in a detector. For a recent study of related issues in connection with AION-10 arising from the unpredictable movements of people and animals (Random Animal Transients, RATs) near an atom interferometer, and possible mitigation strategies, see~\cite{Carlton2023}. Motors and fans have higher frequency components between \SIrange[range-phrase=--]{60}{120}{Hz} in the USA and \SIrange[range-phrase=--]{50}{100}{Hz} in Europe. Simple mitigation measures such as tracking local motion around the atom interferometer and keeping people far away from the instrument during operation can have quite significant effects on noise suppression. An application of such a noise suppression scheme applied to ground-based laser interferometers can be found in~\cite{Coughlin2016}.

Earth-generated sources dominate the lower frequency domain below a few Hz as large-scale effects have wavelengths on the order of \SIrange[range-phrase=--]{1}{1000}{km}. All terrestrial sources are also highly site dependent and vary seasonally \citep{Stutzmann2000,Stutzmann2009}. This requires constant monitoring and calibration. It also makes generating completely general models difficult as a given location for a TVLBAI may have ambient background composed of complicated mixtures of seismic, atmospheric, and oceanic noise coupled together. A pragmatic approach to understanding these sources is to focus on their characteristic differences. Seismic waves come in many varieties including: P-waves (primary longitudinal), S-waves (secondary transverse), Love and Rayleigh surface waves, where the order is with respect to wave speed. Seismicity can also be categorized into long-lived stationary signals and transient signals such as earthquakes and fault slips. Atmospheric noise can be separated into pressure/infrasound waves and advected temperature gradients \citep{Creighton2008}. There also exist transient modes from the atmosphere in the form of sonic booms and shock waves.

\subsubsection{Further Noise Sources}
%Here: Jeremiah / Soumen / Jan, this would be anything else you might want to add
Studies in \citep{Harms2019} (Section 6.5) show that the noise due to an oscillating point mass close to the GW test mass is proportional to the oscillation amplitude and suppressed by the relative distance between the two objects. Hence, the vibration amplitude must be high and/or the object should be located close to the test mass in order to compromise the strain sensitivity. One such study was recently conducted at the KAGRA \citep{Akutsu2021} underground GW facility. The cooling infrastructure is known to generate large mechanical vibrations due to cryocooler operation and structural resonances of the cryostat. As cooling system components are relatively heavy and in close proximity to the test masses, oscillations in the gravitational field induced by their vibrations could contaminate the detector sensitivity. Results show that, while this noise does not limit the current detector inspiral range, it will be an issue in the future when KAGRA improves its sensitivity \citep{Bajpai2023}. Similar studies of the Newtonian Noise (NN) generated by the vibrations of the cryogenic shielding or from boiling cryogenic liquids have been studied by \citep{Bonilla2021} in relation to the development of LIGO Voyager \citep{Adhikari2020}.

Besides these mechanical sources of vibrations, water flow in underground caverns could also be one of the sources of NN. KAGRA removes about 1200 tons of water per year through the drainage pipes near one of the test masses. Although NN produced by water compression can be considered negligible, the water surface profile can vary during its flow, thus leading to a density variation. Preliminary models show that NN from turbulent water flow could limit KAGRA’s sensitivity \citep{Somiya2019}. The proximity of water flow may also be an issue for the Boulby site, see Section~\ref{sec:Boulby}.

\subsubsection{Noise Surveillance}
\label{sec:Mitigation}

%Here: Soumen / Jan, this seems like something you spoke about
The gravitational coupling between the density fluctuation and test masses prevents NN from being mechanically shielded. One of the possible ways of mechanically reducing NN is by reducing the effect of seismic waves on the infrastructure in the vicinity of the test masses. Studies by \citep{Palermo2016} have shown the use of seismic metamaterials to reduce the amplitude of incoming seismic waves. Such a cloaking system can be realized by building periodic structures around the test masses with dimensions of the order of the wavelength of the seismic waves of interest.

However, the most popular schemes for NN cancellation are active noise mitigation systems.
The fundamental concept underlying active noise cancellation is to monitor a noise source with some witness sensors and then use these data to reconstruct the NN dominated strain output. The linear relationship between the observed seismic displacement and the NN makes it possible to design optimal linear filters (Wiener filters) by minimizing the squares of the error between the witness and target channels (strain output) \citep{Wiener1949}. A first such offline scheme for NN cancellation was proposed by Cella \citep{Cella2000}. Later applications to real seismic fields at the LIGO detectors were realized by deploying seismometers near the test masses~\citep{Driggers2012,Coughlin2018}. One of the challenging problems before a noise cancellation system can be deployed is estimating the sensor locations for optimal noise cancellation. The complications arise from inaccuracies in the reconstruction of the seismic cross-correlation field in the vicinity of the test mass and the presence of local solutions to optimization problem. Recent studies at the Advanced Virgo GW detector which makes use of Gaussian Process Regression for cross-correlation modeling and global solvers like the Particle Swarm Optimizer have made successful attempts at designing such optimal seismometer arrays for NN cancellation \citep{Badaracco2019,Badaracco2020}. In cases when the seismic field is dominated by Rayleigh waves, noise cancellation using tiltmeters have also been found to be efficient \citep{Harms2016}. The cancellation performance of such a system relies on the cross-correlation values between the witness channels and the target channel. A noise reduction factor of about 3-5 is expected for future third generation underground GW detectors for frequencies greater than 2 Hz. For frequencies below 1 Hz, when the seismic field is dominated by Rayleigh waves, a greater correlation between the witness channels is expected, and consequently a greater noise reduction factor is foreseen.
\subsubsection{Avoidance of Noise}
%Possibly: Aviodance of Noise Sources by choice of site / envrironment
%Here: Again, Soumen / Jan / Jeremiah, maybe you could enter something here?
An obvious way to reduce the impact of anthropogenic seismic noise to the detector output is to build the detector underground and in a seismically quiet region. Seismic noise below 1 Hz is characterized by surface waves with long wavelengths and hence suffers little attenuation while going underground. However, a significant reduction can be observed for frequencies greater than 1 Hz. The attenuation factor is dependent on the type of seismic sources and the geology at the site. Studies of underground seismic noise have shown that a geological environment with a layer of soft soil on hard rock acts as a low-pass filter, thus attenuating most of the surface noise \citep{Koley2022}. In such geological scenarios, the underground seismic noise above 1 Hz can be described as a mixed stochastic background of body waves. Consequently, the impact of NN on the detector output is also reduced. Although there is some contribution of NN originating from seismic displacement on the surface of the medium, it has been observed that the relatively small wavelengths of the seismic waves near the surface results in a cancellation \citep{Bader2022}. In regions where such a soft soil over hard rock geology is not observed, amplitude reduction of about 5-10 can be expected and little difference in the composition of the seismic wavefield between the surface and underground is observed \citep{DiGiovanni2021}. Consequently, the contribution of NN is greater than soft-soil over hard-rock geology.

For the natural sources of noise contributing below 1 Hz, building the detector farther away from coastal regions can be beneficial. The secondary microseismic peak in the band 0.1 to 1 Hz mostly originates from the nonlinear interaction between the incoming waves and those reflected by the coastal landmass \citep{Ardhuin2015}. Hence the origin is relatively local as compared to the primary microseism that is observed for frequencies between 0.05 and 0.1 Hz, which originates from deep-sea activities \citep{Hasselmann1963} and propagates as diving waves over large distances. Hence, reduction of noise below 0.1 Hz by building the detector away from coastal regions is unlikely. Another approach leverages the correlation between sea weather conditions and infrasound band seismic noise to predict the background seismic noise spectrum around 1 Hz several days in advance \cite{Bertoldi2023}. This prediction can help optimize instrument operation time and potentially facilitate the implementation of thresholding and mitigation measures during data analysis.

\subsection{Cavity Enhancement for Interferometry}
%Covered earlier? 
As discussed above, several different noise sources can impact the measurements of long baseline atom interferometers. 
These need to be distinguished from the desired signal, such as the gravitational wave signal. The difference in wavelength and characteristic length respectively between gravitational waves and gravity gradient noise allows separation between the two. 
This can be achieved by correlating distant  sensors~\cite{Chaibi2016}. By separating the sensors, information on the gravity gradient can be obtained and deployed to discriminate it from the desired gravitational wave signal.

To enhance the interferometric signal, optical cavities can be deployed. In a regular atom interferometric setup, the atoms are subjected to three laser pulses that are separated in time, as depicted in Fig.~\ref{fig:clockai}. For such setups, a mirror is placed perpendicular to the gravitational vector underneath the atom cloud. If the mirror is replaced by a cavity, the atoms are manipulated by the resulting standing light wave. In this case, the interaction between the atoms and the light is not controlled through intensity variations, but by modulating the coupling to the atoms~\cite{Bertoldi2021}.

\begin{figure}[htp]
\centering
\includegraphics[width= \textwidth]{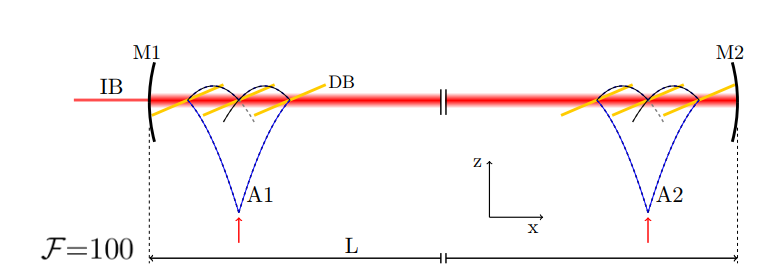}
\caption{Sketch of the deployment of a cavity for atom interferometry. Between the mirrors M1 and M2, a standing light wave is created that manipulates the atom cloud. In the sketch a scheme for two interferometers A1 and A2 separated by length L is depicted. This is a configuration that could be deployed in MIGA or ELGAR: see Sections~\ref{sec:MIGA} and \ref{sec:ELGAR}.}
\label{fig:Cavity_AI}
\end{figure}

Figure~\ref{fig:Cavity_AI} depicts the cavity and the path of the atoms during the interferometric process. 
As depicted, the atoms fall freely in the resonator's volume, which acts as the three individual laser pulses described above. 
In addition to the advantages of using the cavity, this allows multi-loop atom interferometric schemes using the Sagnac effect~\cite{Schubert2021}.
The deploment of an optical resonator gives rise to three main advantages:
\begin{itemize}
    \item Increased laser power at the atoms:
    Laser power is one of the defining factors, as it determines the momentum transfer range. 
Deploying a cavity increases the power and thereby the transferable momentum. 
In turn, the area under the interferometer increases, which enhances the accuracy of the measurement~\cite{Rudolph2020}.
Similarly, the available power and required accuracy determine the pulse length. 
Consequently, with increased power, the pulse length can be shortened while still achieving the same accuracy.
     \item Mode filtering: One of the defining features of an optical resonator is the innate mode filtering. 
     In the framework of atom interferometry this decreases the phase fluctuations and thereby enhances the signal-to-noise ratio in the final interferogram. 
     \item Phase and frequency noise control:
     Finally, the cavity stabilizes the laser to any given frequency. 
     Depending on the finesse, the phase and frequency noise of the laser can be reduced.  
\end{itemize}
In view of these benefits, it has to be mentioned that the achievable finesse of the cavity, especially over long distances, and the resulting pulse length, present challenges for their operation. 
The finesse of the cavity determines the bandwidth of the resulting standing light wave and influences the response time of the light to the auxiliary manipulation process~\cite{Bertoldi2021,Salvi2018}.
If, for example, the ELGAR~\cite{Canuel2020} configuration, with a baseline of $L \sim 10$~km is considered, a cavity with an achievable finesse of $F=100$ gives a bandwidth of $\Delta \nu = 150$~Hz and a pulse length of more than $50$~ms.

The numbers lead to limitations of the configuration~\cite{DovaleAlvarez2017}. If, for instance a cavity is stretched over a long distance, while maintaining the same waist, the system is limited by the small acceptable linewidth and, in consequence, the pulse length. At the other end of the spectrum of possibilities, a smaller cavity results in limitations of the cloud size. Thus, for every configuration, an optimal length exists. This configuration is determined by the geometric limit of the situation. However, the configuration can be optimized by usage of beam-shaping optics, preparing the laser for the atom interrogation optimally~\cite{Riou2017}.

\subsection{Tests of Local Position Invariance using additional Frequency References}
Metric theories of gravity~\cite{Will2014} such as general relativity~\cite{Einstein1911,Einstein1915}, are based on the Einstein Equivalence Principle (EEP) ~\cite{Will2014,DiCasola2015}.
Since atoms couple~\cite{Damour2010} to gravity, they may experience EEP violations, tests of such theories can be performed with the help of atom-based quantum sensors~\cite{Hohensee2013,Delva2018,Asenbaum2020,Lange2021}.
Two main directions have evolved for such experiments, namely atomic clocks~\cite{Brewer2019,McGrew2019} and atom interferometers~\cite{Kasevich1991,Cronin2009}.

Clocks based on superpositions of internal atomic (electronic) states are conventionally used for tests of Local Position Invariance (LPI) ~\cite{Delva2018,Lange2021}, one facet of the EEP.
In order to initialize such internal superpositions, one uses light pulses in the microwave~\cite{Kasevich1989} or optical~\cite{Yudin2010} frequency regime, where recoils can be suppressed either by the Lamb-Dicke~\cite{Lizuain2007} regime for traps or by using effectively recoilless pulses~\cite{Alden2014}. 
In this way, atomic clocks represent a localized frequency reference, reaching sensitivities for LPI violations at the order of $10^{-7}$~\cite{Lange2021}.

In contrast, light-pulse atom interferometers~\cite{Kasevich1991,Cronin2009}, based on delocalized spatial superpositions of atoms, are established inertial sensors~\cite{Riehle1991} able, e.g., to measure gravitational acceleration~\cite{Kasevich1991}.
These spatial superpositions can be created by using the effective recoil of light pulses~\cite{Rasel1995}.
As such, they are natural setups for tests of the universality of free fall as a facet of the EEP that is complementary to LPI.
These tests can be achieved by comparing the gravitational accelerations of two different test masses, represented either by two atomic species~\cite{Schlippert2014}, or two different isotopes~\cite{Asenbaum2020}, or even two different internal states of one atom~\cite{Zhang2020}.

In the proposed terrestrial 100\,m and 1\,km atom-interferometric gravitational-wave and dark-matter detectors, one uses single-photon transitions~\cite{Rudolph2020} to create spatial superpositions, which inherently address two internal states of the atom and induce transitions between them.
These different internal states also couple~\cite{Zych2011} to the centre-of-mass motion of the atom via a relativistic effect, called the mass defect~\cite{Sonnleitner2018,Schwartz2019}, which can encode the proper time of the centre of mass of the atom.
Moreover, it allows for atom interferometers to represent an inertial sensor and a frequency reference at the same time, leading to the concept of quantum-clock interferometry~\cite{Sinha2011,Zych2011,Loriani2019a}. 
As such, quantum-clock interferometry might lead to more abstract interferometer sequences~\cite{DiPumpo2021} than in the case of conventional clocks or atom interferometers.

Consequently, quantum-clock geometries were proposed~\cite{Roura2020,DiPumpo2021} where possible LPI violations are encoded into their signal in complete analogy to atomic clocks.
Normally, however, these atom-interferometric proposals can be estimated~\cite{DiPumpo2021} to be less sensitive to LPI violations compared to clocks when focusing on LPI tests via the gravitational redshift~\cite{Will2014,Delva2018}.
For such kinds of LPI tests large spatial separations between identical frequency references at different heights are crucial, and clocks can reach enormous distances up to thousands of kilometers~\cite{Delva2018}.
In contrast, with current baselines in the few-metre regime~\cite{Schlippert2014,Asenbaum2020} for light-pulse atom interferometers, quantum-clock schemes would be limited both in spatial separations between the branches and in interrogation times by the short free-fall time of about 1\,s to 2\,s in such setups.
Consequently, very-large-baseline experiments~\cite{schlippert2020matter} with longer free-fall times, as discussed in this article, would enhance both spatial separations between the branches and interrogation times of quantum-clock-based LPI tests.

Additionally, one can eliminate the clocks’ advantage of large spatial separations by testing LPI via clock rates~\cite{Will2014}, instead of the redshift.
Here, two different frequency references, implemented e.g., via different species or isotopes, are placed at the same height or are falling in parallel, and it is then measured whether the ticking rate depends on the implementations of the frequency references.

\begin{figure}
    \centering
    \includegraphics{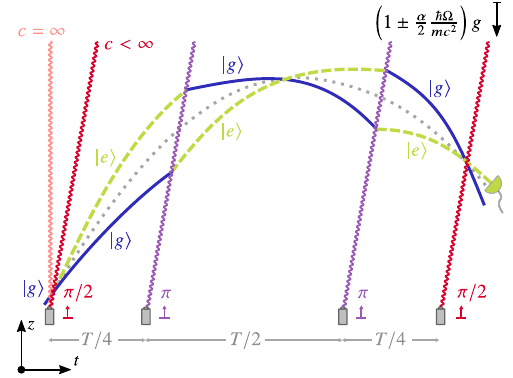}
    \caption{
    Quantum-clock scheme for LPI tests based on internal-state transitions.
    After the atom entered in the ground state $\ket{g}$, a $\pi/2$ pulse (red) brings it into a superposition of ground (blue solid line) and excited state $\ket{e}$ (green dashed line), where the finite speed $c$ of the laser light is depicted by an inclined line.
    The pulse also transfers a momentum $\hbar k$ to the excited state, e.g. induced by single-photon transitions, leading to a spatial superposition of the atom.
    After redirection via two internal-state changing $\pi$ pulses (purple) in time intervals $T/4$ and $3T/4$, the branches are brought to interference by the final $\pi/2$ pulse at interrogation time $T$, and the population in the excited state is detected.
    The experiment is performed in a linear gravitational field with mean acceleration $\textbf{g}$.
    To include possible LPI violations, the acceleration is augmented by the factor $1\pm\alpha\hbar\Omega/(2m c^2)$, including violation parameter $\alpha$, atomic transition frequency $\Omega$, and atomic mass $m$.
    This Figure was taken from~\cite{DiPumpo2023}.}
    \label{fig:t3-geometry}
\end{figure}
Consequently, the quantum-clock scheme depicted in Fig.~\ref{fig:t3-geometry} and testing LPI via clock rates was recently proposed~\cite{DiPumpo2023}, leading to projected sensitivities of the order of $10^{-7}$ for 3\,s of free-fall time, which is comparable to clocks.
Since this scheme is based on a sequence of internal transitions, it would fit perfectly with the proposed single-photon transitions planned for the terrestrial 100\,m and 1\,km setups.
Moreover, it was already estimated~\cite{DiPumpo2023} that effects like the finite speed of light and other deleterious effects impose manageable requirements.
Additionally, the sensitivity to LPI violations scales cubically in time in the proposed scheme, which amounts exactly to the difference between proper time and laboratory time.
Hence, due to the very long free-fall time of roughly 4\,s in 100\,m and 14\,s in 1\,km setups, the sensitivity would increase rapidly up to the order of $10^{-8}$ to $10^{-9}$, thus possibly outperforming all current tests.

However, a single experimental run of the proposed quantum-clock scheme does not lead to the desired clock-rate-based LPI test, since two different frequency references have to be compared.
The LPI test is provided by introducing an additional frequency reference, such as another atomic isotope or species, and performing the quantum-clock experiment with two frequency references simultaneously.
As proposed in~\cite{DiPumpo2023}, one could use different strontium isotopes with similar experimental requirements, fitting perfectly with the projected strontium sources for the 100\,m and 1\,km setups.

In summary, LPI tests with cubic time scaling would  supplement significantly the other science cases for terrestrial very-long-baseline experiments without requiring major additional technology development.

\subsection{Applications to Geodetic Research}
\label{Sec:geodres}
%\subsubsection{Understanding of Earths Gravitational Field and Dynamics}
%Here: We did not have the talk, but I still think it is an important science case to be gained from a TVLBAI (network)
% Jeremiah M: Could tie in here the Cambridge workshop investigating atom interferometry and Earth science
%\subsubsection{Early Warning System Potential}
%Here: I think adding some lines on the possibility of using a TVLBAI (network) as an Early Warning system for Earthquakes and / or Floods might be worth some space
As described at length earlier, atom interferometers are sensitive to accelerations. 
Prior to this Section, the usage of these systems primarily for gravitational wave detection and tests of fundamental physics have been discussed. 
In the spirit of this Section, here we note the application to geodetic research, studying the Earth's gravitational field and its dynamics.
The local gravitational field is relevant for monitoring the developments of the local height and the atmosphere.
If the quality of the recorded data is precise enough, it can also be used to determine SI units. 
The realization of the \si{\kilo\g} by means of a Watt balance requires precise knowledge of absolute \emph{g} in combination with a precise modelling of the local environment as well as relative gravimetric measurements to transfer the measured value of \emph{g} to the location of the test mass inside the Watt balance~\cite{Liard2014,Merlet2008}. 

With many applications of knowledge of the (local) gravitational field, cold atom interferometry has been investigated as a novel technique to improve on the quality of the available data. 
One of the major advantages of cold atoms is the (almost complete) absence of drift in the measurement, allowing for absolute comparability of experiments over time. 
Here, only those with relevance to terrestrial very long baseline atom interferometers are discussed.

To investigate the local gravitational field, small, transportable apparatuses were developed~\cite{McGuirk2002,Hu2017a,Farah2014}, with some commercial products being made available.  However, the sensitivity of the apparatus scales with the separation of the two measured atom clouds or the space traveled in the gravitational potential. Consequently, long baseline atom interferometers, such as the very long baseline atom interferometer, have spiked the interest of geological as well as fundamental researchers~\cite{Schilling2020}.

In order to deploy individual terrestrial very long baseline atom interferometers for geodesy or geophysical research, intimate knowledge of the system and its surroundings are necessary. This includes intricate modelling of the environment and the expected gravitational field. In addition to the quality of the model, the quality of the measurement also depends on the knowledge of changes to the surroundings.~\footnote{The movement of scientists was identified earlier as a noise source.} Here, in addition, knowledge of the presence of any mass, stationary or dynamical, is important, as it changes the environment and results in a deviation from the original model.

As an example of applications on a local scale, we now describe the operation of a vertical VLBAI (cf., Section~\ref{sec:vertical}) as a gravimeter. The current method of measuring \emph{g} and achieving SI traceability requires group comparisons of absolute gravimeters \cite{Schilling2016}, in which metrological institutes and users of absolute gravimeters meet and compare a series of measurements following a predefined protocol~\cite{Palinkas2021}. This method is necessary because users and metrological  institutes use the same instruments and no ``gravity standard'' of higher order exits. Regular participation in such comparisons every two to four years enables the users to monitor the stability of their instruments, which are typically deployed on measurement campaigns on a national to continental scale. The combination of different instruments across decades makes the close monitoring of the individual gravimeters performance mandatory~\cite{Olsson2019,BilkerKoivula2021}.

A higher-order gravity standard, even in stationary form, providing \emph{g} on demand (potentially as a service) for users in geodesy and geophysics, would directly contribute to the scientific goals and achievements of these users. Potentially, a vertical atom interferometer, such as devices like the \SI{10}{\meter} VLBAI in Hannover (cf., Section~\ref{Sec:VLBAI}) or the similar device in Wuhan (cf. Section~\ref{Sec:Wuhan10m}), could provide such a gravity reference.

However, the local gravity field needs to be well understood to provide a gravity reference for visiting absolute gravimeters. The simplest comparison between a VLBAI and a transportable absolute gravimeter would be the assumption of a constant offset stable over time (multiple visits over several years) between these devices. This would require a constant reference height of the VLBAI, as gravity changes with height, and the identical effect on gravity due to, e.g., changes in local hydrology, at the VLBAI and the instrument under test. Depending on the size of the VLBAI and the local conditions, the assumption of such a constant offset between two or more locations for the VLBAI and transportable absolute gravimeters is unlikely to be valid. Thus, the local gravity field needs to be modelled considering changes in the environment, e.g., in hydrology but also of heavy equipment in the laboratories. For the instrument in Hannover, the local gravity field is mapped with repeated surveys using transportable relative gravimeters and absolute gravity measurements provided by the local FG5X-220~\cite{Schilling2020}.
A model of the building and its interior was created and updated according to the progress of the installation of the VLBAI and other equipment in the building. Ground water gauges directly at the building monitor the variations of the local groundwater level so they can be considered in the gravity field model. The aforementioned gravimetric measurement campaigns are also used to validate this model. As a result of the modelling a height- and location-dependent gravity effect of groundwater changes due to the design of the building and the basements is expected. Due to a lack of groundwater recharge % The groundwater body has not rebuilt in recent years and is at an all time low 
in previous years, this effect has not been validated by measurements of the gravity field yet.

Once the system on a local scale is fully understood, it allows one to monitor the local gravitational field precisely over long periods of time. 
This includes measurements of tides, atmospheric mass changes, and dynamic processes in the Earth's mantle and the deep interior of the Earth.
The latter applications related to the internal structure of the Earth are currently the domain of superconducting gravimeters \cite{Goodkind1999}, which are the most precise relative gravimeters to date.
Superconducting gravimeters are commonly used in small networks, e.g. monitoring a volcano~\cite{Carbone2019} or networking on a global scale, e.g. in the search for dark matter~\cite{Hu2020}.
However, a lot of phenomena, e.g., the detection of Slichter modes, inner-core wobble or free core nutation, either require a large excitation of the Earth's body via a large-amplitude earthquake~\cite{Milyukov2023} or are beyond the current sensitivity of these instruments \cite{Rekier2022}.
Data from VLBAI's operated, for example, at national metrology institutes to realize the SI unit \emph{g}, could be reevaluated for these geodetic and geophysical applications.

With strategic site choices, already individual systems, can be deployed as early warning systems. 
Due to the repetition rate of cold atom interferometers, these are particularly sensitive to events during which the local gravitational field changes on time scales $\geq 1$~s. 
This category includes  flooding events and the monitoring of volcanic activities.
While probably not the primary area of application, it is possible to envisage such systems to predict avalanches and land slides. 
Due to the long integration time, a single system will not be capable of acting effectively  as an early warning system for earthquakes and similar events.

Such devices can in addition be deployed to monitor effects of climate change, such as sea level rises and glacial melting, and help fight the effects of climate change by monitoring ground water levels or the structural integrity of critical infrastructures.

With a large network of atom interferometers, such as proposed for ELGAR, some of the limitations on detectable frequencies and events could be lifted. 
Finally, a network of several independent atom interferometers could be capable of studying global effects with local precision. 
Differently from satellite-based geodesy, possible studies would benefit from the achievable accuracy of the local systems, while receiving data from a global network. 
Of course, no network of atom interferometers can ever achieve the global coverage a satellite has, but it benefits from the small distance to the target, stationary experiments, and higher separation baseline of the atoms.

Most likely, experimental data taken in the search for gravitational waves could be used for geophysical research, allowing a more in-depth investigation of the local environment.

\subsection{Molecular Interference}

Molecular interference is performed with a diferent aim from atom interference. 
Of course, these experiments could in principle, like cold atom interferometry, enable gravitational wave research, but the ultimate goal of molecular interferometry is mainly in the area of decoherence and exploring the limits of quantum mechanics~\cite{Carlesso2019,Bassi2017,Donadi2020,Carlesso2022}.~\footnote{See Section~\ref{sec:physics-qm} for a discussion of the interest in probes of quantum mechanics.}
Consequently, the mass of molecules being studied in interference experiments has increased over the past century.
The latest experiment in the sequence are molecules with more than 2000 individual atoms with a mass of around $27$~kDa~\cite{Fein2019}.

Differently from the atom interferometric experiments, molecular interference usually follows the logic of Young's double slit experiment. State-of-the art experiments, such as the Kapitza-Dirac-Talbot-Lau interferometer (KDTLI)~\cite{Eibenberger2013}, the optical time-domain ionizing matter-wave (OTIMA) interferometer~\cite{Haslinger2013}, and the long-baseline matter-wave interferometer (LUMI)~\cite{Fein2019a}, make use of the self-replicating effect of waves behind a grating, the so-called Talbot carpet~\cite{Talbot2009}.
This allows for much more compact setups and, in consequence, through the relation between wavelength and impulse, an increase in particle mass.
To make the most of the Talbot effect, these setups usually consist of three gratings equidistant in time or space. 
The first grating prepares the planar waves, the second grating is used for diffraction and the third grating for scanning through the Talbot carpet.

To increase the particle mass beyond $10^{6}$~Da, further tricks have to be deployed. 
At these masses, the residence time inside the interferometer expands into several seconds, during which the particles fall out of the active volume. 
This can be counteracted by designing missions for space, such as the MAQRO proposal~\cite{Kaltenbaek2023}. 
In the microgravity environment of space, the relative displacement of the particles to the interrogation lasers remains manageable, enabling research on large nanoparticles.

Alternatively, molecular fountains or freely-falling apparatuses could allow more prolonged free-fall times in the gravitational potential and thereby enable the study of more massive particles~\cite{Kialka2022}. Here, we focus on fountain-based setups and ignore the possibility of executing molecular interferometry in facilities such as the drop tower in Bremen or the Einstein Elevator in Hannover. A molecular fountain with a maximal height of $25$~m equipped with gratings produced by pulsed ultraviolet lasers at $175$~nm would enable the study of particles with a mass in the range of $10^{8}$~Da.

In order to ensure that any observed decoherence is caused by underlying physics, other decoherence channels have to be reduced. This includes the thermal and structural noise floor of the tower, external magnetic fields, the vacuum quality, and the relaxation of possible internal states. While these effects cannot be eliminated completely, proper engineering based on the experience and knowledge gained by the cold atom community, coupled with the deployment of advanced source techniques can reduce the impact. We note also that recent advances in cooling of the external states support the experimental road towards decoherence studies using molecules~\cite{Asenbaum2013, Delic2020}. 

Other applications of long-baseline molecular interferometry includes particle metrology and investigation of particle structure inaccessible to classical means~\cite{Eibenberger2013}. 
By applying external electric or magnetic fields, the resulting interferogram allows to distinguish between properties, such as dipoles, in otherwise identical particles. 
These leads to applications in chemistry, biology, and chemistry. 
Finally, it shall be mentioned, that people have been discussing to deploy interferometric decoherence to study further fundamental phenomena and processes~\cite{Kialka2022}. 

\subsection{Summary}
The topics covered in this section act as additional information to the topics discussed in the other sections of this paper. 
It links lessons learned from other long-baseline experiments to the presented ideas, describes additional means to improve the scientific signal, details other uses for the obtained data, and explains further developments, that benefit from the technology evolution expected from the operation of terrestrial long baseline atom interferometers. 
Consequently, the topics in this section are not connected to each other, but act as additional information to support a road map. 
Additionally, these are chosen to draw attention to necessary developments, opportunities, and challenges involved in constructing and operating complex systems.

\section{Conclusion}
\label{sec:con}
% \todo{OB: First draft -- this needs more work and expansion. }
As summarized here, the TVLBAI workshop has highlighted the emerging potential of Atom Interferometry (AI) as a highly promising field in fundamental physics and related applications. AI harnesses the principles of superposition and interference of atomic wave packets to achieve remarkable sensitivity to inertial and gravitational effects. In recent years the landscape of AI experiments has undergone rapid expansion and development, including advances in ultra-sensitive setups, portable devices and commercially available gravimeters.

The development of large-scale AI projects, including ongoing prototype detectors at the ${\cal O}(10)$-m scale, the construction of ${\cal O}(100)$-m experiments and proposals for kilometer-scale detectors, reflects the growing international interest and investment in this field. These initiatives aim to demonstrate the feasibility of implementing AI on a larger scale and lay the foundation for future exploration. Kilometer-scale detectors hold significant promise for sensitive investigations of gravitational waves in the unexplored deciHz frequency band and shedding light on the nature of dark matter, while simultaneously serving as technology testbeds for space-based AI missions.

The TVLBAI workshop brought together experts from diverse disciplines including cold atom physics, fundamental physics, astrophysics and cosmology, with a shared vision to drive the progress of large-scale AI projects. The TVLBAI Workshop facilitated successfully the convergence of these communities, and fostered discussions on establishing a roadmap for advancing this vision. This roadmap encompasses well-defined technological milestones and refined scientific goals, and underscores the importance of forging a proto-collaboration that works together and speaks with a common voice. This proto-collaboration will be essential for planning, promoting and realizing the world landscape of large-scale AI projects, enabling a structured approach to the realization of kilometer-scale detectors by the mid-2030s.

%\section{Road-Map}
%\label{sec:roadmap}
\begin{comment}

\section{Declarations}
 \subsection{Ethical Approval and Consent to participate}
 Not applicable 
 \subsection{Consent for publication}
Consents given by the authors at the time of submission
\subsection{Availability of supporting data}
Not applicable. For all requests relating to the paper, please contact author Oliver Buchmueller.
\subsection{Competing interests}
The authors declare that they have no competing interests.
\subsection{Funding}
\subsection{Authors' contributions}
The author Oliver Buchmueller is the contact person for this paper. The following authors are Section Editors and/or Workshop Organisers: 

Angelo~Bassi, Kai~Bongs, Philippe~Bouyer, Oliver~Buchmueller, Luigi~Cacciapuoti, Olivier~Carraz, Maria~Luisa~Chiofalo, A. Michael Cruise, Albert~De~Roeck, Michael~Doser, John~Ellis, Ren\'e~Forsberg, Naceur~Gaaloul, Richard~Hobson, Sebastian~Koke, Timothy~L.~Kovachy, Thomas~L\'ev\`eque, Christian~Lisdat, Federica~Migliaccio, Eamonn~Murphy, Ernst~M.~Rasel, Dennis~Schlippert, Stephan~Schiller, Ulrich~Schneider, Christian~Schubert, Carla~Signorini, Guglielmo~M.~Tino, Wolf~von~Klitzing, Eric~Wille, Peter~Wolf, Lisa~W\"orner.

All authors discussed the content of the paper and contributed to the writing of the manuscript. All authors read and approved the final manuscript.
\subsection{Acknowledgements}
We thank the CERN Quantum Technology Initiative for their support of the workshop that laid the basis for this paper.

\end{comment}

\bibliographystyle{JHEP}
%\bibliography{AION,library,bibsec9,Wuhan,NNC,Intro,main}
\bibliography{main}

\end{document}